\documentclass[preprint,12pt]{elsarticle}
\usepackage{subfigure}
\usepackage{lineno,hyperref}
\usepackage{indentfirst}
\usepackage{bm}
\usepackage{threeparttable}
\usepackage{multirow}
\usepackage{mathrsfs}
\usepackage{amsmath}
\usepackage{array}
\usepackage{amssymb}
\usepackage{graphicx}
\usepackage{url}
\usepackage{float}
\usepackage{epstopdf}
\usepackage{booktabs}
\usepackage{color}
\usepackage{lscape}
\usepackage{fancyhdr}
\usepackage{dsfont}
\usepackage[top=2.5cm, bottom=2.5cm, left=2cm, right=2cm]{geometry}
\modulolinenumbers[5]

\journal{?}

\biboptions{numbers,sort&compress}

\begin{document}

\begin{frontmatter}

\title{A unified gas-kinetic wave-particle method for multiscale gas-mixture flow with an elementary chemical reaction}

\author{Junzhe Cao$^a$}
\ead{jcaobb@connect.ust.hk}

\author{Yufeng Wei$^a$}
\ead{yweibe@connect.ust.hk}

\author{Wenpei Long$^a$}
\ead{wlongab@connect.ust.hk}

\author{Chengwen Zhong$^b$}
\ead{zhongcw@nwpu.edu.cn}

\author[]{Kun Xu$^{a,}$$^{c,}$$^{d,}$$^*$\corref{mycorrespondingauthor}}
\ead{makxu@ust.hk}

\address{$^a$Department of Mathematics, Hong Kong University of Science and Technology, Hong Kong, China\\
$^b$School of Aeronautics, Northwestern Polytechnical University, Xi'an, Shaanxi 710072, China\\
$^c$Department of Mechanical and Aerospace Engineering, Hong Kong University of Science and Technology, Hong Kong, China\\
$^d$HKUST Shenzhen Research Institute, Shenzhen, 518057, China}

\begin{abstract}
Hypersonic flows in the near space often couple continuum-rarefied multiscale effect with finite-rate chemistry. The linear constitutive relation in Navier--Stokes equations cannot describe the strong nonequilibrium effect, while transport-collision splitting Boltzmann solvers become inefficient in the near-continuum region, which necessitates the development of multiscale numerical methods. This paper extends the unified gas-kinetic wave-particle (UGKWP) method to multiscale gas mixture flows with a single elementary reaction. In the UGKWP method, hydrodynamic waves are employed to describe near-equilibrium distribution functions, and numerical particles are used for the evolution of nonequilibrium ones. The adaptive conversion between waves and particles, guided by the characteristic integral solution, together with the introduction of $\Delta t$ into the flux as an observation scale, has enabled the UGKWP method to succeed in many multiscale problems involving complex physics. In this work, rather than relying on a comprehensive reactive kinetic model for the entire distribution function, chemical source terms are first evaluated at the macroscopic level and then incorporated into the wave-particle update, while free-transport particles are kept chemically inactive in the monatomic setting considered here. This approach leverages the modeling advantages of wave-particle decoupling, facilitating extension to more complex chemical reactions. Moreover, an approximate extension of an advanced multispecies kinetic model is developed in this work for multispecies effect with species number larger than two. The present UGKWP method is assessed for the Zeldovich-type reaction $O_2+N\rightleftharpoons NO+O$ through hypersonic cylinder flows over a wide Knudsen number range, covering chemically inert, forward exothermic, forward endothermic and $\Delta E=0$ conditions, and through shock structures with hot upstream/downstream equilibrium states. Agreement with DSMC is obtained for gas mixture flow fields, species mole fractions and wall quantities. A three-dimensional side jet flow over a blunt cone is further simulated to demonstrate the three-dimensional capability of the present code. The present framework lays the foundation for further extensions that simultaneously account for multiple complex chemical reactions and for rotational and vibrational energy relaxation.
\end{abstract}

\begin{keyword}
Unified gas-kinetic wave-particle method; Multiscale flow; Gas mixture; Elementary chemical reaction
\end{keyword}

\end{frontmatter}

\section{Introduction}\label{Sec:introduction}
Chemical reactions are crucial to hypersonic flow, driving continuous innovation in this cutting-edge field~\cite{nrl,zxk,zzm,dins}. In the near space, chemical reactions are tightly coupled with the continuum-rarefied multiscale effect, posing further challenges for the accurate and efficient numerical simulation~\cite{martin,pgl,hqz,zsh,pfeiffer}. For instance, the strong nonequilibrium phenomenon introduced by rarefaction effects cannot be adequately described by the linear constitutive relations in the Navier--Stokes (NS) equations, necessitating the more fundamental Boltzmann equations. Meanwhile, for traditional transport-collision decoupling methods for solving Boltzmann equations, represented by the direct simulation Monte Carlo (DSMC) method~\cite{dsmc} and discrete velocity method (DVM)~\cite{dvm}, mesh size $h$ and time step $\Delta t$ are constrained by the microscopic molecular mean free path $\lambda$ and mean collision time $\tau$, leading to significant inefficiency in the near-continuum region. Consequently, it is essential to develop multiscale numerical methods for simulating the complex hypersonic flow coupling the chemical reaction effect together with the continuum-rarefied multiscale effect. The multiscale unified gas-kinetic scheme (UGKS)~\cite{ugks} and unified gas-kinetic wave-particle (UGKWP) method~\cite{wp1,wp2} have been successfully developed in multiple complex physics in recent years~\cite{wp-rad,wp-plasma,wp-turb,ugks-rad,ugks-reaction}. The former is a deterministic method and the latter is a stochastic particle version. In these methods, the characteristic integral solution is utilized to construct the flux in the finite volume method (FVM) framework~\cite{muscl}, where the collision is accurately coupled with the molecular transport, and the time step $\Delta t$ is introduced into the flux as an observation scale, achieving the direct modeling of physical laws in the discretized space~\cite{dm1,dm2}. When $\Delta t$ is much larger than $\tau$, both schemes recover the gas-kinetic scheme (GKS)~\cite{gks}, which is a NS solver and can overcome the limitation from $\lambda$ and $\tau$. When $\Delta t$ is much smaller than $\tau$, they become microscopic Boltzmann solvers. Because stochastic particles in the UGKWP method give a highly adaptive description of microscopic velocity space, it has significant advantages in efficiency and memory consumption when simulating high-speed flows. As a result, in this work we preliminarily extend the UGKWP method to the chemical reaction flow.

The construction of multiscale numerical methods for chemical reaction flow is an active and advanced research frontier. Most work is based on kinetic models, in which multispecies and chemical reaction effects are considered in a coupled approach. The recent extensions of both UGKS~\cite{ugks-reaction} and gas kinetic unified algorithm (GKUA)~\cite{gkua1,gkua2} are inherently reliant on the kinetic model proposed by Groppi et al.~\cite{groppi1,groppi2,groppi3}. Employing a comprehensive kinetic model offers advantages in achieving consistency across various physical treatments at the numerical method level, and these models are under continuous development~\cite{bisi1,bisi2}. However, it remains very challenging to generalize to more complex chemical reactions, including the simultaneous consideration of multiple kinds of reactions, coupling with rotational and vibrational effects of polyatomic molecules, and integration with novel multi-species models. This work uses a simpler approach, similar to Ref.~\cite{tuttas}. Our present method is not explicitly dependent on chemical-reaction kinetic models, and does not operate on the whole microscopic gas distribution function for the chemical reaction effect. Instead, the effects of chemical reactions are first applied to macroscopic variables. Then, leveraging the rich microscopic information and flexibility contained in particles, we consider the further impact of chemical reactions through aspects such as particle free transport time, and whether particles are kept or deleted. The benefit of this approach is that chemical reactions can be applied once key parameters, such as reaction rates, are available. These rates may be obtained from the classical Arrhenius equation or from more advanced models that incorporate particle non-equilibrium information. In this way, the treatment of chemical reactions is simplified, without requiring a detailed kinetic model to be developed in advance. As a preliminary investigation, this work considers only a single elementary reaction and treats all gases as monatomic, while substantial potential remains for extension to more realistic and complex scenarios. For the multispecies effect, the UGKWP method proposed in Ref.~\cite{wp-binary} is used with an advanced multispecies kinetic model~\cite{groppi-2,todorova-2}, which is further approximately extended in this work from binary species to more species.

There are two other classes of approaches to simulate multiscale chemical reaction flow. The first one includes domain decoupling methods between DSMC and NS solvers~\cite{mpc1,mpc2}. This approach is well-developed and can effectively take advantage of the strengths of both NS and DSMC. However, the buffer zone needs to achieve accuracy for both DSMC and NS simultaneously, which requires DSMC to use excessively fine mesh. The continuum-rarefied criterion is also difficult to satisfy appropriately, further increasing the computational cost of DSMC. The second approach is the direct intermittent GSIS-DSMC coupling (DIG)~\cite{dig}, which uses NS-like inner iterations to accelerate DSMC. It offers significant potential for improving the efficiency of particle-based methods. However, overcoming the statistical noise of particles is a challenge when providing the microscopic nonequilibrium information of particles to the NS solver.

The remainder of this paper is organized as follows. Section~\ref{sec:ugkwp} introduces the formulation of the UGKWP method for multiscale gas-mixture flow with an elementary chemical reaction. Numerical test cases are presented in Sec.\ref{sec:cases}, and concluding remarks are provided in Sec.\ref{sec:conclusion}.

\section{UGKWP method}\label{sec:ugkwp}
\subsection{Discretized governing equations in the finite volume method framework}\label{sec:discretization}
In the general FVM framework, for the microscopic gas distribution function, the discretized governing equation is as follows,
\begin{equation}\label{eq:conservation-micro}
f^{n+1}_{\alpha,i} = f^{n}_{\alpha,i} - \frac{1}{\Omega_i}\sum\limits_{j\in \mathcal{M}\left(i\right)}\int^{\Delta t}_0\boldsymbol{u}\cdot\boldsymbol{n}_jf_{\alpha,j}s_j{\rm d}t + \int^{\Delta t}_0\mathcal{J}_{\alpha,i}{\rm d}t,
\end{equation}
where $\alpha\in\left\{1,2,3,4\right\}$ is the label of species, ``$i$'' is the label of cell, $\mathcal{M}\left(i\right)$ is a collection of interfaces surrounding cell ``$i$'', $\Omega_i$ is the area or volume of cell ``$i$'', $s_j$ is the length or area of interface ``$j$'', $\boldsymbol{n}_j$ is the outer unit normal vector, $\Delta t$ is the time step. $f_{\alpha}\left(\rho,\boldsymbol{x},\boldsymbol{u},\boldsymbol{\xi},t\right)$ is the microscopic gas distribution function of species $\alpha$, where $\boldsymbol{x}$ denotes the position, $t$ denotes the time, $\boldsymbol{u}$ denotes the molecular velocity and internal energy is represented by the equivalent velocity $\boldsymbol{\xi}$ in length $D$. $\mathcal{J}_{\alpha}$ is the microscopic collision term considering both multispecies effect and chemical reaction. The macroscopic conserved variable vector $\boldsymbol{W}_{\alpha}=\left(\rho_{\alpha}, \rho_{\alpha} \boldsymbol{U}_{\alpha}, \rho_{\alpha} E_{\alpha}\right)^T$ can be directly obtained from the distribution function $f_{\alpha}$ by calculating moments as follows, where $\rho$ is density, $\boldsymbol{U}$ is macroscopic velocity, $E$ is specific total energy.
\begin{equation}
\begin{aligned}
&\boldsymbol{W}_{\alpha} = \int_{\mathbb{R}^D}\int_{\mathbb{R}^3} \boldsymbol{\Psi}f_{\alpha} {\rm d}\boldsymbol{u} {\rm d}\boldsymbol{\xi},\\
&\boldsymbol{\Psi}=\left({ 1,\boldsymbol{u}, {1\over 2}\left|\boldsymbol{u}\right|^2+{1\over 2}\left|\boldsymbol{\xi}\right|^2 }\right)^T.
\end{aligned}
\nonumber
\end{equation}
The heat flux vector $\boldsymbol{Q}_{\alpha}$ can also be obtained by the following moments,
\begin{equation}
{\boldsymbol{Q}_{\alpha}} = \int_{\mathbb{R}^D}\int_{\mathbb{R}^3} \left(\boldsymbol{u}-\boldsymbol{U}_{\alpha}\right)\left[{ {1\over 2}\left({ \left|\boldsymbol{u}-\boldsymbol{U}_{\alpha}\right|^2 + \left|\boldsymbol{\xi}\right|^2 }\right)f_{\alpha} }\right] {\rm d}\boldsymbol{u} {\rm d}\boldsymbol{\xi}.
\nonumber
\end{equation}
On the other hand, regarding the macroscopic governing equation, an operator splitting is implemented before the discretization as follows, for a better consistency of microscopic particles with macroscopic variables, which will be illustrated in Sec.\ref{sec:source}.
\begin{equation}
\begin{aligned}
&\frac{\partial\boldsymbol{W}_{\alpha}}{\partial t}+\nabla\cdot\boldsymbol{F}_{\alpha}= {\boldsymbol{S}}_{\alpha}^{C}+{\boldsymbol{S}}_{\alpha}^{S}\\
\Rightarrow&\left\{ \renewcommand{\arraystretch}{2}\begin{array}{l}
\dfrac{\partial\boldsymbol{W}_{\alpha}}{\partial t}+\nabla\cdot\boldsymbol{F}_{\alpha}= \boldsymbol{0},\\
\dfrac{\partial\boldsymbol{W}_{\alpha}}{\partial t}= {\boldsymbol{S}}_{\alpha}^{C},\\
\dfrac{\partial\boldsymbol{W}_{\alpha}}{\partial t}= {\boldsymbol{S}}_{\alpha}^{S},
\end{array}\right.\renewcommand{\arraystretch}{1}
\end{aligned}
\nonumber
\end{equation}
where source term vector ${\boldsymbol{S}}_{\alpha}^{C}$ denotes the effect of chemical reaction, and ${\boldsymbol{S}}_{\alpha}^{S}$ is for the exchanged momentum and energy between different species. Their discretized governing equations are,
\begin{equation}\label{eq:conservation-macro}
\boldsymbol{W}^{\ast}_{\alpha,i} = \boldsymbol{W}^{n}_{\alpha,i} - \frac{1}{\Omega_i}\sum\limits_{j\in \mathcal{M}\left(i\right)}\boldsymbol{F}_{\alpha,j}s_j,
\end{equation}
\begin{equation}\label{eq:conservation-macrosource-c}
\frac{\partial\boldsymbol{W}_{\alpha,i}}{\partial t}= {\boldsymbol{S}}_{\alpha,i}^{C},
\end{equation}
and,
\begin{equation}\label{eq:conservation-macrosource}
\frac{\partial\boldsymbol{W}_{\alpha,i}}{\partial t}= {\boldsymbol{S}}_{\alpha,i}^{S},
\end{equation}
where macroscopic flux $\boldsymbol{F}_{\alpha,j}$ is calculated by moments of $f_{\alpha,j}$ at the interface,
\begin{equation}\label{eq:macrof}
\boldsymbol{F}_{\alpha,j} = \int_{\mathbb{R}^D} \int_{\mathbb{R}^3}\int^{\Delta t}_0\boldsymbol{u}\cdot\boldsymbol{n}_jf_{\alpha,j}\boldsymbol{\Psi} {\rm d}t{\rm d}\boldsymbol{u}{\rm d}\boldsymbol{\xi}.
\end{equation}
Eq.~\eqref{eq:conservation-macro}, Eq.~\eqref{eq:conservation-macrosource-c}, and Eq.~\eqref{eq:conservation-macrosource} will be solved sequentially for the remainder of Sec.~\ref{sec:ugkwp}, together with the microscopic gas distribution function.

\subsection{Kinetic model for multispecies flow}
In this section, the kinetic model is introduced for further calculating the source term and the fluxes of gas distribution function $f_{\alpha,j}$ at the interface as Eq.~\eqref{eq:conservation-micro} and Eq.~\eqref{eq:macrof}. We employ the single-collision operator in the type of Bhatnagar--Gross--Krook (BGK) model~\cite{bgk} in this study as,
\begin{equation}\label{eq:bgk}
\begin{aligned}
& \frac{\partial f_{\alpha}}{\partial t}+\boldsymbol{u}\cdot\frac{\partial f_{\alpha}}{\partial \boldsymbol{x}}=\frac{g_{\alpha}-f_{\alpha}}{\tau_0},\\
& g_{\alpha}=g^M_{\alpha}C^S_{\alpha},\\
& g^{M}_{\alpha}=\rho_{\alpha}\left( {\frac{1}{{2\pi R_{\alpha}\tilde{T}_{\alpha}}}} \right)^{\frac{3}{2}}\exp \left(  - \frac{\left|\boldsymbol{u}-\tilde{\boldsymbol{U}}_{\alpha}\right|^2}{2R_{\alpha}\tilde{T}_{\alpha}} \right),\\
& C^S_{\alpha}={1+(1-{\rm{Pr_{0}}})\frac{\left(\boldsymbol{u}-\tilde{\boldsymbol{U}}_{\alpha}\right)\cdot\boldsymbol{Q}_{\alpha}}{5\tilde{p}_{\alpha}R_{\alpha}\tilde{T}_{\alpha}}\left( {\frac{\left|\boldsymbol{u}-\tilde{\boldsymbol{U}}_{\alpha}\right|^2}{R_{\alpha}\tilde{T}_{\alpha}}-5} \right)},
\end{aligned}
\end{equation}
where $R_{\alpha}=k_B/m_{\alpha}$ is the specific gas constant of species $\alpha$, $k_B=1.380649\times10^{-23}J/K$ is Boltzmann constant, $m_{\alpha}$ is the molecular mass, $T$ is the macroscopic temperature, $p_{\alpha}=\rho_{\alpha} R_{\alpha} T_{\alpha}$ is the species pressure, $\tau$ is the relaxation time and $\mu$ is the viscosity coefficient, which will be introduced together with the heat conduction coefficient and ${\rm{Pr}}_0$ in detail from Eq.~\eqref{eq:coeffs}. The target near-equilibrium gas distribution function $g_{\alpha}$ contains two parts. The first part $g^{M}_{\alpha}$ is in the form of Maxwellian distribution. In this work, we consider only monatomic gas in three dimensional physical space so internal energy is not introduced. Different from variables at the current state, we use tilde to denote the target macroscopic variables $\tilde{\boldsymbol{U}}_{\alpha}$ and $\tilde{T}_{\alpha}$, which approximately recover the diffusion effect. The second part of $g_{\alpha}$ in Eq.~\eqref{eq:bgk}, $C^S_{\alpha}$, is for the correction of ${\rm{Pr}}_0$ in the approach of Shakhov model~\cite{shakhov}. For binary-species gas mixture, by introducing a second relaxation parameter, the target macroscopic variables $\tilde{\boldsymbol{U}}_{\alpha}$ and $\tilde{T}_{\alpha}$ are modeled by Groppi et al.~\cite{groppi-2}, allowing the kinetic model to recover both the shear viscosity and diffusion coefficients in the continuum flow regime. On this base, Todorova et al.~\cite{todorova-2} further add the $C^S_{\alpha}$ term in the approach of Shakhov model, to enable the heat conduction coefficient derived by Chapman--Enskog expansion consistent with the NS equations. The kinetic model in Ref.~\cite{todorova-2} has been developed in the UGKWP method with detailed validations~\cite{wp-binary}. In this work, the model is extended to a larger species number ($C>2$) under the condition that the mass ratio is not extreme. The consistency of transport coefficients are analyzed in Appendix A, and hereby we introduce the model first. The target macroscopic variables $\tilde{\boldsymbol{U}}_{\alpha}$ and $\tilde{T}_{\alpha}$ are calculated by,
\begin{equation}\label{eq:groppi}
\begin{aligned}
& \tilde{\boldsymbol{U}}_{\alpha}=\left(1-\vartheta_{\alpha}\right)\boldsymbol{U}_{\alpha}+\vartheta_{\alpha}\hat{\boldsymbol{U}}_{0},\\
& \tilde{T}_{\alpha}=T_{0}-\frac{1}{3n_0k_B}\sum\limits_{\beta=1}^C\rho_{\beta}\left|\tilde{\boldsymbol{U}}_{\beta}-\boldsymbol{U}_{0}\right|^2,
\end{aligned}
\end{equation}
where $\beta$ is also a label of species, to differentiate from $\alpha$, $n_{\alpha}=\rho_{\alpha}/m_{\alpha}$ is the number density. The subscript ``$0$'' denotes the macroscopic variable of gas mixture, such as,
\begin{equation}\label{eq:gasmixture}
\begin{aligned}
& n_{0}=\sum\limits_{\alpha=1}^Cn_{\alpha},\quad \rho_{0}=\sum\limits_{\alpha=1}^C\rho_{\alpha}, \quad m_0=\frac{\rho_0}{n_0}=\sum\limits_{\alpha=1}^C\chi_{\alpha}m_{\alpha},\\
& \rho_{0}\boldsymbol{U}_{0}=\sum\limits_{\alpha=1}^C\rho_{\alpha}\boldsymbol{U}_{\alpha},\quad \rho_{0}E_{0}=\sum\limits_{\alpha=1}^C\rho_{\alpha}E_{\alpha},\\
&
\nonumber
\frac{3}{2}n_0k_BT_{0}=\sum\limits_{\alpha=1}^C{\frac{3}{2}n_{\alpha}k_BT_{\alpha}}+\frac{1}{2}\sum\limits_{\alpha=1}^C{\rho_{\alpha}\left|\boldsymbol{U}_{\alpha}-\boldsymbol{U}_{0}\right|^2},
\end{aligned}
\end{equation}
where $\chi_{\alpha}=\frac{n_{\alpha}}{n_0}$ is the mole fraction, $C$ is the species number which is set to be $4$ in this work, $T_{\alpha}$ is the species temperature calculated as,
\begin{equation}
T_{\alpha}=\frac{1}{3\rho_{\alpha}R_{\alpha}}\int_{\mathbb{R}^3} \left|\boldsymbol{u}-\boldsymbol{U}_{\alpha}\right|^2f_{\alpha}{\rm d}\boldsymbol{u}=\frac{2E_{\alpha}-\left|\boldsymbol{U}_{\alpha}\right|^2}{3R_{\alpha}},
\nonumber
\end{equation}
and $\hat{\boldsymbol{U}}_{0}$ is specially defined by,
\begin{equation}\label{eq:u2}
\hat{\boldsymbol{U}}_{0}=\frac{1}{n_0}\sum\limits_{\alpha=1}^Cn_{\alpha}\boldsymbol{U}_{\alpha}.
\end{equation}
Then, the diffusion coefficient is approximately recovered by setting,
\begin{equation}\label{eq:theta}
\vartheta_{\alpha}=\frac{5}{6A^{\ast}}\frac{m_0}{m_{\alpha}},
\end{equation}
where the values of $A^{\ast}$ between different gases are given in Ref.~\cite{weia} and a suggested value is $1.11$~\cite{todorova-2}. In this kinetic model, each species shares the same $\tau$ and ${\rm{Pr}}$, which is calculated by,
\begin{equation}\label{eq:coeffs}
\tau_0=\frac{\mu_0}{n_0k_BT_0},\quad {\rm{Pr}}_0=\frac{C_{p,0}\mu_0}{k_0},
\end{equation}
where $C_{p,0}=\frac{5}{2}\frac{k_B}{m_0}$ is the specific heat capacity of gas mixture at constant pressure~\cite{todorova-2} and $k$ is the heat conduction coefficient. The viscosity and heat conduction coefficients of gas mixture are calculated by Wilke's model and Wassiljewa's model respectively as follows~\cite{wilke,was1,was2,coeff},
\begin{equation}
\begin{aligned}
\mu_0=\sum\limits_{\alpha=1}^C{\frac{\chi_{\alpha}\mu_{\alpha}}{\sum\limits_{\beta=1}^C{\chi_{\beta}\frac{\left(1+\sqrt{\frac{\mu_{\alpha}}{\mu_{\beta}}}\sqrt[4]{\frac{m_{\beta}}{m_{\alpha}}}\right)^2}{\sqrt{8\left(1+\frac{m_{\alpha}}{m_{\beta}}\right)}}}}},\quad k_0=\sum\limits_{\alpha=1}^C{\frac{\chi_{\alpha}k_{\alpha}}{\sum\limits_{\beta=1}^C{\chi_{\beta}\frac{\left(1+\sqrt{\frac{k_{\alpha}}{k_{\beta}}}\sqrt[4]{\frac{m_{\beta}}{m_{\alpha}}}\right)^2}{\sqrt{8\left(1+\frac{m_{\alpha}}{m_{\beta}}\right)}}}}}.
\nonumber
\end{aligned}
\end{equation}
For each species the viscosity and heat conduction coefficients are calculated as,
\begin{equation}
\begin{aligned}
\mu_{\alpha}=\mu_{\alpha,{\rm{ref}}}\left(\frac{T_{\alpha}}{T_{\alpha,{\rm{ref}}}}\right)^{\omega},\quad k_{\alpha}=\frac{C_{p,\alpha}\mu_{\alpha}}{{\rm{Pr}_{\alpha}}},
\nonumber
\end{aligned}
\end{equation}
where subscript ``$\rm{ref}$'' denotes the reference value, $\omega$ is the viscosity index corresponding to variable hard sphere (VHS) model, $C_{p,\alpha}=\frac{5}{2}\frac{k_B}{m_{\alpha}}$ and for the monatomic gas in this work ${\rm{Pr}_{\alpha}}$ is set to be $2/3$.

\subsection{Integral solution along the characteristic line}
Considering the coordinate origin set at the center of an interface, we can derive the integral solution along the characteristic line as,
\begin{equation}\label{eq:integral}
f_{\alpha}\left( {\boldsymbol{0},t} \right) = \frac{1}{\tau_0}\int^t_0 g_{\alpha}\left[{ -\boldsymbol{u}\left({ t-\hat{t} }\right),\hat{t} }\right]e^{\frac{\hat{t}-t}{\tau_0}} d\hat{t} + e^{-t/\tau_0}f_{\alpha}\left( {-\boldsymbol{u}t,0} \right),
\end{equation}
where the free transport is coupled with the collision. The first term denotes the accumulation of target near-equilibrium distribution. The second term denotes the transport and decay of initial non-equilibrium state. Because operator splitting is used to derive $g_{\alpha}$, the chemical reaction effect is considered within the integral solution as well, whose details will be discussed in Sec.~\ref{sec:source-c} and Sec.~\ref{sec:source}. Expanding $g_{\alpha}$ and $f_{\alpha}$ as,
\begin{equation}\label{eq:taylor}
\begin{aligned}
g_{\alpha}\left({ \boldsymbol{x},t }\right)&=g_{\alpha}\left({ \boldsymbol{0},0 }\right)+\frac{\partial g_{\alpha}}{\partial \boldsymbol{x}}\cdot\boldsymbol{x}+\frac{\partial g_{\alpha}}{\partial t}t,\\
f_{\alpha}\left({ \boldsymbol{x},0 }\right)&=f_{\alpha}\left({ \boldsymbol{0},0 }\right)+\frac{\partial f_{\alpha}}{\partial \boldsymbol{x}}\cdot\boldsymbol{x},
\nonumber
\end{aligned}
\end{equation}
it can be derived from Eq.~\eqref{eq:integral} that,
\begin{equation}\label{eq:integrala1}
f_{\alpha}\left( {\boldsymbol{0},t} \right) = \varepsilon_ag_{\alpha}\left({ \boldsymbol{0},0 }\right) + \varepsilon_b\frac{\partial g_{\alpha}}{\partial \boldsymbol{x}}\cdot\boldsymbol{u}+\varepsilon_c\frac{\partial g_{\alpha}}{\partial t}+\varepsilon_df_{\alpha}\left({ \boldsymbol{0},0 }\right) + \varepsilon_e\frac{\partial f_{\alpha}}{\partial \boldsymbol{x}}\cdot\boldsymbol{u},
\end{equation}
where,
\begin{equation}\label{eq:integrala2}
\begin{aligned}
\varepsilon_a &= 1-e^{-t/\tau_0},\\
\varepsilon_b &= te^{-t/\tau_0}-\tau_0\left(1-e^{-t/\tau_0}\right),\\
\varepsilon_c &= t-\tau_0\left(1-e^{-t/\tau_0}\right),\\
\varepsilon_d &= e^{-t/\tau_0},\\
\varepsilon_e &= -te^{-t/\tau_0}.
\nonumber
\end{aligned}
\end{equation}
Then by substituting Eq.~\eqref{eq:integrala1} into Eq.~\eqref{eq:macrof}, the macroscopic flux can be derived as,
\begin{equation}\label{eq:integralb1}
\boldsymbol{F}_{\alpha} = \boldsymbol{F}_{\alpha}^{eq}+\boldsymbol{F}_{\alpha}^{fr},
\end{equation}
where,
\begin{equation}\label{eq:integralb2}
\begin{aligned}
\boldsymbol{F}_{\alpha}^{eq} =& \int_{\mathbb{R}^3}{\boldsymbol{\Psi}\left(\delta_ag_{\alpha}\left({ \boldsymbol{0},0 }\right)+\delta_b\frac{\partial g_{\alpha}}{\partial \boldsymbol{x}}\cdot\boldsymbol{u}+ \delta_c\frac{\partial g_{\alpha}}{\partial t} \right)\left(\boldsymbol{u}\cdot\boldsymbol{n}\right) {\rm d}\boldsymbol{u}},\\
\boldsymbol{F}_{\alpha}^{fr} =& \int_{\mathbb{R}^3}{ \boldsymbol{\Psi}\left(\delta_df_{\alpha}\left({ \boldsymbol{0},0 }\right)+\delta_e\frac{\partial f_{\alpha}}{\partial \boldsymbol{x}}\cdot\boldsymbol{u} \right)} \left(\boldsymbol{u}\cdot\boldsymbol{n}\right) {\rm d}\boldsymbol{u},
\nonumber
\end{aligned}
\end{equation}
and,
\begin{equation}\label{eq:integralb3}
\begin{aligned}
\delta_a &= \Delta t-\tau_0\left( {1-e^{-\Delta t/\tau_0}} \right),\\
\delta_b &= 2\tau_0^2\left( {1-e^{-\Delta t/\tau_0}} \right)-\tau_0\Delta t-\tau_0\Delta te^{-\Delta t/\tau_0},\\
\delta_c &= \Delta t^2/2-\tau_0\Delta t+\tau_0^2\left( {1-e^{-\Delta t/\tau_0}} \right),\\
\delta_d &= \tau_0\left( {1-e^{-\Delta t/\tau_0}} \right),\\
\delta_e &= \tau_0\Delta te^{-\Delta t/\tau_0}-\tau_0^2\left( {1-e^{-\Delta t/\tau_0}} \right).
\nonumber
\end{aligned}
\end{equation}
In the continuum flow regime, the $\boldsymbol{F}^{eq}_{\alpha}$ term in Eq.~\eqref{eq:integralb1} occupies a dominant position, where the gas evolution is constructed by deterministic hydrodynamic waves. When the flow gets rarefied, the $\boldsymbol{F}^{fr}_{\alpha}$ term gradually plays the main role, and particles are employed for this part of simulation. The adaptive decomposition of waves and particles not only offers an accurate and efficient scheme for the numerical simulation of multiscale physics, but also provides a deeper perspective for analyzing and constructing multiscale kinetic equations~\cite{wp-6,ugkf,wpd}.

\subsection{Wave evolution}\label{sec:macro}
For the first term in Eq.~\eqref{eq:integralb1}, $\boldsymbol{F}^{eq}$, the method in Ref.~\cite{awp2} is used to improve the accuracy and robustness in the flow field regions of drastic scale variation. The scale-related coefficient $\delta$ is separated into left and right sides of the interface ``$j$'', by using the cell-center $\tau_{L/R}$. Because $\delta_{a,L/R}$, $\delta_{b,L/R}$ and $\delta_{c,L/R}$ used by the near-equilibrium flux are calculated within the cell instead of at the interface, they are more consistent with the scale-related coefficient $e^{-\Delta t/\tau}$ used in the particle sampling, which is calculated at the same location. Further illustrations can be found in Ref.~\cite{awp2}, and the resultant equation is as follows,
\begin{equation}\label{eq:feq}
\begin{aligned}
\boldsymbol{F}^{eq}_{\alpha,j} =& \int_{\boldsymbol{u}\cdot\boldsymbol{n}_j>0}{\boldsymbol{\Psi}\left(\delta_{a,L}g_{\alpha}\left({ \boldsymbol{0},0 }\right)+\delta_{b,L}\frac{\partial g_{\alpha}}{\partial \boldsymbol{x}}\cdot\boldsymbol{u}+ \delta_{c,L}\frac{\partial g_{\alpha}}{\partial t} \right)\left(\boldsymbol{u}\cdot\boldsymbol{n}_j\right) {\rm d}\boldsymbol{u}}\\
+& \int_{\boldsymbol{u}\cdot\boldsymbol{n}_j<0}{\boldsymbol{\Psi}\left(\delta_{a,R}g_{\alpha}\left({ \boldsymbol{0},0 }\right)+\delta_{b,R}\frac{\partial g_{\alpha}}{\partial \boldsymbol{x}}\cdot\boldsymbol{u}+ \delta_{c,R}\frac{\partial g_{\alpha}}{\partial t} \right)\left(\boldsymbol{u}\cdot\boldsymbol{n}_j\right) {\rm d}\boldsymbol{u}}.
\end{aligned}
\end{equation}
Further details about $g_{\alpha}\left({ \boldsymbol{0},0 }\right)$, $\frac{\partial g_{\alpha}}{\partial \boldsymbol{x}}$, $\frac{\partial g_{\alpha}}{\partial t}$, and moment calculation can be found in Refs.~\cite{gks,wp-binary}.

\subsection{Particle evolution}\label{sec:micro}
In the UGKWP method, we employ numerical particles to describe the free-transport part of gas distribution function. The particle parameters are: species label $\alpha$, particle mass $m$, location $\boldsymbol{x}$ and velocity $\boldsymbol{u}$. As Eq.~\eqref{eq:integral}, the distribution function after $\Delta t$ is a combination of the initial state $f_{\alpha}$ and target near-equilibrium state $g_{\alpha}$. Firstly, for a particle from the initial state, the probability for keeping free-transport is $e^{-\frac{\Delta t}{\tau_0}}$, otherwise it collides with other particles and is relaxed into $g_{\alpha}$. As a result, the cumulative distribution of a free-transport particle is $e^{-\frac{\Delta t}{\tau_0}}$. By taking a uniform random number $\epsilon\in\left(0,1\right)$, the free-transport time $t_{f,k}$ of a particle labeled by ``$k$'' can be calculated by,
\begin{equation}\label{eq:mic1}
t_{f,k} = {\rm{min}}\left(-\tau_{0,k}{\rm{ln}}\left(\epsilon\right),\Delta t\right).
\nonumber
\end{equation}
Then the location of the particle can be renewed by,
\begin{equation}\label{eq:mic2}
\boldsymbol{x}_k \Rightarrow \boldsymbol{x}_k + \boldsymbol{u}_kt_{f,k},
\end{equation}
and the contribution of all particles to the macroscopic conserved variables is,
\begin{equation}\label{eq:mic3}
\boldsymbol{W}_{\alpha,i}^{fr,p} = \boldsymbol{W}_{\alpha,i}^{p,+}-\boldsymbol{W}_{\alpha,i}^{p,-},
\end{equation}
and,
\begin{equation}\label{eq:mic4}
\boldsymbol{W}_{\alpha,i}^{p,+/-} = \frac{1}{\Omega_i}\sum\limits_{k\in \mathcal{N}^{+/-}_{\alpha}\left(i\right)}m_{p,k}\boldsymbol{\Psi}_k,
\nonumber
\end{equation}
where $\mathcal{N}_{\alpha}\left(i\right)$ is a collection of particles of species $\alpha$ within cell ``$i$''. Superscripts ``$-$'' and ``$+$'' denote the time before and after the free transport, respectively. Then, after calculating $\boldsymbol{W}_{\alpha,i}^{fr,p}$, if $t_{f,k}<\Delta t$, the particle is deleted, relaxing into $g_{\alpha}$, called collisional particles. Otherwise, if $t_{f,k}=\Delta t$, the particle will be kept, called collisionless particles.

Furthermore, when calculating $t_{f,k}$, there is more freedom to improve the microscopic model in the premise of conservation. In this study, the free-transport time is reduced for high-speed particles as Ref.~\cite{wp-binary}, which can recover the single-species case in Ref.~\cite{xu-tau}. This primarily breaks through the limitation that particles with different speeds have the same collision frequency. Particularly when the thermal speed of the particles is large, their collision frequency should significantly increase. The following treatment on calculating free transport time can improve high-Ma case results such as the temperature profile in the pre-shock location and the heat flux at the stagnation point.
\begin{equation}\label{eq:taustar}
t_{f,k,\alpha}={\rm{min}}\left(-\tau^{\ast}_{\alpha,k}{\rm{ln}}\left(\epsilon\right),\Delta t\right),
\end{equation}
and,
\begin{equation}
\begin{aligned}
\tau^{\ast}_{\alpha,k}=\left\{ \begin{array}{ll}
\tau_{0,k}, & \left|\boldsymbol{u}_{k}-\boldsymbol{U}_{\alpha}\right|\leq b\sqrt{R_{\alpha}T_{\alpha}},\\
\frac{1}{1+\sum\limits_{\beta=1}^C{c_{\alpha\beta}}}\tau_{0,k}, & \left|\boldsymbol{u}_{k}-\boldsymbol{U}_{\alpha}\right|> b\sqrt{R_{\alpha}T_{\alpha}},
\end{array}\right.
\end{aligned}
\nonumber
\end{equation}
where,
\begin{equation}
\begin{aligned}
c_{\alpha\beta}=\left\{ \begin{array}{ll}
0, & \left|\boldsymbol{u}_{k}-\boldsymbol{U}_{\beta}\right|\leq b\sqrt{R_{\beta}T_{\beta}},\\
a\chi_{\beta}\frac{\left|\boldsymbol{u}_{k}-\boldsymbol{U}_{\beta}\right|}{\sqrt{R_{\beta}T_{\beta}}}, & \left|\boldsymbol{u}_{k}-\boldsymbol{U}_{\beta}\right|> b\sqrt{R_{\beta}T_{\beta}},
\end{array}\right.
\end{aligned}
\nonumber
\end{equation}
where $a$ and $b$ are set to be $0.1$ and $5$ respectively as validated in Ref.~\cite{xu-tau}.

Secondly, according to Eq.~\eqref{eq:integral}, particles in proportion $e^{-\frac{\Delta t}{\tau_{0,i}}}$ should be sampled from the hydrodynamic wave part, $\boldsymbol{W}_{\alpha,i}^{h}=\boldsymbol{W}_{\alpha,i}-\boldsymbol{W}_{\alpha,i}^{p}$, and their free-transport time is set to be $t_{f,k}=\Delta t$.  The density of newly sampled particles is,
\begin{equation}\label{eq:mic5}
\rho_{\alpha,i}^{hp} = e^{-\frac{\Delta t}{\tau_{0,i}}}\rho_{\alpha,i}^{h}.
\end{equation}
Unless no particle will be sampled as $\rho_{\alpha,i}^{h}=0$, the newly-sampled particle number is calculated by,
\begin{equation}\label{eq:mic6}
N^{hp}_{\alpha,i} = \lceil{\frac{\rho^{hp}_{\alpha,i}}{\rho_{\alpha,i}}{\rm{min}}\left(\chi_{\alpha,i}N^{hp,{\rm{ref1}}},N^{hp,{\rm{ref2}}}\right)}\rceil,
\end{equation}
where $N^{hp,{\rm{ref1}}}$ is a reference total particle number for the gas mixture. It is set to be $800$ in this work, except that in the shock structure case it is set to be $5000$ to prevent the shock from moving; $N^{hp,{\rm{ref2}}}$ is a cut-off for the case with very few but nonnegligible chemical products. A suggested value is $N^{hp,{\rm{ref2}}}=40$. Then the mass for a newly-sampled particle in species $\alpha$ is,
\begin{equation}\label{eq:mic7}
m_{p,k} = \frac{\rho_{\alpha,i}^{hp}\Omega_i}{N^{hp}_{\alpha,i}}.
\nonumber
\end{equation}
The particle position is set to be uniformly distributed within cell ``$i$''. Their velocity is achieved by acceptance-rejection sampling, as follows. Firstly, the velocity according to the Maxwellian distribution function is calculated by,
\begin{equation}\label{eq:maxw}
\begin{aligned}
u_{k,1} &= \tilde{U}_{\alpha,i,1} + \sqrt{2R_{\alpha}\tilde{T}_{\alpha,i}}{\rm{cos}}(2\pi \epsilon_{a1})\sqrt{-{\rm{ln}}(\epsilon_{a2})},\\
u_{k,2} &= \tilde{U}_{\alpha,i,2} + \sqrt{2R_{\alpha}\tilde{T}_{\alpha,i}}{\rm{cos}}(2\pi \epsilon_{b1})\sqrt{-{\rm{ln}}(\epsilon_{b2})},\\
u_{k,3} &= \tilde{U}_{\alpha,i,3} + \sqrt{2R_{\alpha}\tilde{T}_{\alpha,i}}{\rm{cos}}(2\pi \epsilon_{c1})\sqrt{-{\rm{ln}}(\epsilon_{c2})},
\nonumber
\end{aligned}
\end{equation}
where $\epsilon\in\left(0,1\right)$ is a uniformly distributed random number. Then according to $C^S_{\alpha}$ in Eq.~\eqref{eq:bgk}, the criterion of the acceptance-rejection method is suggested to be:
\begin{equation}\label{eq:ajshak}
\frac{1+(1-{\rm{Pr}}_{0,i})\frac{\left(\boldsymbol{u}_k-\tilde{\boldsymbol{U}}_{\alpha,i}\right)\cdot\boldsymbol{Q}_{\alpha,i}}{5\tilde{p}_{\alpha,i}R_{\alpha}\tilde{T}_{\alpha,i}}\left( {\frac{|\boldsymbol{u}_k-\tilde{\boldsymbol{U}}_{\alpha,i}|^2}{R_{\alpha}\tilde{T}_{\alpha,i}}-5} \right)}{1+(1-{\rm{Pr}}_{0,i})\frac{20|\boldsymbol{Q}_{\alpha,i}|}{\tilde{p}_{\alpha,i}\sqrt{R_{\alpha}\tilde{T}_{\alpha,i}}}},
\end{equation}

Additionally, because the analytical macroscopic flux of newly-sampled particles corresponds to the second-order DVM, the free transport fluxes contributed from the collisional particles of $\left(\boldsymbol{W}_{\alpha}^h-\boldsymbol{W}_{\alpha}^{hp}\right)$ can be calculated as,
\begin{equation}
\begin{aligned}
&\boldsymbol{F}_{\alpha}^{fr,wave}=\boldsymbol{F}^{fr}_{\rm{UGKS}}\left(\boldsymbol{W}_{\alpha}^h\right)-\boldsymbol{F}^{fr}_{\rm{DVM}}\left(\boldsymbol{W}_{\alpha}^{hp}\right)\\
=&\int_{\mathbb{R}^3}\boldsymbol{\Psi}\left(\delta_dg_{\alpha}^{h}\left({ \boldsymbol{0},0 }\right)+\delta_e\frac{\partial g_{\alpha}^h}{\partial \boldsymbol{x}}\cdot\boldsymbol{u}\right) \left(\boldsymbol{u}\cdot\boldsymbol{n}\right) {\rm d}\boldsymbol{u}\\
-&e^{-\frac{\Delta t}{\tau_0}}\int_{0}^{\Delta t}\int_{\mathbb{R}^3}\boldsymbol{\Psi}\left(g_{\alpha}^h\left({ \boldsymbol{0},0 }\right)-t\frac{\partial g_{\alpha}^h}{\partial \boldsymbol{x}}\cdot\boldsymbol{u}\right) \left(\boldsymbol{u}\cdot\boldsymbol{n}\right) {\rm d}\boldsymbol{u} {\rm d}t\\
=&\int_{\mathbb{R}^3}\boldsymbol{\Psi}\left[\left(\delta_d-\Delta te^{-\frac{\Delta t}{\tau_0}}\right)g_{\alpha}^h\left({ \boldsymbol{0},0 }\right)+\left(\delta_e+\frac{\Delta t^2}{2}e^{-\frac{\Delta t}{\tau_0}}\right)\frac{\partial g_{\alpha}^h}{\partial \boldsymbol{x}}\cdot\boldsymbol{u}\right] \left(\boldsymbol{u}\cdot\boldsymbol{n}\right) {\rm d}\boldsymbol{u}.
\nonumber
\end{aligned}
\end{equation}
The same as in Eq.~\eqref{eq:feq}, the scale-related coefficient $\delta$ is separated into left-side and right-side cells of the interface ``$j$'' for a better consistency with sampled particles, as follows,
\begin{equation}\label{eq:ffrwave}
\begin{aligned}
\boldsymbol{F}^{fr,wave}_{\alpha,j}=&\int_{\boldsymbol{u}\cdot\boldsymbol{n}_j>0}\boldsymbol{\Psi}\left[\left(\delta_{d,L}-\Delta te^{-\frac{\Delta t}{\tau_{0,L}}}\right)g_{\alpha}^h\left({ \boldsymbol{0},0 }\right)+\left(\delta_{e,L}+\frac{\Delta t^2}{2}e^{-\frac{\Delta t}{\tau_{0,L}}}\right)\frac{\partial g_{\alpha}^h}{\partial \boldsymbol{x}}\cdot\boldsymbol{u}\right] \left(\boldsymbol{u}\cdot\boldsymbol{n}_j\right) {\rm d}\boldsymbol{u}\\
+&\int_{\boldsymbol{u}\cdot\boldsymbol{n}_j<0}\boldsymbol{\Psi}\left[\left(\delta_{d,R}-\Delta te^{-\frac{\Delta t}{\tau_{0,R}}}\right)g_{\alpha}^h\left({ \boldsymbol{0},0 }\right)+\left(\delta_{e,R}+\frac{\Delta t^2}{2}e^{-\frac{\Delta t}{\tau_{0,R}}}\right)\frac{\partial g_{\alpha}^h}{\partial \boldsymbol{x}}\cdot\boldsymbol{u}\right] \left(\boldsymbol{u}\cdot\boldsymbol{n}_j\right) {\rm d}\boldsymbol{u}.\\
\end{aligned}
\end{equation}
As a result, the macroscopic update equation is renewed from Eq.~\eqref{eq:conservation-macro} as follows.
\begin{equation}\label{eq:update1}
\boldsymbol{W}^{\ast}_{\alpha,i} = \boldsymbol{W}^{n}_{\alpha,i} - \frac{1}{\Omega_i}\sum\limits_{j\in \mathcal{M}\left(i\right)}\boldsymbol{F}_{\alpha,j}^{eq}s_j - \frac{1}{\Omega_i}\sum\limits_{j\in \mathcal{M}\left(i\right)}\boldsymbol{F}_{\alpha,j}^{fr,wave}s_j+\boldsymbol{W}_{\alpha,i}^{fr,p}.
\end{equation}

\subsection{Chemical reaction source term}\label{sec:source-c}
For Eq.~\eqref{eq:conservation-macrosource-c}, considering the chemical reaction effect, macroscopic variables are updated from $\boldsymbol{W}^{\ast}_{\alpha}$ to $\boldsymbol{W}^{\ast\ast}_{\alpha}$. The variation of species number density and gas-mixture energy are firstly calculated in the approach of NS solvers,
\begin{equation}\label{eq:chemical}
\begin{aligned}
n_\alpha^{\ast\ast} = n_\alpha^{\ast}-k_c\Delta t,\alpha\in CR,\\
n_\alpha^{\ast\ast} = n_\alpha^{\ast}+k_c\Delta t,\alpha\in CP,\\
\rho_0^{\ast\ast}E_0^{\ast\ast} = \rho_0^{\ast}E_0^{\ast} + k_c\Delta t\Delta E,
\end{aligned}
\end{equation}
where ``CR'' and ``CP'' are sets of chemical reactants and products respectively. $\Delta E$ denotes the released energy from the reaction ($\Delta E>0$ for exothermic reaction, and $\Delta E<0$ for endothermic reaction). Because there is only one elementary chemical reaction considered in this work and it is slow, a simple explicit forward Euler method is used. Reaction rate coefficient $k_c$ is calculated by Arrhenius equation as follows,
\begin{equation}\label{eq:arrhenius}
\begin{aligned}
k_c = k_f\prod\limits_\beta^{CR}n_\beta^{\ast} - k_b\prod\limits_\beta^{CP}n_\beta^{\ast},\\
k_f = A_f\left(T_0^{\ast}\right)^{B_f}\exp\left(-\frac{E_{a,f}}{k_BT_0^{\ast}}\right),\\
k_b = A_b\left(T_0^{\ast}\right)^{B_b}\exp\left(-\frac{E_{a,b}}{k_BT_0^{\ast}}\right),
\end{aligned}
\end{equation}
where subscripts ``$f$'' and ``$b$'' denote forward and backward parameters respectively. Constants include pre-exponential factor $A_{f/b}$, temperature exponent $B_{f/b}$, activation energy $E_{a,f/b}$ and Boltzmann constant $k_B$. Furthermore, the variation in gas-mixture is detailed into each species. For reactants $\alpha\in CR$, macroscopic variables are proportionally decreased,
\begin{equation}\label{eq:chemical-r}
\begin{aligned}
&\rho_\alpha^{\ast\ast}\boldsymbol{U}_\alpha^{\ast\ast}=\rho_\alpha^{\ast\ast}\boldsymbol{U}_\alpha^{\ast},\\
&\rho_\alpha^{\ast\ast}E_\alpha^{\ast\ast}=\rho_\alpha^{\ast\ast}E_\alpha^{\ast}.
\nonumber
\end{aligned}
\end{equation}
While for products $\alpha\in CP$, both macroscopic momentum and internal energy of reactants are equally divided as follows,
\begin{equation}\label{eq:chemical-p}
\begin{aligned}
&\rho_\alpha^{\ast\ast}\boldsymbol{U}_\alpha^{\ast\ast}=\rho_\alpha^{\ast}\boldsymbol{U}_\alpha^{\ast}+m_\alpha\frac{\sum\limits_{\beta=1}^{CR}\left(\rho_\beta^{\ast}
-\rho_\beta^{\ast\ast}\right)\boldsymbol{U}_\beta^{\ast}}{\sum\limits_{\beta=1}^{CP}m_\beta},\\
&\rho_\alpha^{\ast\ast}E_\alpha^{\ast\ast}=\rho_\alpha^{\ast}E_\alpha^{\ast}+m_\alpha\frac{\sum\limits_{\beta=1}^{CR}\left(\rho_\beta^{\ast}-\rho_\beta^{\ast\ast}\right)
E_\beta^{\ast}+k_c\Delta t\Delta E}{\sum\limits_{\beta=1}^{CP}m_\beta}.
\end{aligned}
\end{equation}
As an intermediate step, the second equation in Eq.~\eqref{eq:chemical-p} is derived from,
\begin{equation}
\begin{aligned}
\underbrace{\frac{3}{2}\rho_\alpha^{\ast\ast}R_\alpha T_\alpha^{\ast\ast}}_{\text{Renewed internal energy}}=&\underbrace{\frac{3}{2}\rho_\alpha^{\ast}R_\alpha T_\alpha^{\ast}}_{\text{Original internal energy}}+\underbrace{\frac{1}{2}\left(\rho_\alpha^{\ast}\left|\boldsymbol{U}_\alpha^{\ast}\right|^2
-\rho_\alpha^{\ast\ast}\left|\boldsymbol{U}_\alpha^{\ast\ast}\right|^2\right)}_{\substack{\text{Contributed from velocity}\\\text{change of products}}}+\underbrace{\frac{1}{2}\rho_{\alpha,CP}^{\ast\ast}\left|\boldsymbol{U}_{CR}^{\ast\ast}\right|^2}_{\substack{\text{Contributed from kinetic}\\\text{energy of reactant gas mixture}}}\\
+&\frac{m_\alpha}{\sum\limits_{\beta=1}^{CP}m_\beta}\left[\underbrace{\sum\limits_{\beta=1}^{CR}\left(\rho_\beta^{\ast}-\rho_\beta^{\ast\ast}\right)\left(\frac{3}{2}\rho_\beta^{\ast}R_\beta T_\beta^{\ast}+\frac{\left|\boldsymbol{U}_\beta^{\ast}\right|^2-\left|\boldsymbol{U}_{CR}^{\ast\ast}\right|^2}{2}\right)}_{\substack{\text{Contributed from reactants' internal energy}\\\text{and velocity difference between each species and gas mixture}}}+k_c\Delta t\Delta E\right],
\nonumber
\end{aligned}
\end{equation}
where,
\begin{equation}
\begin{aligned}
&\rho_{\alpha,CP}^{\ast\ast}=\frac{m_\alpha}{\sum\limits_{\beta=1}^{CP}m_\beta}\sum\limits_{\beta=1}^{CR}\left(\rho_\beta^{\ast}-\rho_\beta^{\ast\ast}\right),\\
&\boldsymbol{U}_{CR}^{\ast\ast}=\frac{\sum\limits_{\beta=1}^{CR}\left(\rho_\beta^{\ast}-\rho_\beta^{\ast\ast}\right)\boldsymbol{U}_\beta^{\ast}}{\sum\limits_{\beta=1}^{CR}
\left(\rho_\beta^{\ast}-\rho_\beta^{\ast\ast}\right)}.
\nonumber
\end{aligned}
\end{equation}
Notably, our choice of equipartitioned internal energy, rather than total energy or others, is not only consistent with DSMC but also avoids the emergence of negative temperatures.

After that the effects of chemical reactions are imposed on the macroscopic variables of each species, their influence on the microscopic distribution function is then accounted for. Because particles carry information such as whether a collision occurs and the free-transport time, the particle and wave can be treated separately according to collision participation and the collision timing. Since all gases are treated as monatomic in the present work, this step remains relatively simple and can serve as a basic test. If thermal non-equilibrium is further considered, particles will also carry rotational and vibrational energies, which will make this step more involved. In this work, we note that the geometrically effective collision condition for chemical reactions is nested within that for elastic collisions. Analogously, in the DSMC method, the steric factor is defined as the ratio of the reaction cross section to the elastic cross section~\cite{dsmc}. As a result, in the present UGKWP method, free-transport particles are assumed not to participate in chemical reactions, and particle collision timings are not advanced by chemical reactions: their parameters remain unchanged, including the free-transport time and microscopic velocity. The entire influence of chemical reactions is instead absorbed into the wave part, manifested in $\boldsymbol{W}_{\alpha,i}^{\ast\ast}-\boldsymbol{W}_{\alpha,i}^{p,\ast}$, with $\boldsymbol{W}_{\alpha,i}^{p,\ast}$ unchanged.

\subsection{Multispecies source term}\label{sec:source}
By calculating moments in the kinetic model as Eq.~\eqref{eq:bgk}, the source term in Eq.~\eqref{eq:conservation-macrosource} can be derived as,
\begin{equation}\label{eq:source1}
\frac{\partial\boldsymbol{W}_{\alpha,i}}{\partial t}= \frac{\tilde{\boldsymbol{W}}^{\ast\ast}_{\alpha,i}-\boldsymbol{W}_{\alpha,i}}{\tau_0},
\nonumber
\end{equation}
whose initial condition is $\boldsymbol{W}^{\ast\ast}_{\alpha,i}$ and the target value $\tilde{\boldsymbol{W}}^{\ast\ast}_{\alpha,i}$ is also calculated from $\boldsymbol{W}^{\ast\ast}_{\alpha,i}$. The integral solution is,
\begin{equation}\label{eq:source2}
\boldsymbol{W}^{n+1}_{\alpha,i}= e^{-\frac{\Delta t}{\tau_0}}\boldsymbol{W}^{\ast\ast}_{\alpha,i}+\left(1-e^{-\frac{\Delta t}{\tau_0}}\right)\tilde{\boldsymbol{W}}^{\ast\ast}_{\alpha,i},
\end{equation}
In contrast to the Andries--Aoki--Perthame (AAP) model~\cite{aap}, because the target velocity is obtained by a convex combination in Eq.~\eqref{eq:groppi}, an extra implicit module~\cite{dugks-aap2} is not necessary. Moreover, by calculating target values in this step, the consistency between wave and particles is achieved. Specifically, when constructing flux and source term, both particle sampling and wave evolution use consistent target values, $g_{\alpha,i}$ and $\tilde{\boldsymbol{W}}^{\ast\ast}_{\alpha,i}$. The time coefficients are consistent as well, for example, the exponential form in the macroscopic source term Eq.~\eqref{eq:source2} and particle resample term Eq.~\eqref{eq:mic5}, $e^{\frac{-\Delta t}{\tau_0}}$. Additionally, both chemical reaction effect and multispecies effect are considered in calculating $g_{\alpha,i}$ and $\tilde{\boldsymbol{W}}^{\ast\ast}_{\alpha,i}$, as in Sec.\ref{sec:source-c} and Sec.\ref{sec:source}. These values are used in calculating fluxes as well as source terms.

\subsection{Algorithm}\label{sec:sum1}
\begin{figure}[H]
	\centering
	\subfigure[]{\label{fig1a}
			\includegraphics[width=0.22 \textwidth]{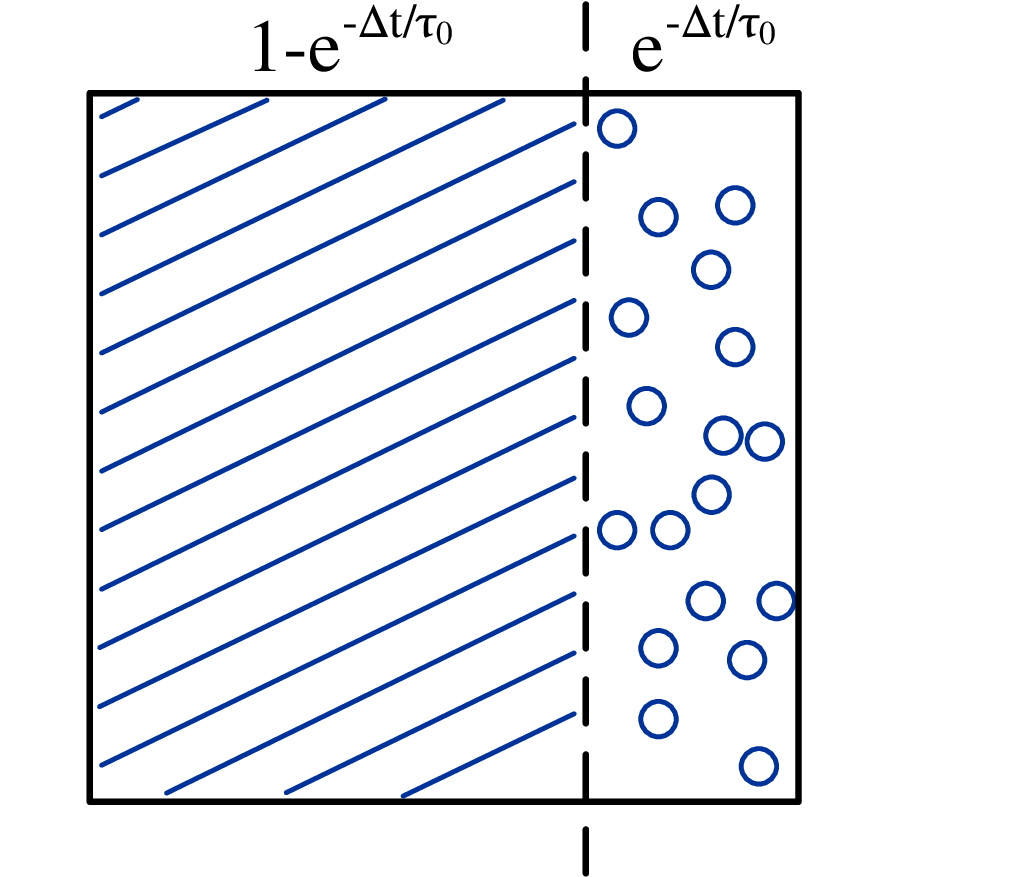}
		}
    \subfigure[]{\label{fig1b}
    		\includegraphics[width=0.22 \textwidth]{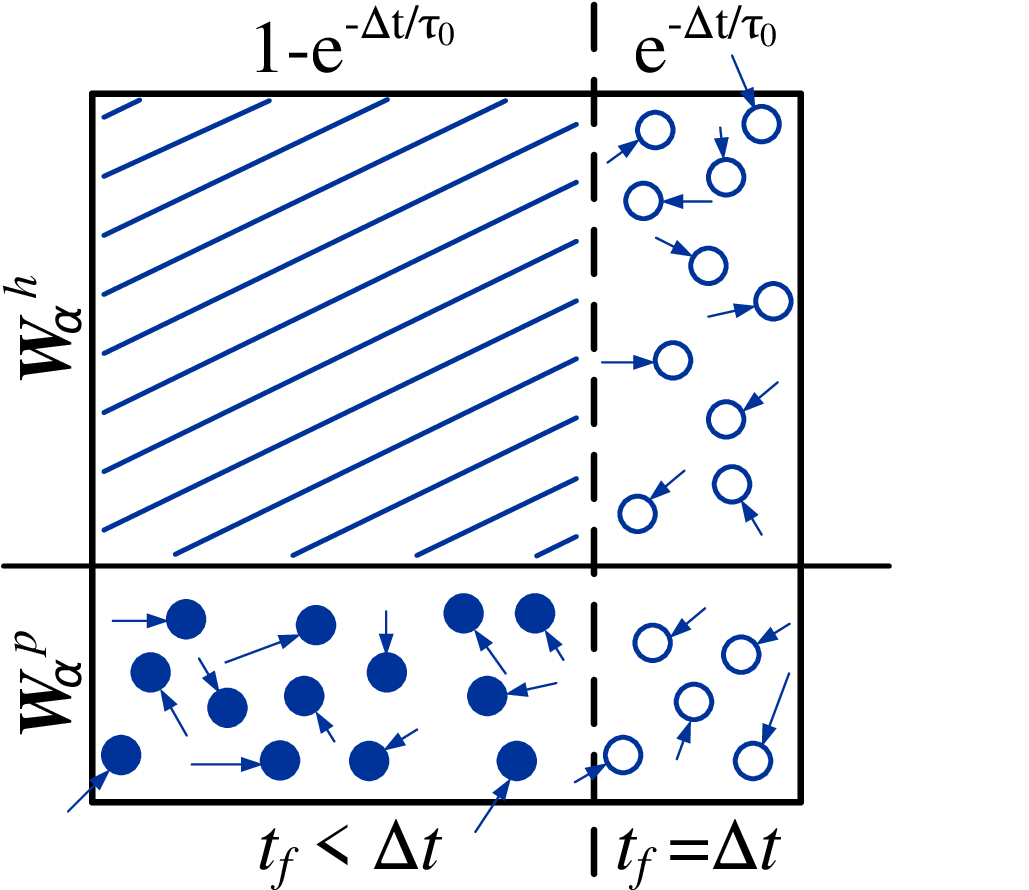}
    	}
    \subfigure[]{\label{fig1c}
    		\includegraphics[width=0.24 \textwidth]{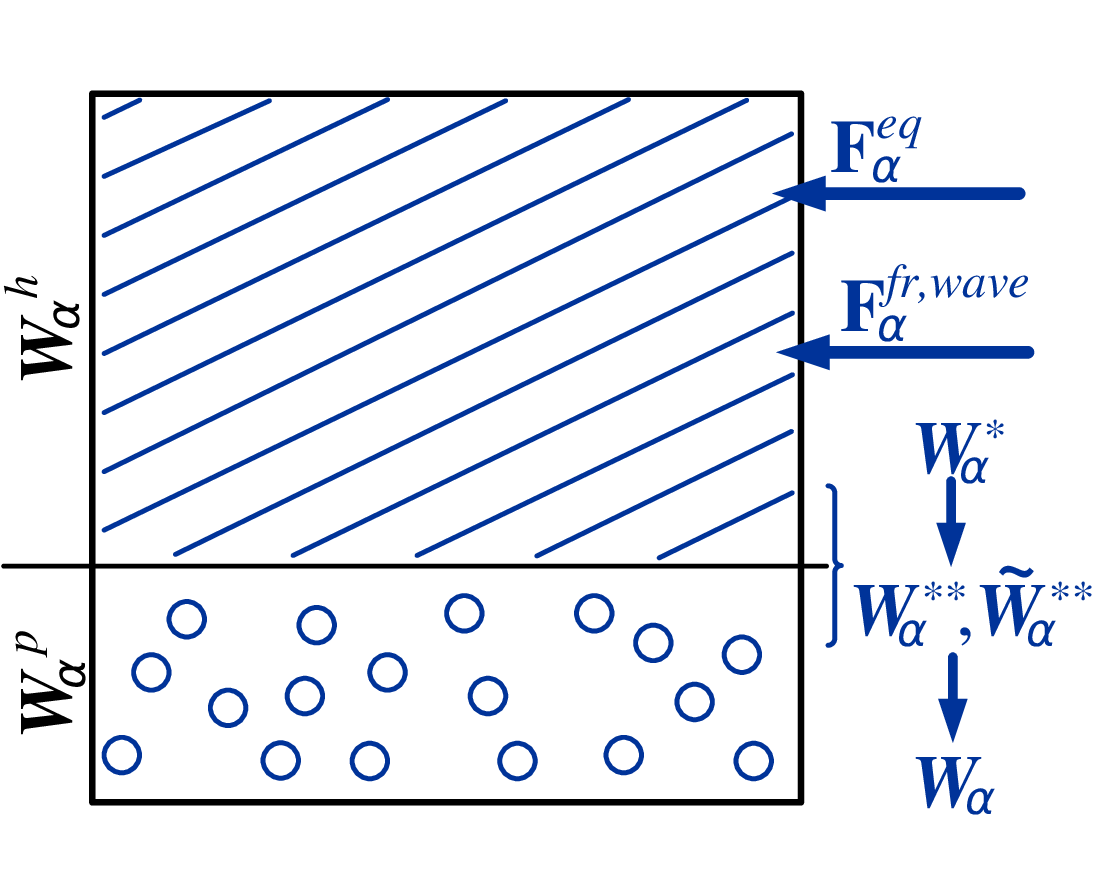}
    	}
    \subfigure[]{\label{fig1d}
    		\includegraphics[width=0.215 \textwidth]{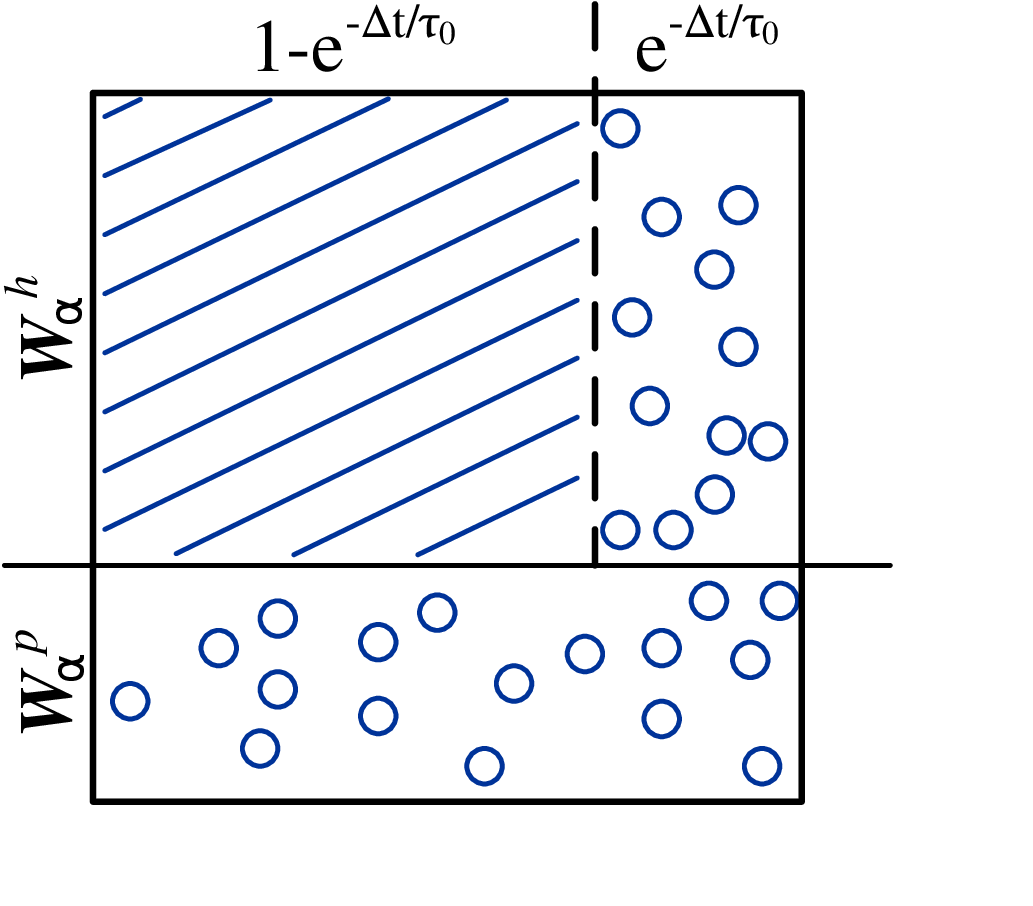}
    	}
	\caption{\label{fig1} Diagram to illustrate the algorithm of UGKWP method: (a) Initial state at $n=0$ step, (b) step $2$, (c) step $3$, (d) initial state at other steps.}
\end{figure}

Regarding Fig.~\ref{fig1}, an algorithm summary is introduced for the proposed UGKWP method for multiscale gas-mixture flow with an elementary chemical reaction.
\begin{description}
    \item[Step (1)] Initial state. As shown in Fig.~\ref{fig1d}, in the $n-1$ time step, particles $\boldsymbol{W}^p_{\alpha}$ coexist with hydrodynamic wave $\boldsymbol{W}^h_{\alpha}$. We sample collisionless particles $\boldsymbol{W}^{hp}_{\alpha}$ as Eq.~\eqref{eq:mic5}, whose parameters are calculated as equations from Eq.~\eqref{eq:mic6} to Eq.~\eqref{eq:ajshak}. For the first step $n=0$, $\boldsymbol{W}^p_{\alpha}=\boldsymbol{0}$, as shown in Fig.~\ref{fig1a}.
    \item[Step (2)] Free transport. First, divide particles $\boldsymbol{W}^p_{\alpha}$ into two parts as Eq.~\eqref{eq:taustar}. As in Fig.~\ref{fig1b}, collisionless particles are denoted by hollow circles, and collisional ones are denoted by solid circles. Then transport all particles as Eq.~\eqref{eq:mic2}, and sum their contribution to macroscopic conservative variables as Eq.~\eqref{eq:mic3}. Meanwhile, free transport fluxes $\boldsymbol{F}_{\alpha,j}^{fr,wave}$ contributed from collisional particles of $\left(\boldsymbol{W}^h_{\alpha}-\boldsymbol{W}^{hp}_{\alpha}\right)$ are calculated as Eq.~\eqref{eq:ffrwave}. Finally, delete the collisional particles denoted by solid circles.
    \item[Step (3)] Collision. Firstly, calculate the equilibrium flux $\boldsymbol{F}^{eq}_{\alpha,j}$ as Eq.~\eqref{eq:feq}. Secondly, get $\boldsymbol{W}^{\ast}_{\alpha}$ as Eq.~\eqref{eq:update1}. Thirdly, considering the chemical reaction effect as equations from Eq.~\eqref{eq:chemical} to Eq.~\eqref{eq:chemical-p}, update $\boldsymbol{W}^{\ast}_{\alpha}$ to $\boldsymbol{W}^{\ast\ast}_{\alpha}$. Then get the target macroscopic variables $\tilde{\boldsymbol{W}}^{\ast\ast}_{\alpha}$ from $\boldsymbol{W}^{\ast\ast}_{\alpha}$ by Eq.~\eqref{eq:groppi}, along with other variables such as relaxation time $\tau_0$ and ${\rm{Pr}}_0$ in Eq.~\eqref{eq:coeffs}. Finally, for the multispecies effect, update $\boldsymbol{W}^{n+1}_{\alpha}$ as Eq.~\eqref{eq:source2}.
    \item[Step (4)] If the simulation continues, go to Step $\left(1\right)$, where collisionless particles $\boldsymbol{W}^{hp}_{\alpha}$ will be sampled.
\end{description}

\section{Numerical cases}\label{sec:cases}
In this section, the capability of the proposed UGKWP method is verified through a set of numerical tests. Firstly, a classical hypersonic flow past a cylinder is simulated over a wide range of ${\rm{Kn}}_{\infty}$. Cases ranging from chemically inert flow to forward endothermic, forward exothermic, and $\Delta E=0$ reactive conditions are compared with DSMC, including both stagnation-line profiles and wall quantities. Then shock structures are simulated at ${\rm{Ma}}=3.0$ and ${\rm{Ma}}=5.0$. This test case differs from the hypersonic cylinder flow. In the cylinder case, the freestream temperature is low and chemical reaction rates remain small. Reactions are triggered by the high post-shock temperature. While in the shock structure case, the upstream and downstream temperatures are sufficiently high, and asymptotic equilibrium conditions are prescribed through the law of mass action, $k_b/k_f=\prod\limits_\beta^{CR}n_\beta\left.\Big/\right.\prod\limits_\beta^{CP}n_\beta$. For the comparison with DSMC, DS2V is used, and parameters of gases and chemical reactions are consistent. In addition, a three-dimensional hypersonic side-jet flow is constructed to examine the code's capability for more complex cases. In $2D$ and $3D$ cases, reference particle number per cell $N^{hp,\rm{ref1}}$ is set to be $800$, while in the shock structure case, $N^{hp,\rm{ref1}}$ is set to be $5000$ for avoiding the shock moving. In all cases, the Courant-Friedrichs-Lewy (CFL) number is set to be $0.8$. The full accommodation is applied for the isothermal walls as the boundary condition.

As in Refs~\cite{ugks-reaction,gkua1}, the Zeldovich-type reaction,
\begin{equation}
O_2 + N \rightleftharpoons NO + O,
\end{equation}
is implemented in all cases. Gas parameters are taken from the DS2V code~\cite{dsmc}, as Tab.~\ref{tab-gas},
\begin{table}[h]
\caption{\text{Gas parameters}}\label{tab-gas}
\centering
\begin{tabular}{*{6}{c}}
\toprule
    Gas &$m$ $\left({\rm{kg}}\right)$ &$d$ $\left({\rm{m}}\right)$ at $273{\rm{K}}$ &$\mu$ $\left({\rm{Pa\cdot s}}\right)$ at $273{\rm{K}}$ &$\omega$ in VHS model &$\rm{Pr}$\\
\midrule
	$O_2$ &$5.312\times 10^{-26}$ &$4.07\times 10^{-10}$ &$1.9133\times 10^{-5}$ &$0.77$ &$2/3$  \\
	$N$   &$2.325\times 10^{-26}$ &$3.00\times 10^{-10}$ &$2.3972\times 10^{-5}$ &$0.8$  &$2/3$  \\
    $NO$  &$4.980\times 10^{-26}$ &$4.20\times 10^{-10}$ &$1.7730\times 10^{-5}$ &$0.79$ &$2/3$  \\
    $O$   &$2.656\times 10^{-26}$ &$3.00\times 10^{-10}$ &$2.5622\times 10^{-5}$ &$0.8$  &$2/3$  \\
\bottomrule
\end{tabular}
\end{table}

Chemical reaction parameters are also taken from the DS2V code, as Tab.~\ref{tab-chemical},
\begin{table}[h]
\caption{\text{Chemical reaction parameters}}\label{tab-chemical}
\centering
\begin{tabular}{*{6}{c}}
\toprule
    Parameter &Value &Parameter &Value &Parameter &Value  \\
\midrule
	$A_f$    &$1.598\times 10^{-18}$    &$B_f$        &$0.5$    &$E_{a,f}$    &$5.0\times 10^{-20}{\rm{J}}$    \\
	$A_b$    &$5.279\times 10^{-21}$    &$B_b$        &$1.0$    &$E_{a,b}$    &$2.2\times 10^{-19}{\rm{J}}$    \\
\bottomrule
\end{tabular}
\end{table}
The original reaction is forward exothermic $\Delta E=2.2\times 10^{-19}{\rm{J}}$, but for different cases, it may be set to be forward endothermic or $\Delta E=0{\rm{J}}$.

\subsection{Hypersonic flow around a cylinder}\label{sec:cylinder}
\begin{figure}[H]
	\centering
	\subfigure[]{
			\includegraphics[width=0.3 \textwidth]{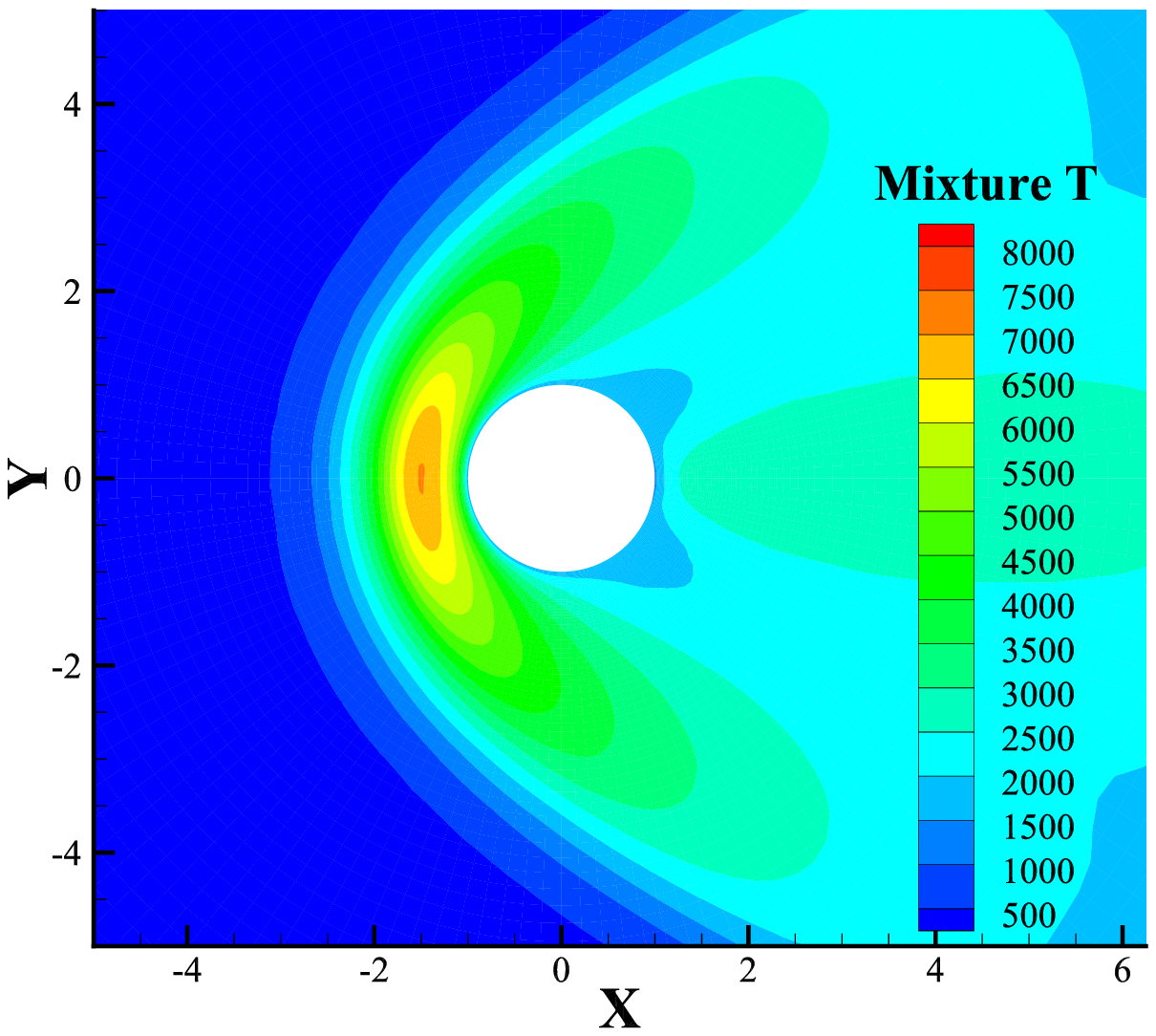}
		}
    \subfigure[]{
    		\includegraphics[width=0.3 \textwidth]{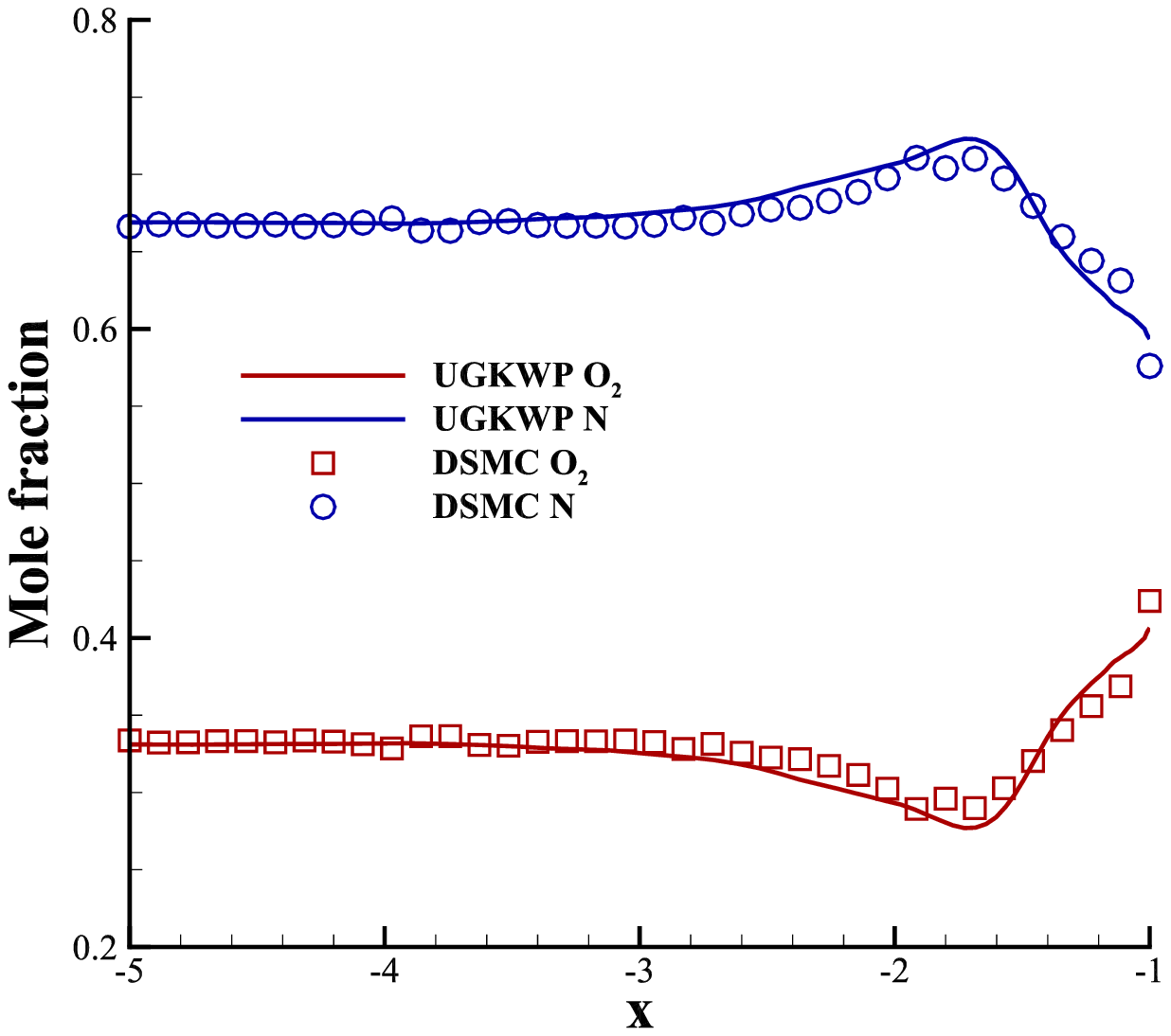}
    	}
    \subfigure[]{
    		\includegraphics[width=0.3 \textwidth]{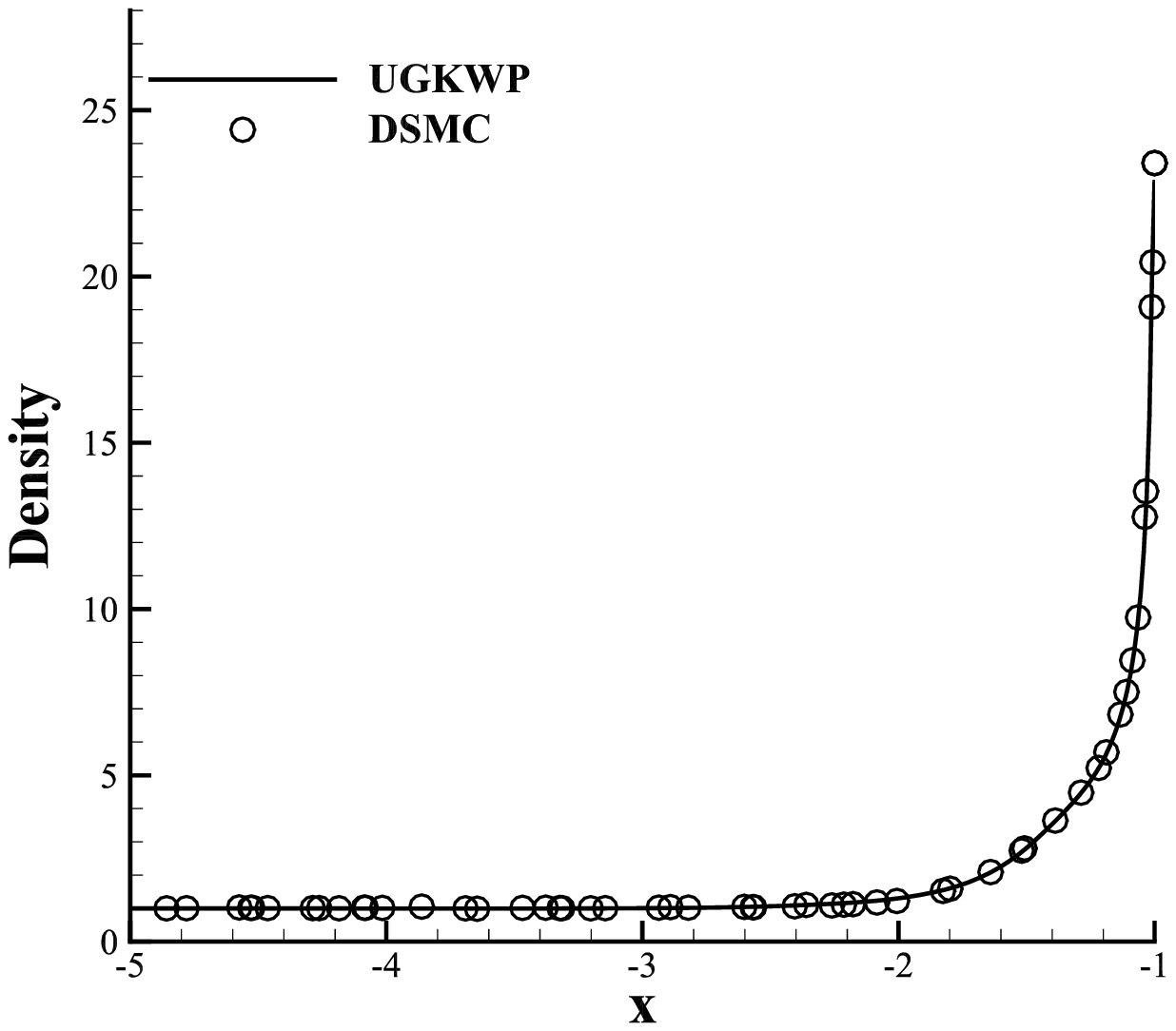}
    	}
    \\
    \subfigure[]{
			\includegraphics[width=0.3 \textwidth]{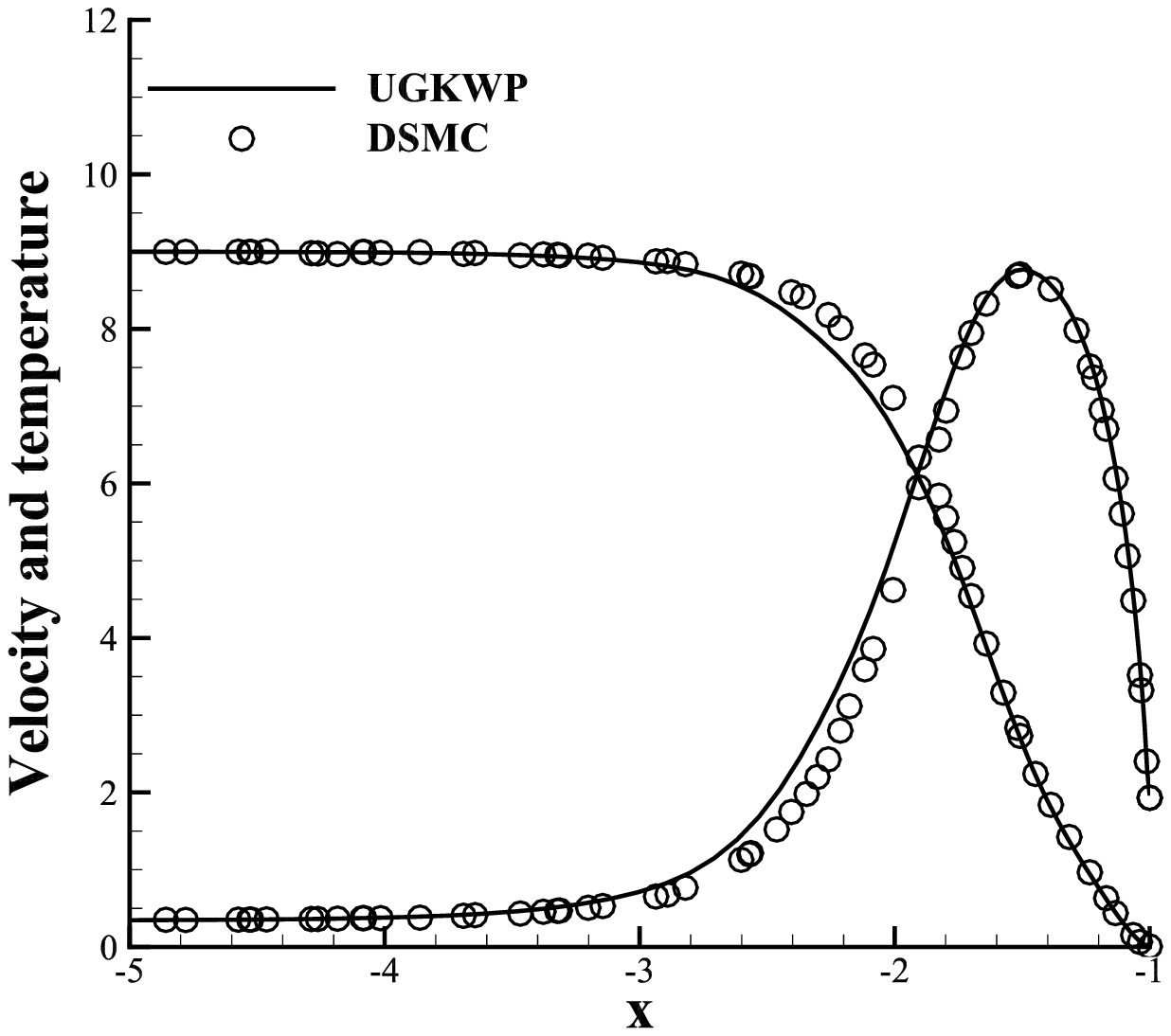}
		}
    \subfigure[]{
    		\includegraphics[width=0.3 \textwidth]{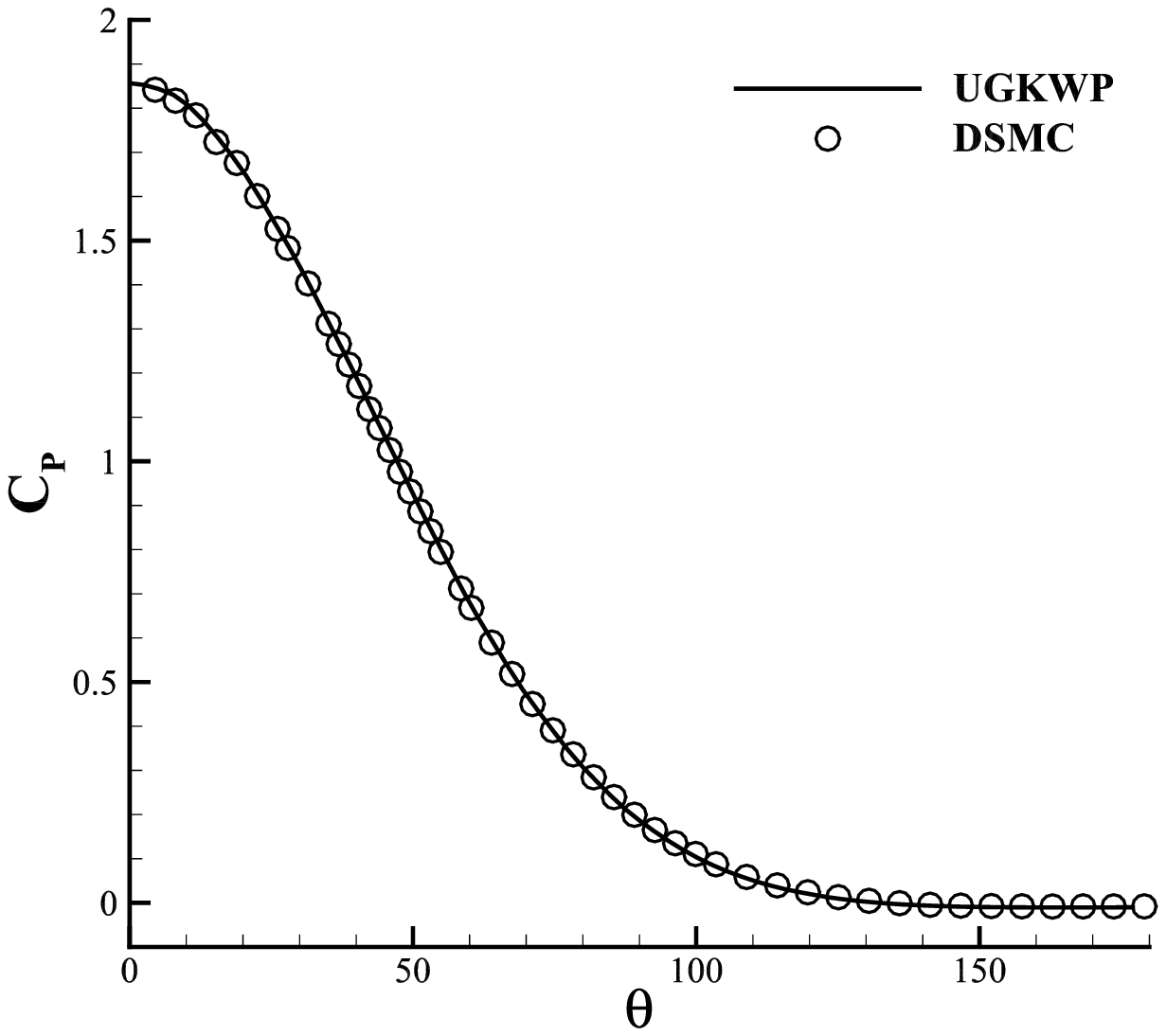}
    	}
    \subfigure[]{
    		\includegraphics[width=0.3 \textwidth]{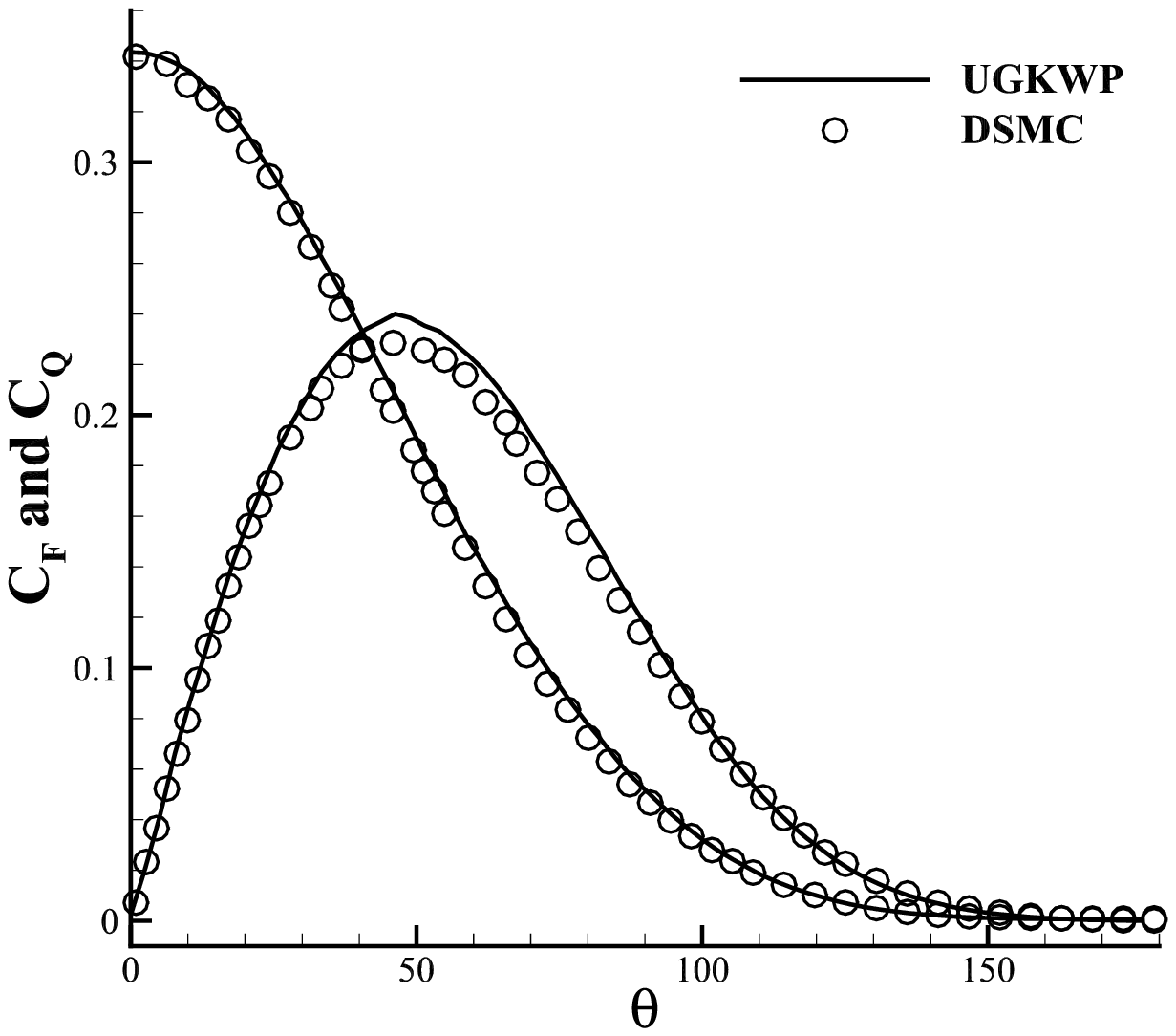}
    	}
	\caption{\label{cylinder-kn0.1-binary} Hypersonic flow around a cylinder at ${\rm{Ma}}_{\infty}=9$, ${\rm{Kn}}_{\infty}=0.1$, and the chemical reaction is not considered: (a) Gas mixture temperature contour, (b) mole fraction along the stagnation line, (c) gas mixture density along the stagnation line, (d) gas mixture velocity and temperature along the stagnation line, (e) pressure coefficient at the wall, (f) shear stress and heat flux coefficients at the wall.}
\end{figure}

\begin{figure}[H]
	\centering
	\subfigure[]{
			\includegraphics[width=0.3 \textwidth]{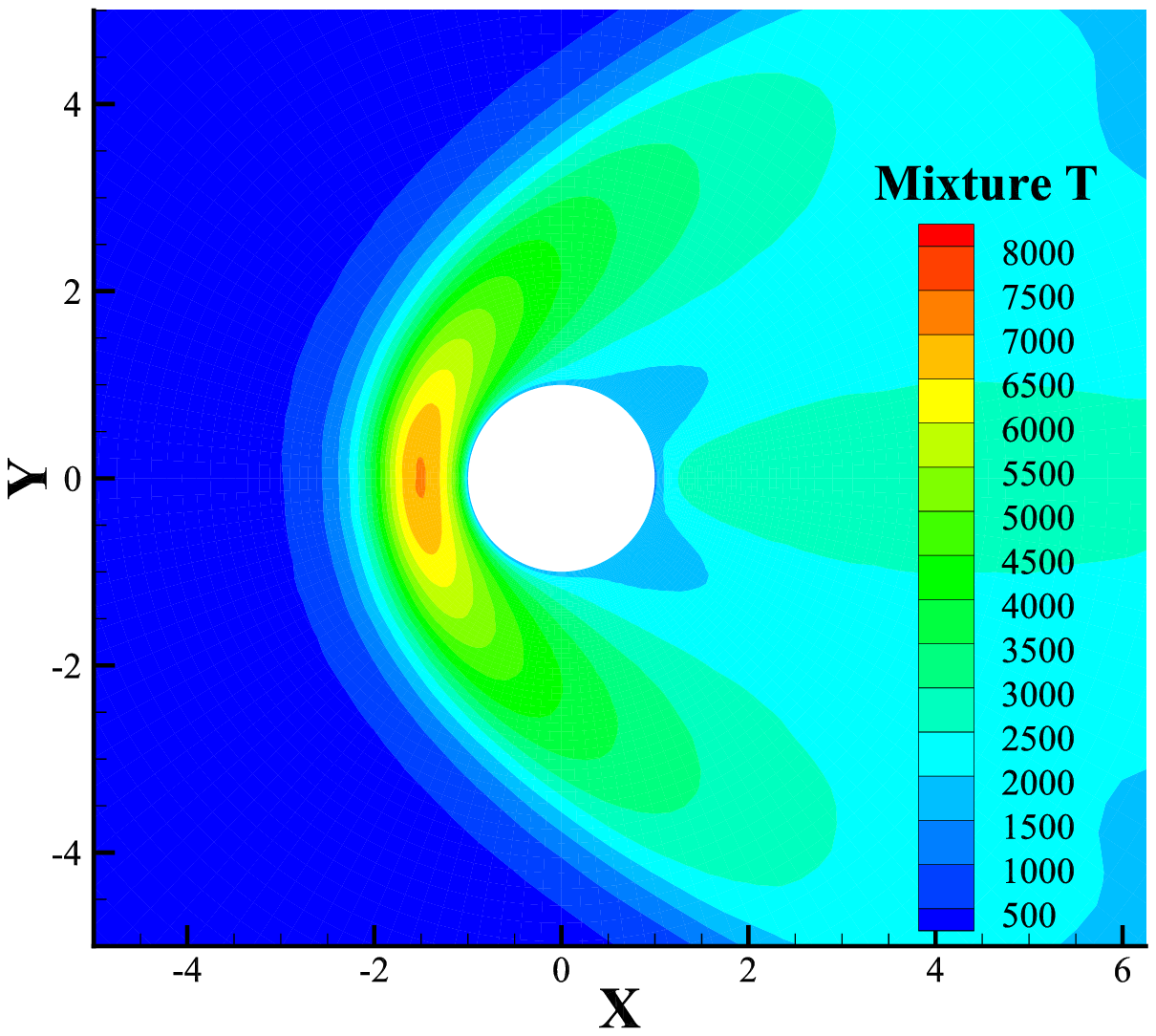}
		}
    \subfigure[]{
    		\includegraphics[width=0.3 \textwidth]{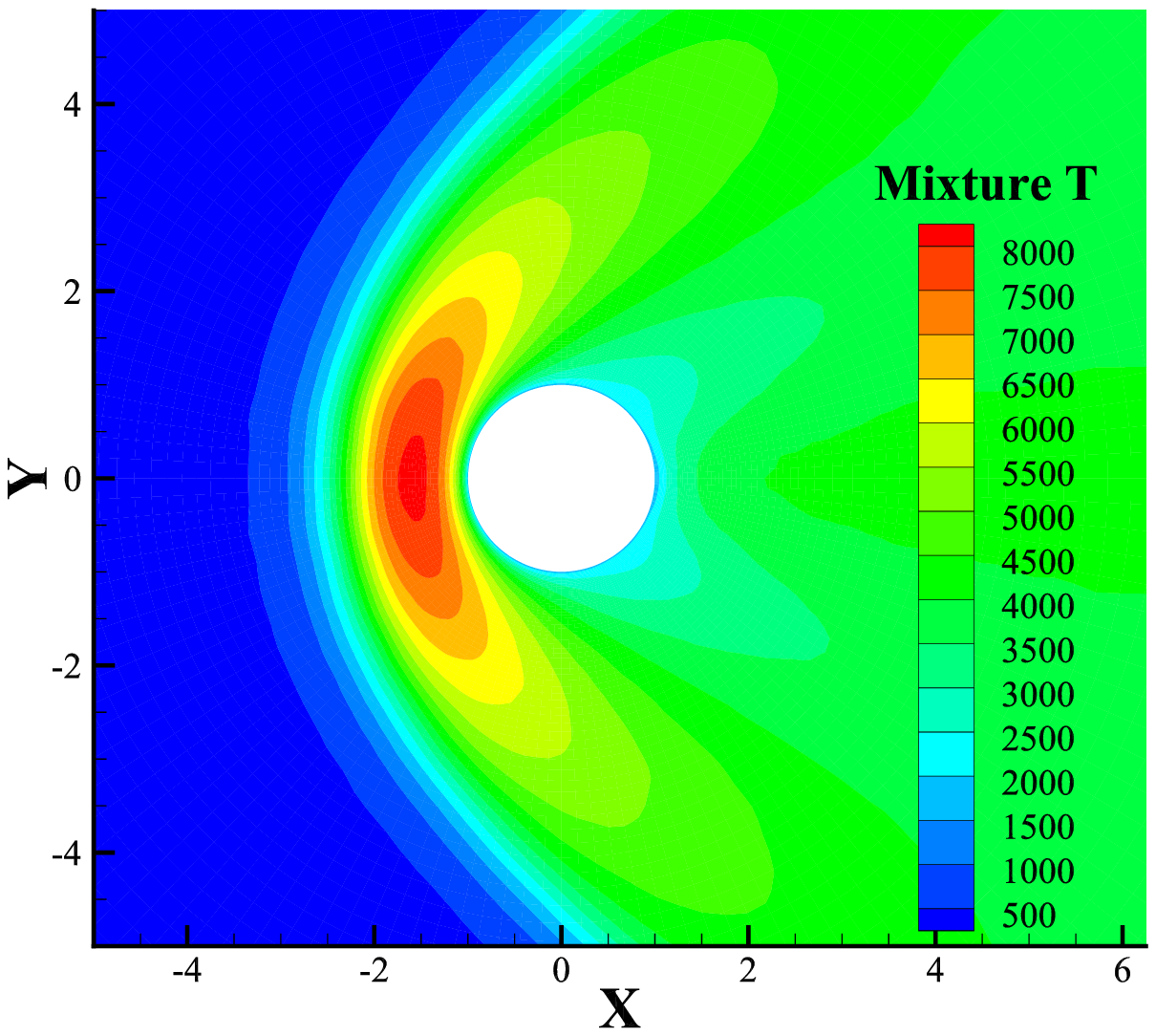}
    	}
    \subfigure[]{
    		\includegraphics[width=0.3 \textwidth]{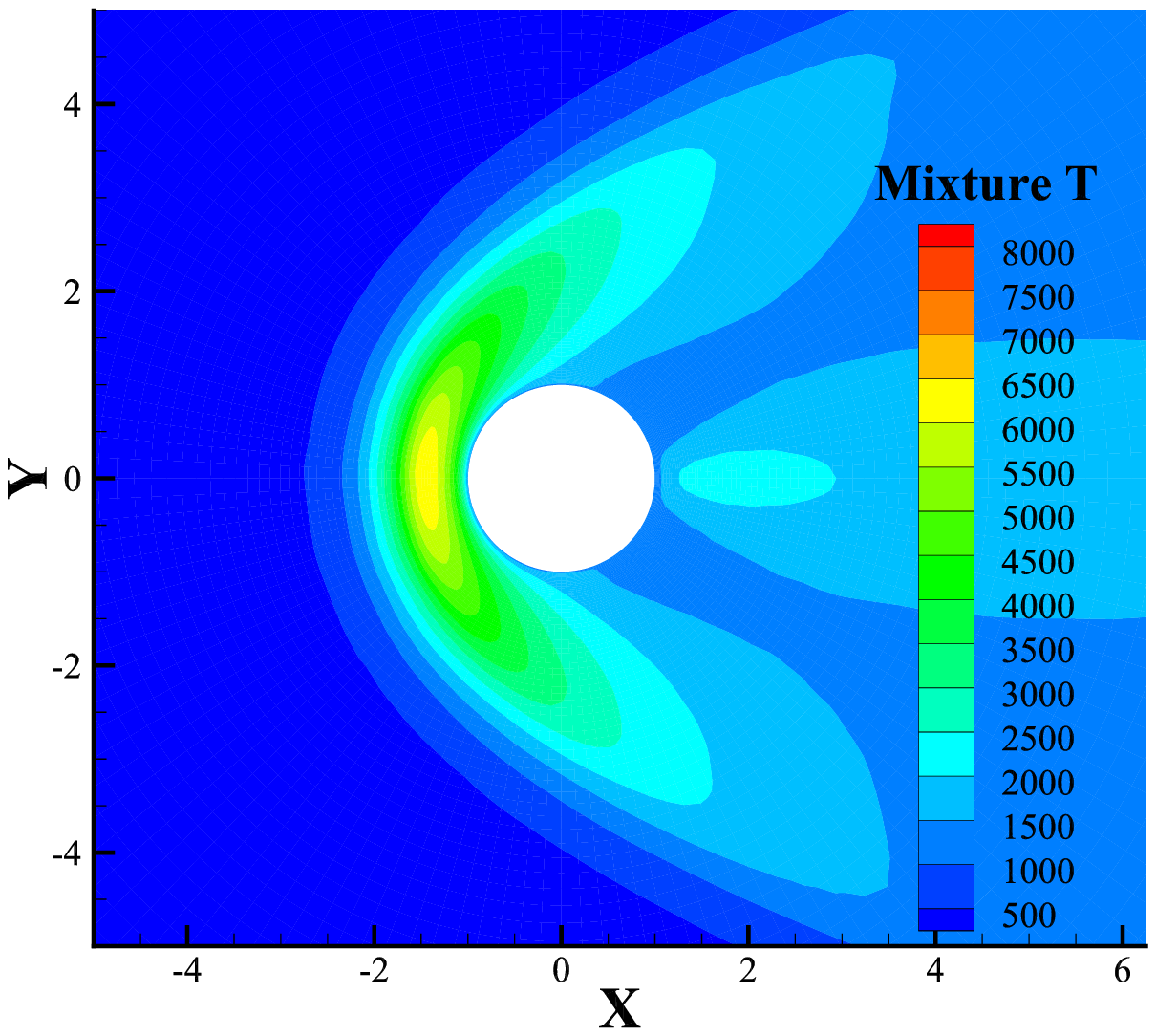}
    	}
    \\
    \subfigure[]{
			\includegraphics[width=0.3 \textwidth]{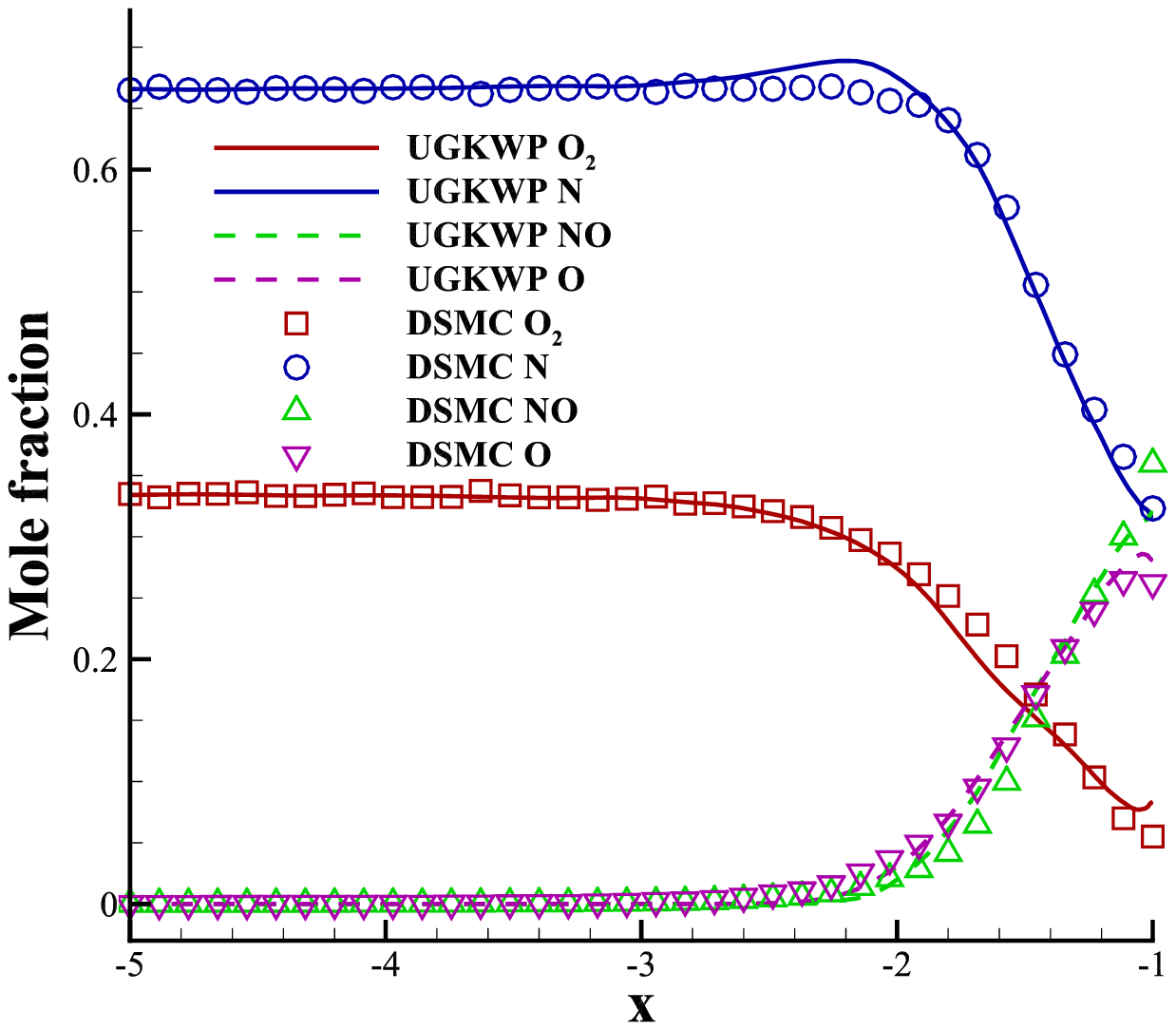}
		}
    \subfigure[]{
    		\includegraphics[width=0.3 \textwidth]{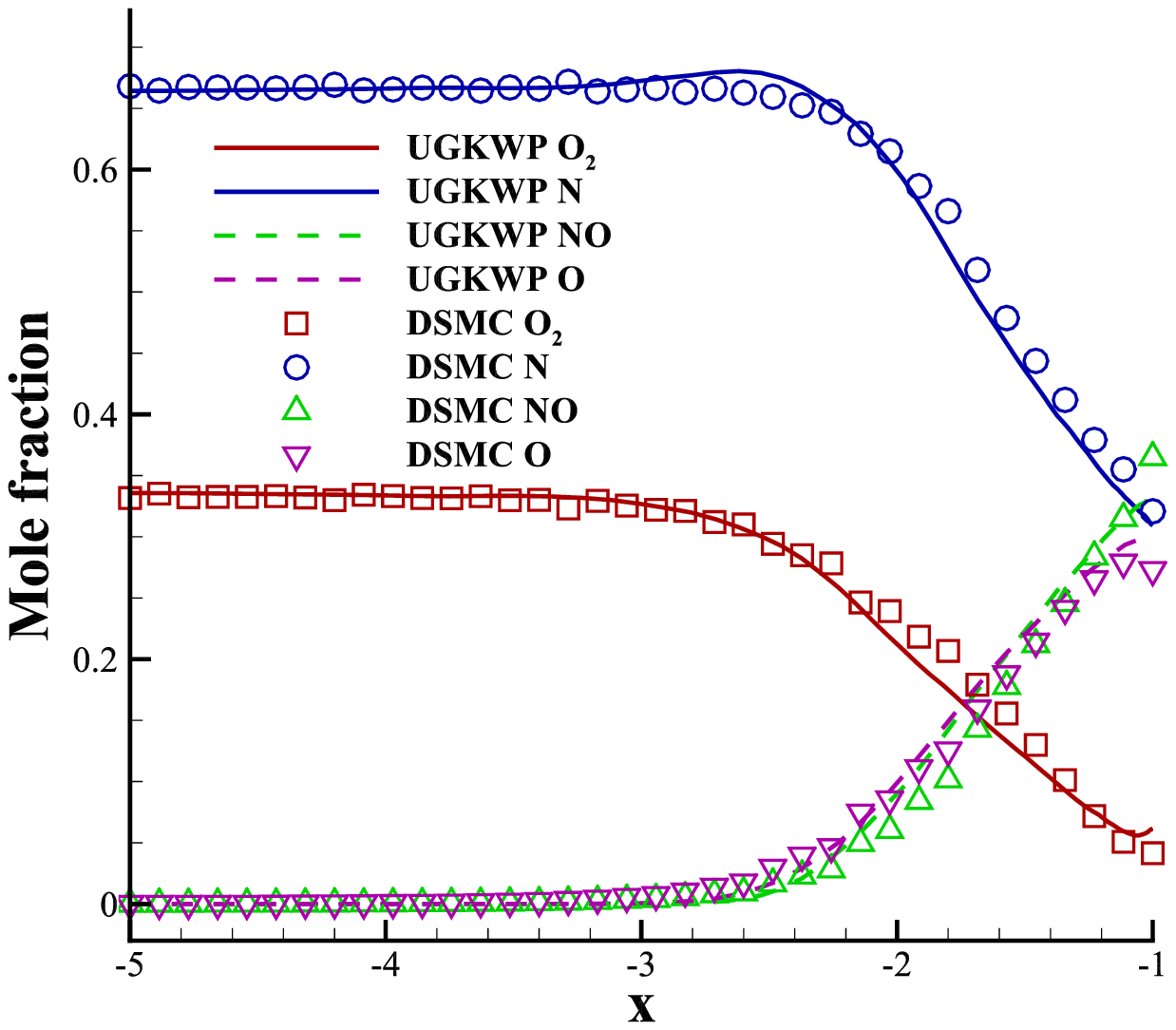}
    	}
    \subfigure[]{
    		\includegraphics[width=0.3 \textwidth]{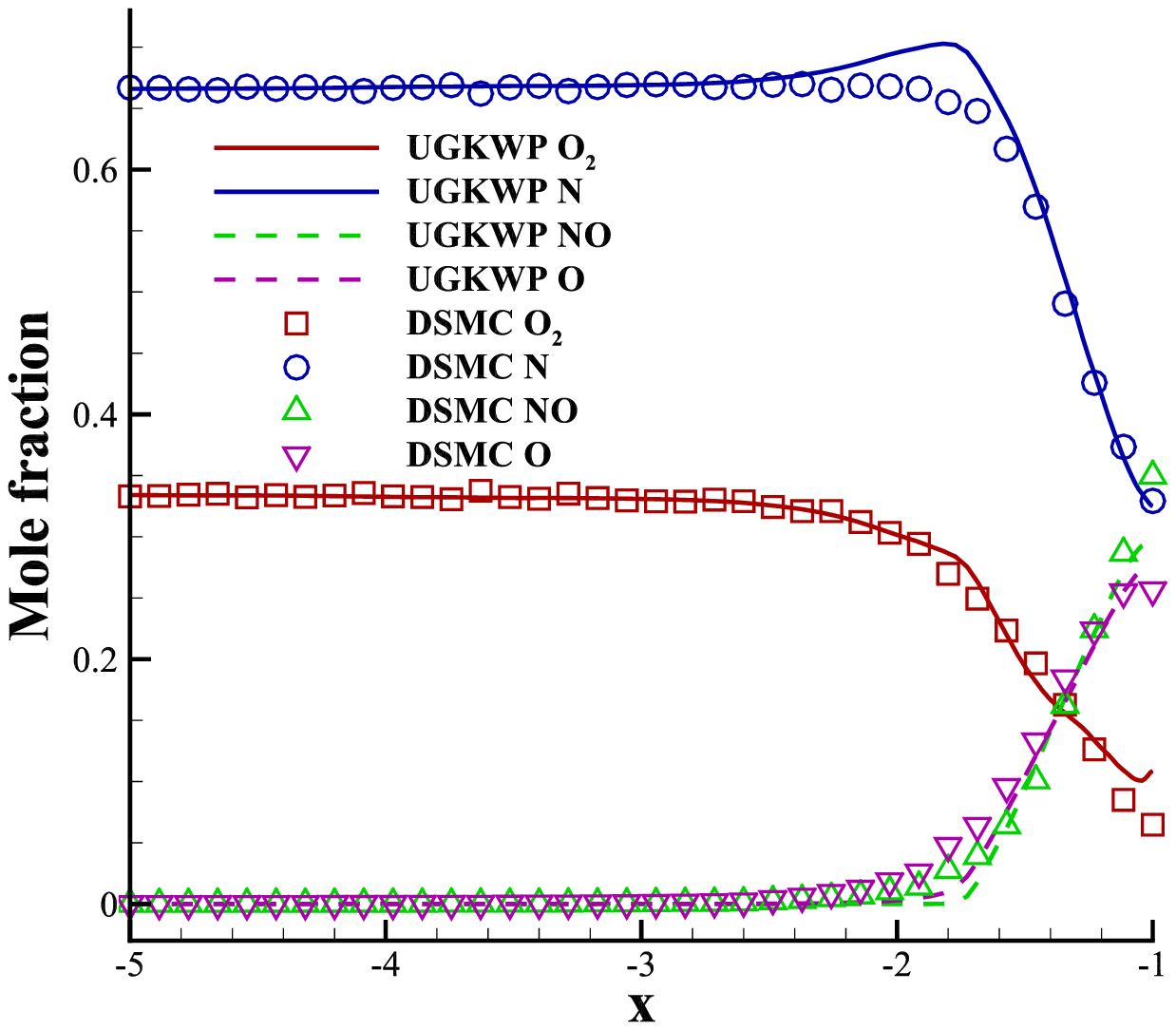}
    	}
	\caption{\label{cylinder-kn0.1-tx} Hypersonic flow around a cylinder at ${\rm{Ma}}_{\infty}=9$, ${\rm{Kn}}_{\infty}=0.1$, and chemical reaction $O_2+N\rightleftharpoons NO+O$ is considered: (a) Gas mixture temperature contour for $\Delta E=0{\rm{J}}$ case, (b) gas mixture temperature for forward exothermic reaction, (c) gas mixture temperature for forward endothermic reaction, (d) mole fraction along the stagnation line for $\Delta E=0{\rm{J}}$ case, (e) mole fraction along the stagnation line for forward exothermic reaction, (f) mole fraction along the stagnation line for forward endothermic reaction.}
\end{figure}

\begin{figure}[H]
	\centering
	\subfigure[]{
			\includegraphics[width=0.3 \textwidth]{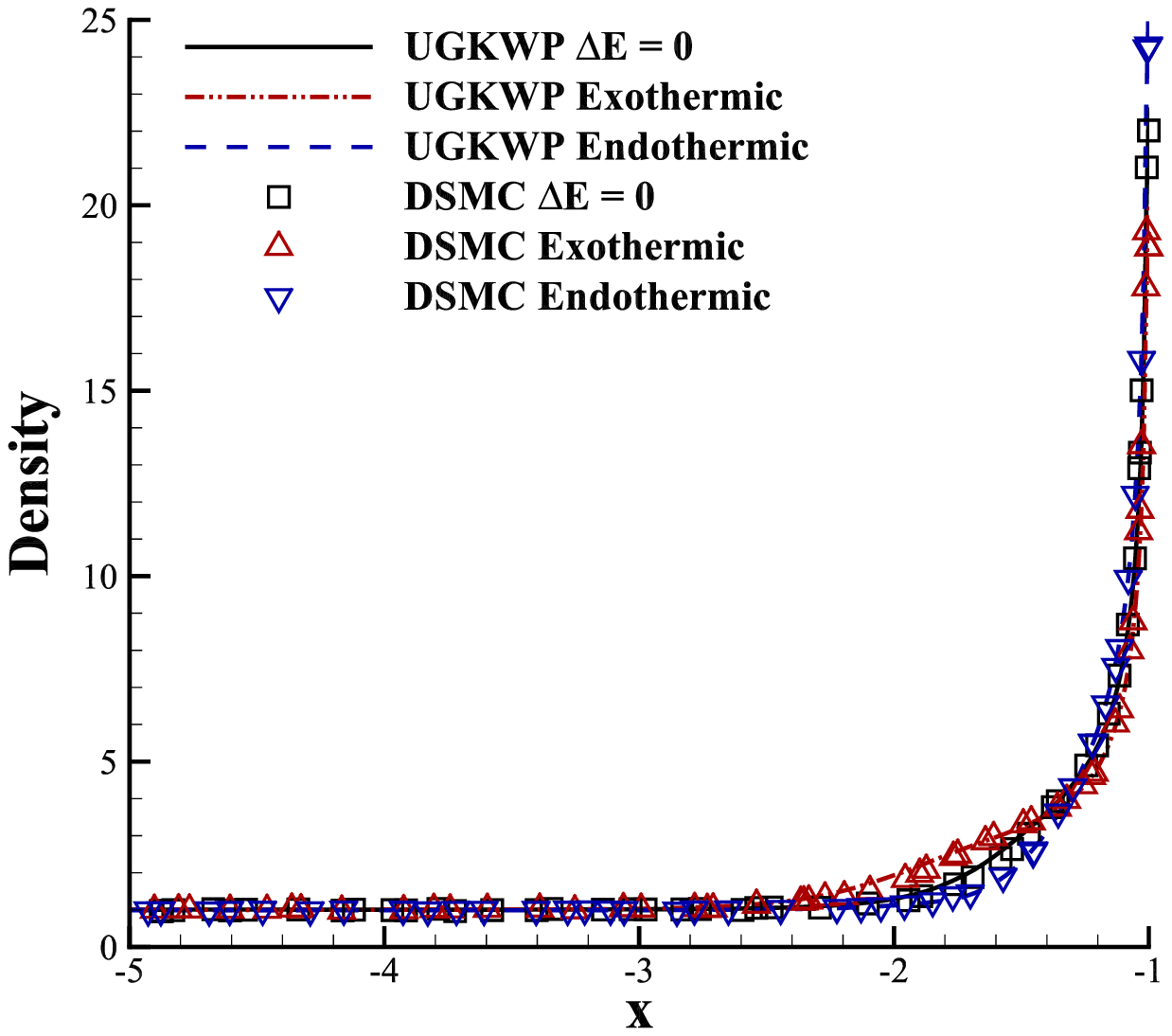}
		}
    \subfigure[]{
    		\includegraphics[width=0.3 \textwidth]{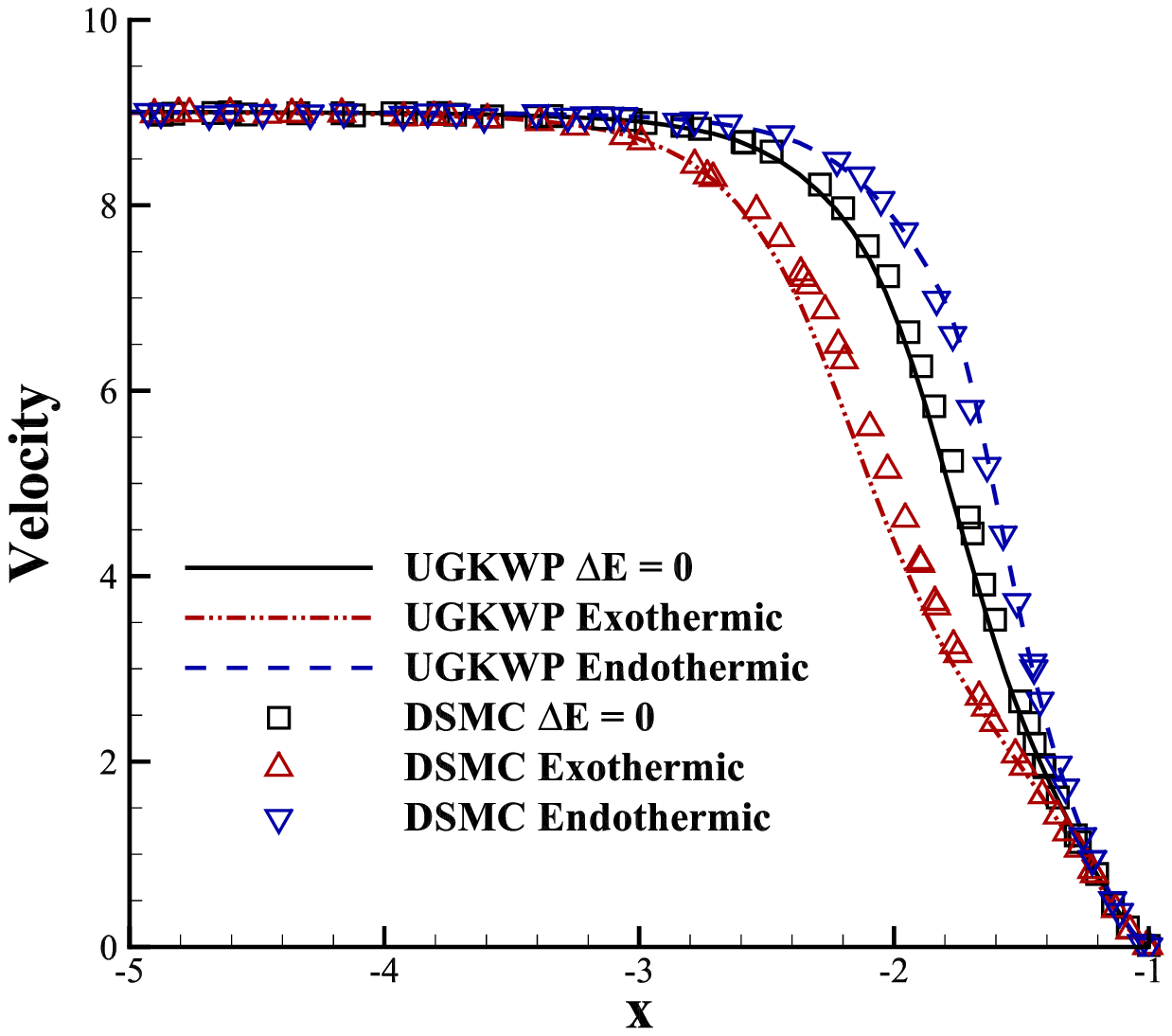}
    	}
    \subfigure[]{
    		\includegraphics[width=0.3 \textwidth]{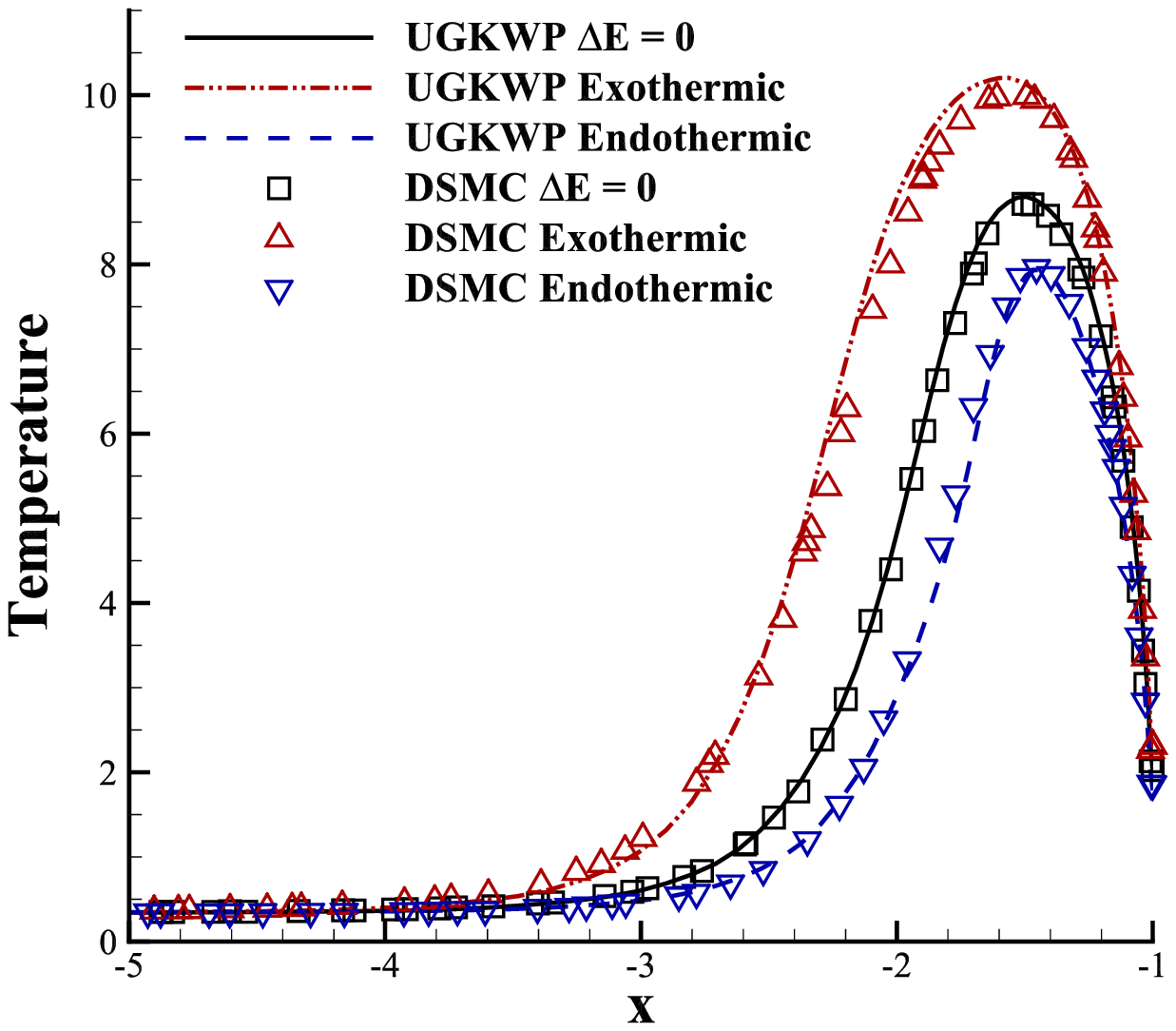}
    	}
    \\
    \subfigure[]{
			\includegraphics[width=0.3 \textwidth]{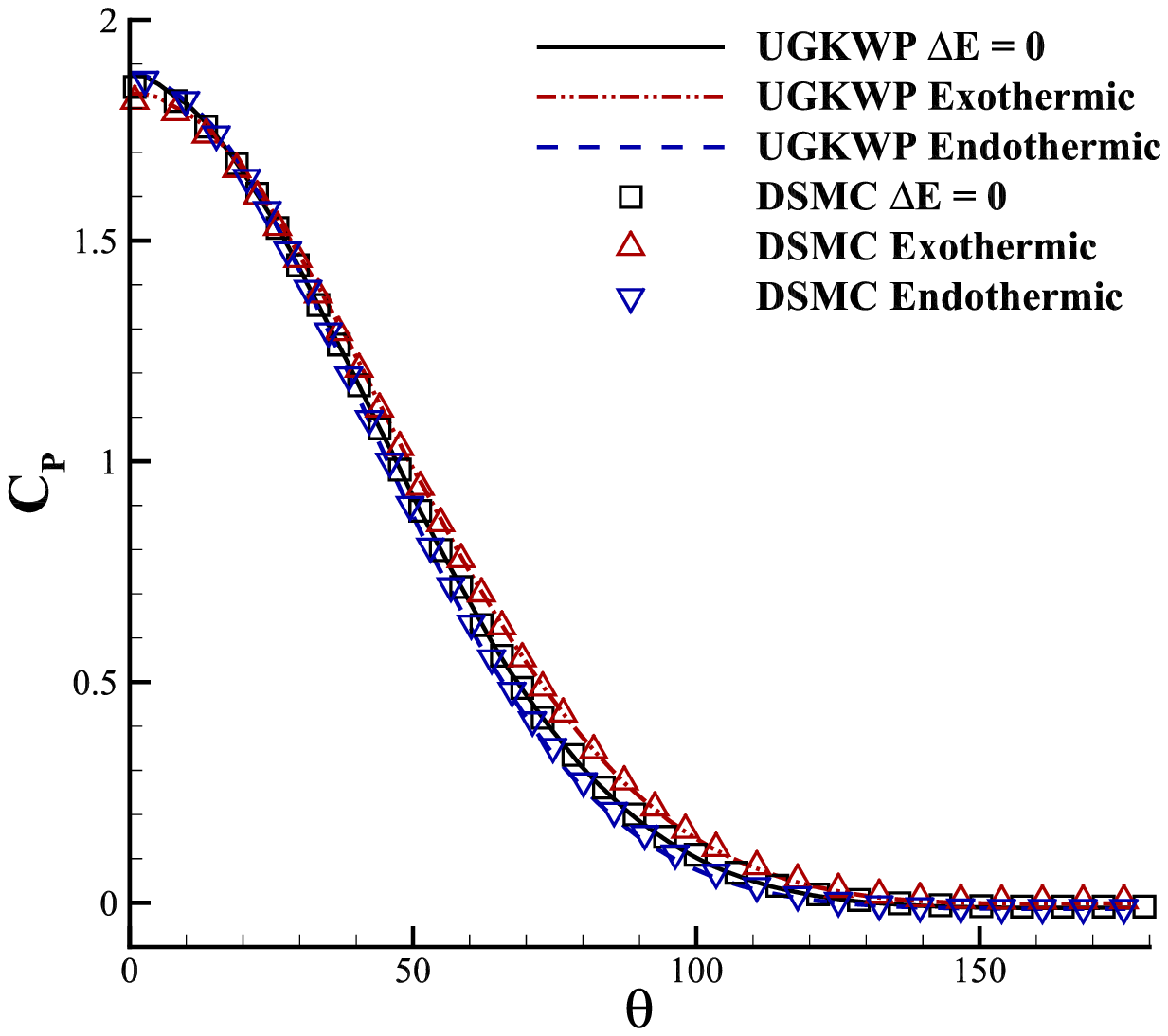}
		}
    \subfigure[]{
    		\includegraphics[width=0.3 \textwidth]{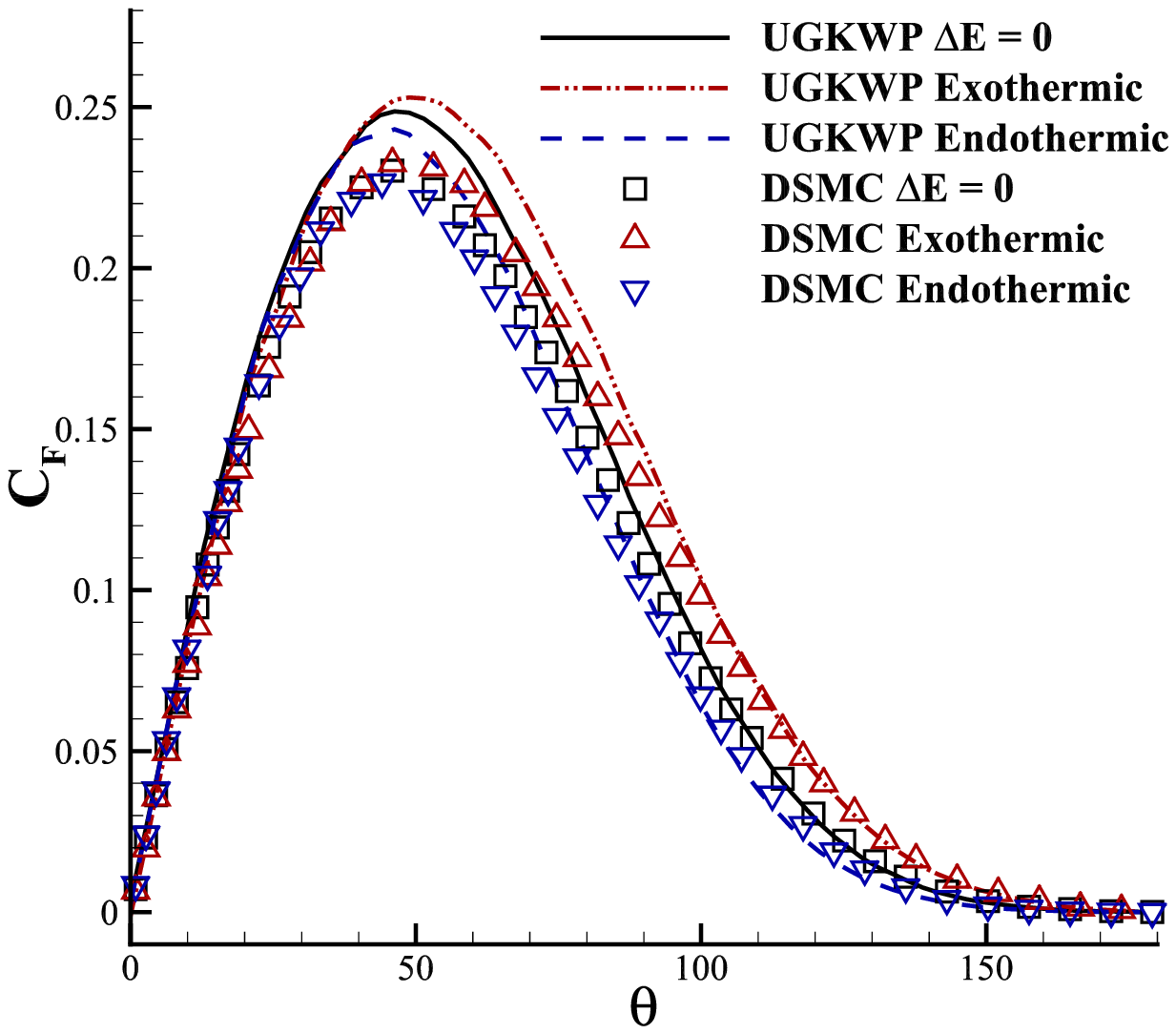}
    	}
    \subfigure[]{
    		\includegraphics[width=0.3 \textwidth]{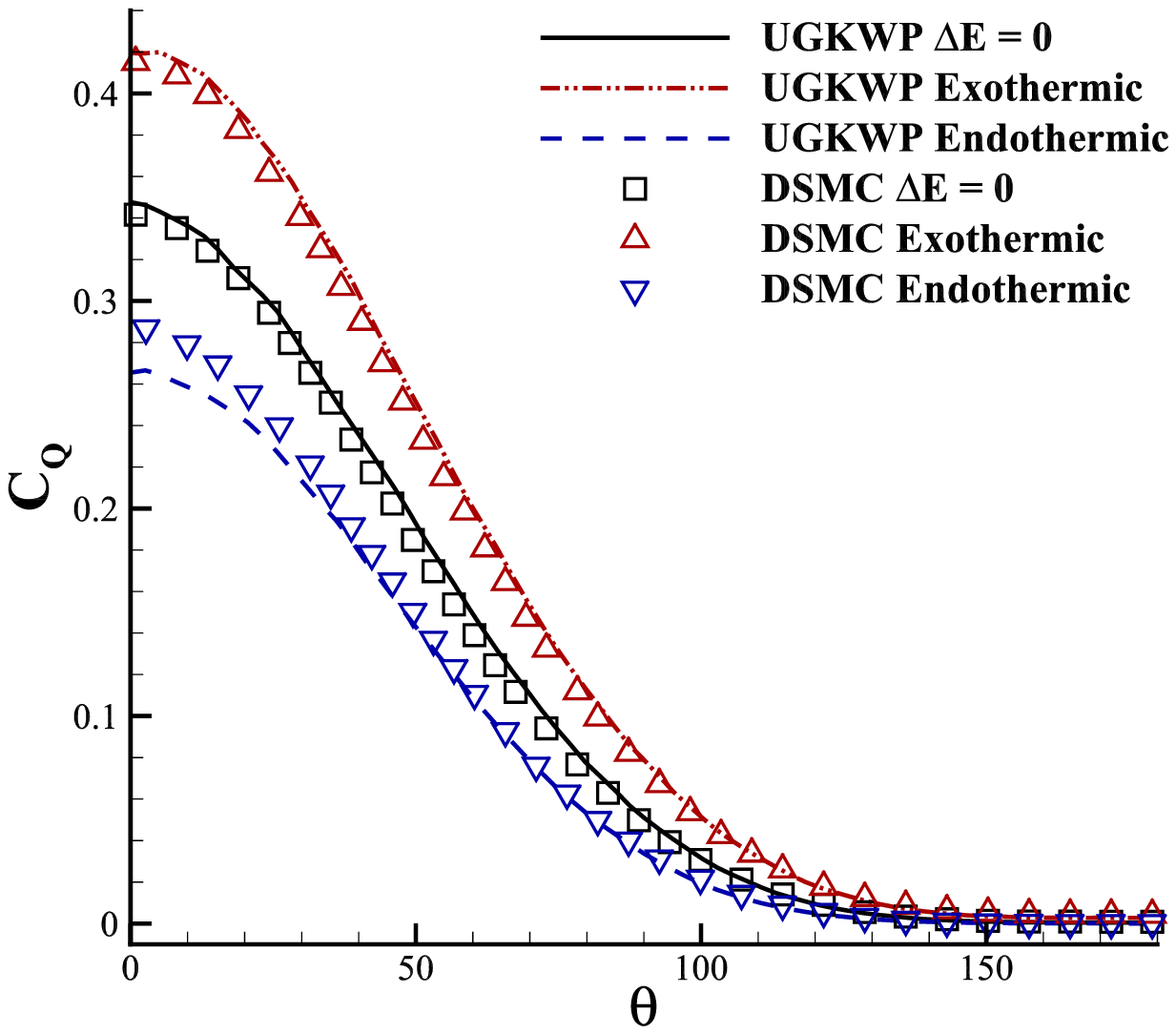}
    	}
	\caption{\label{cylinder-kn0.1-rhoutwall} Hypersonic flow around a cylinder at ${\rm{Ma}}_{\infty}=9$, ${\rm{Kn}}_{\infty}=0.1$, and chemical reaction $O_2+N\rightleftharpoons NO+O$ is considered: (a) Gas mixture density along the stagnation line, (b) gas mixture velocity along the stagnation line, (c) gas mixture temperature along the stagnation line, (d) pressure coefficients at the wall, (e) shear stress coefficients at the wall, (f) heat flux coefficients at the wall.}
\end{figure}

\begin{figure}[H]
	\centering
	\subfigure[]{
			\includegraphics[width=0.3 \textwidth]{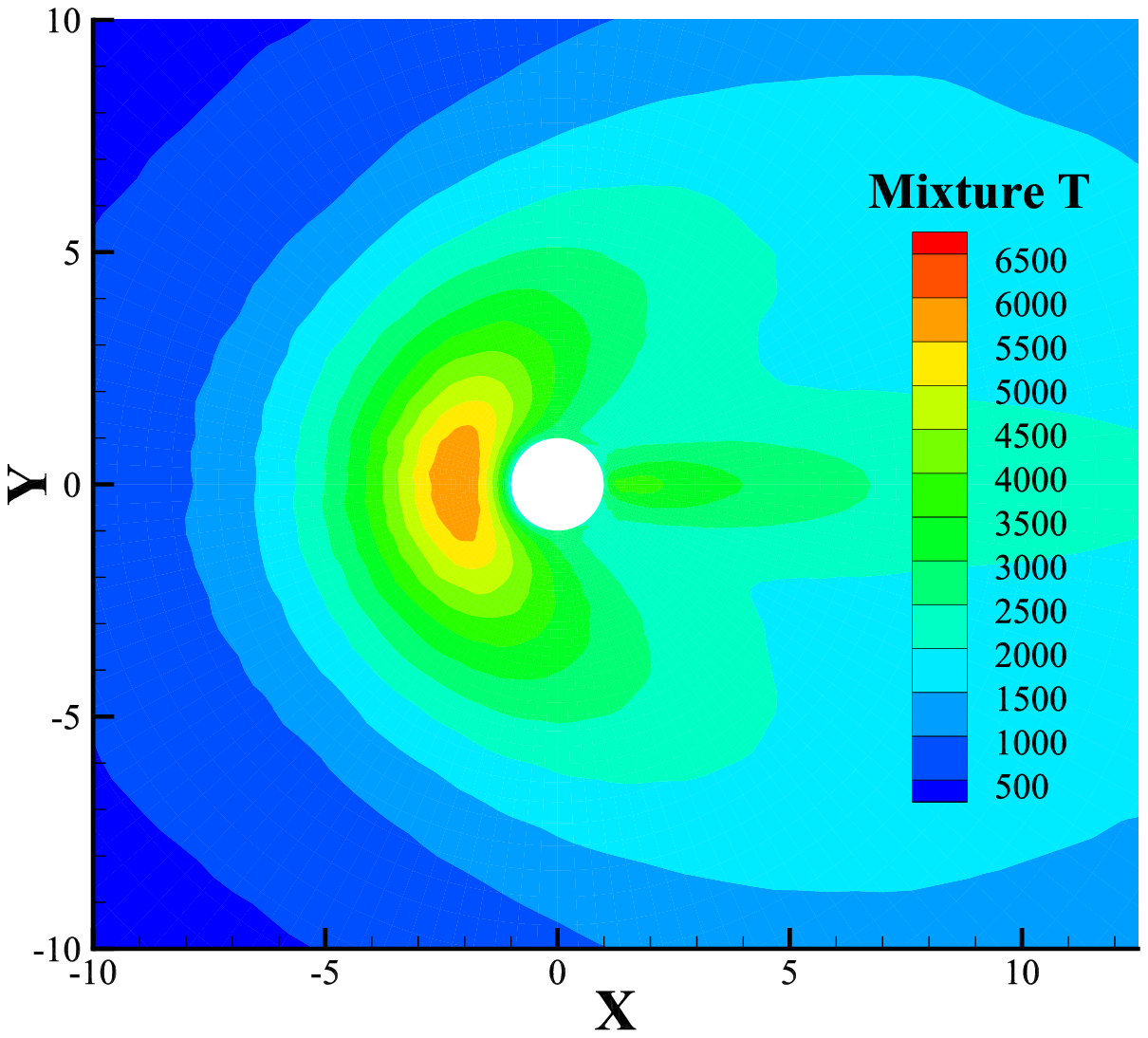}
		}
    \subfigure[]{
    		\includegraphics[width=0.3 \textwidth]{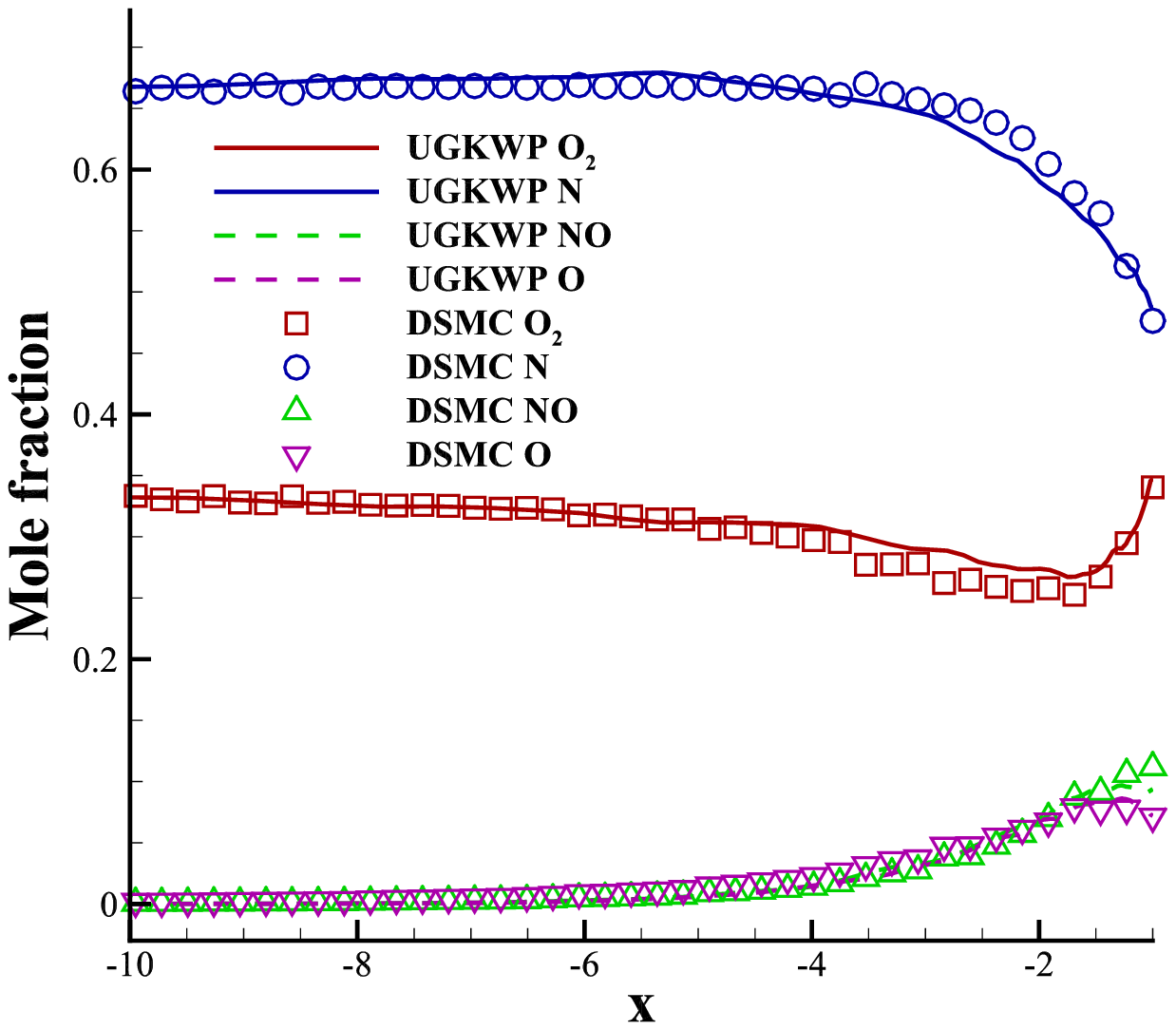}
    	}
    \subfigure[]{
    		\includegraphics[width=0.3 \textwidth]{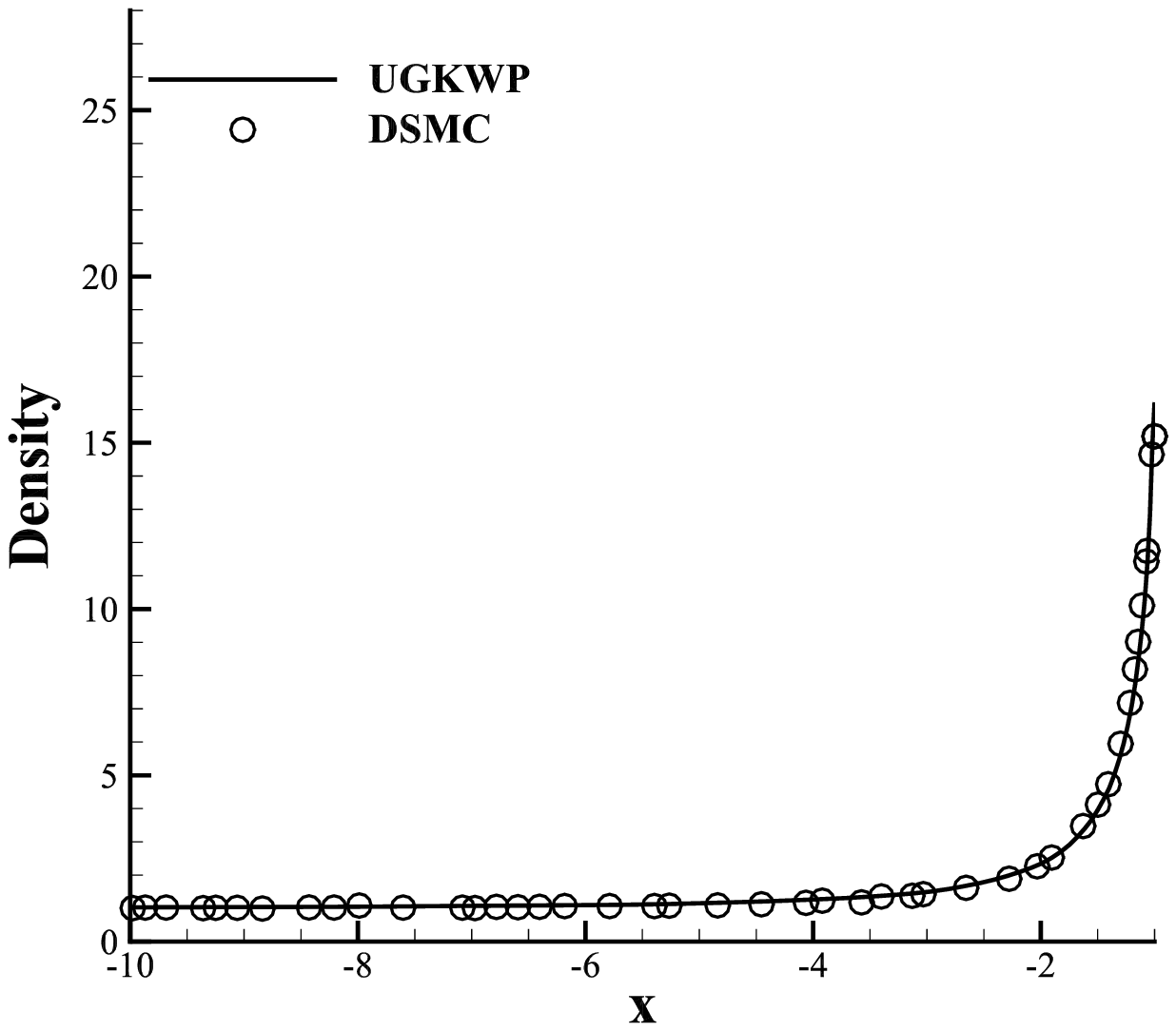}
    	}
    \\
    \subfigure[]{
			\includegraphics[width=0.3 \textwidth]{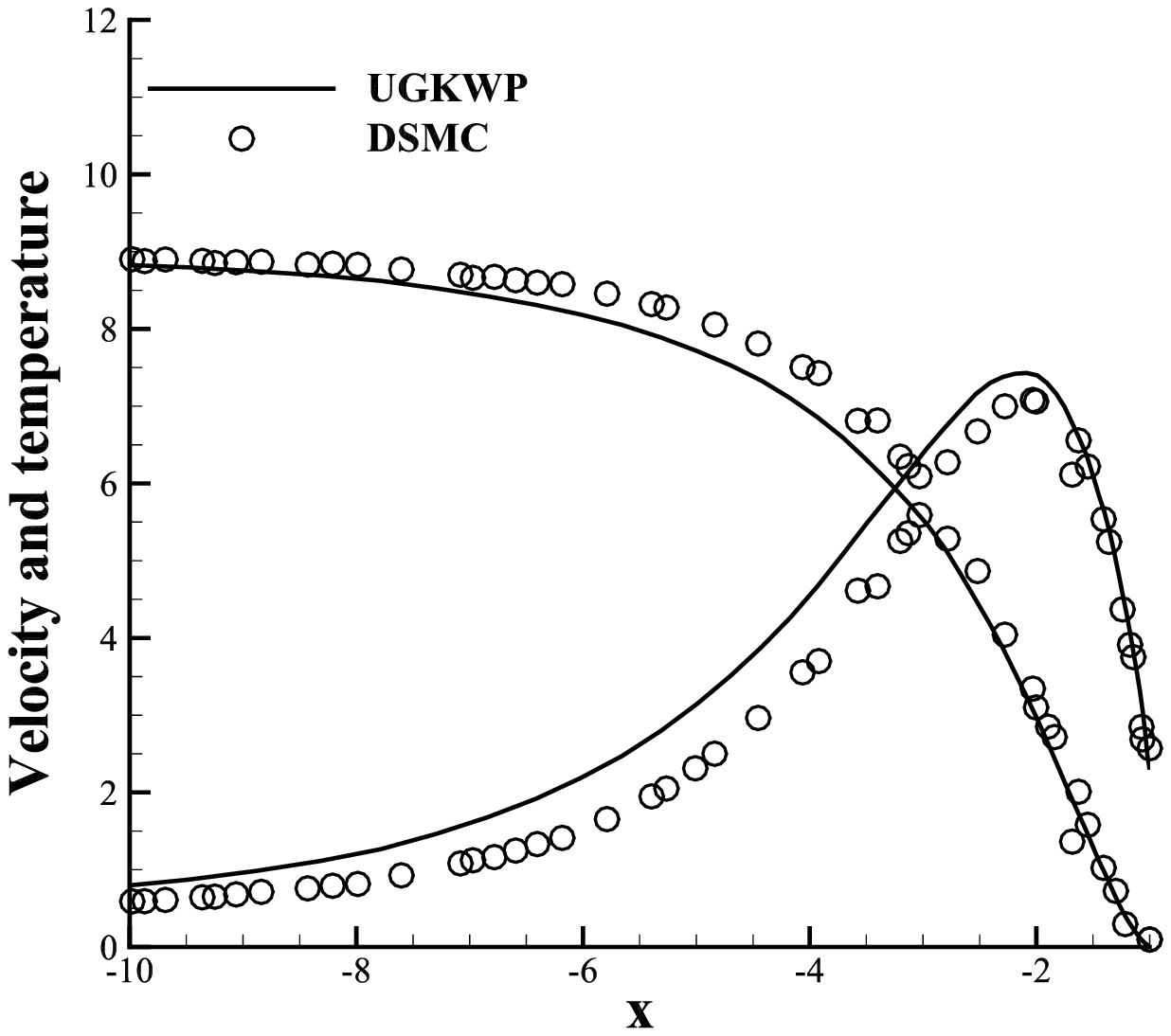}
		}
    \subfigure[]{
    		\includegraphics[width=0.3 \textwidth]{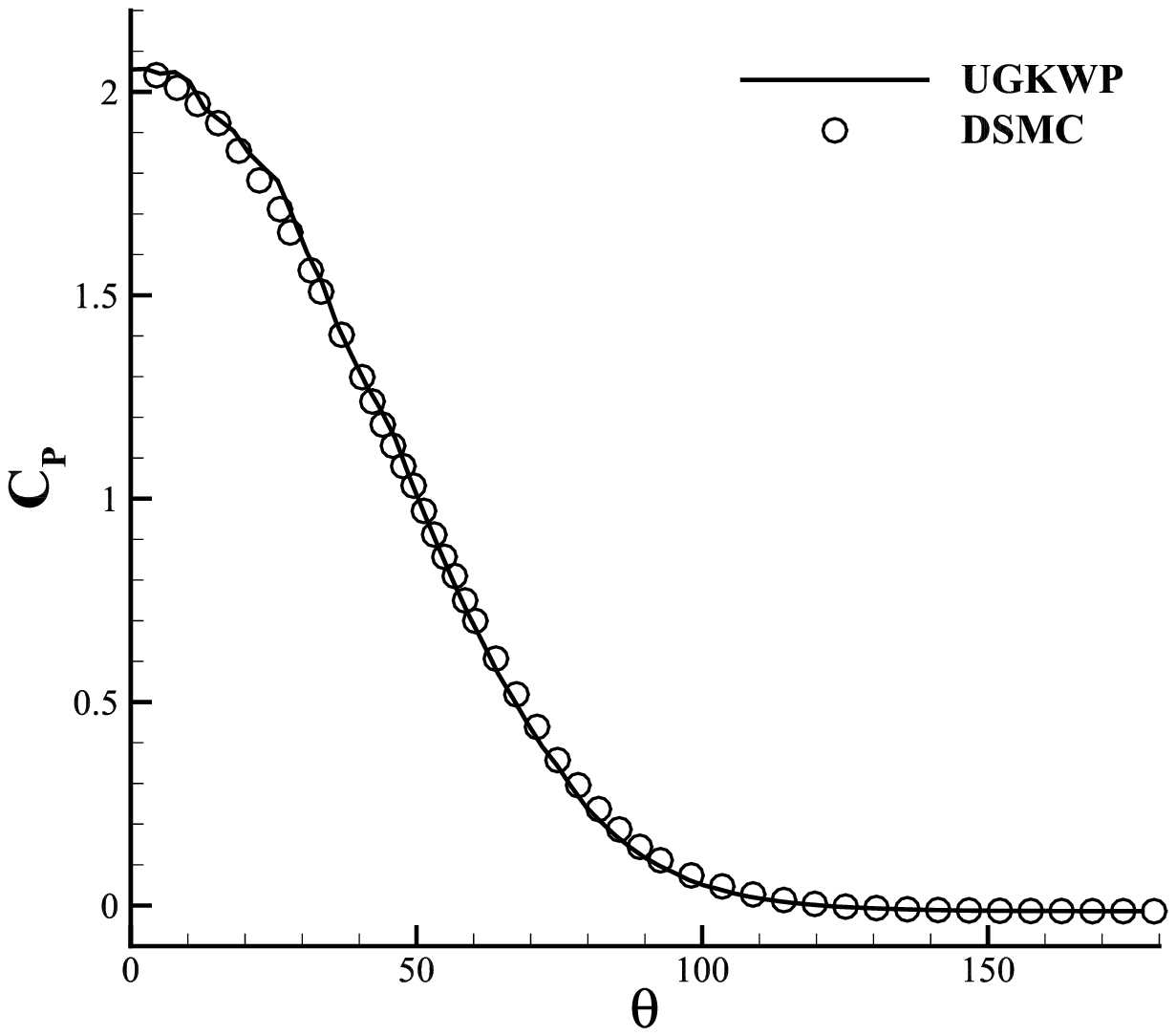}
    	}
    \subfigure[]{
    		\includegraphics[width=0.3 \textwidth]{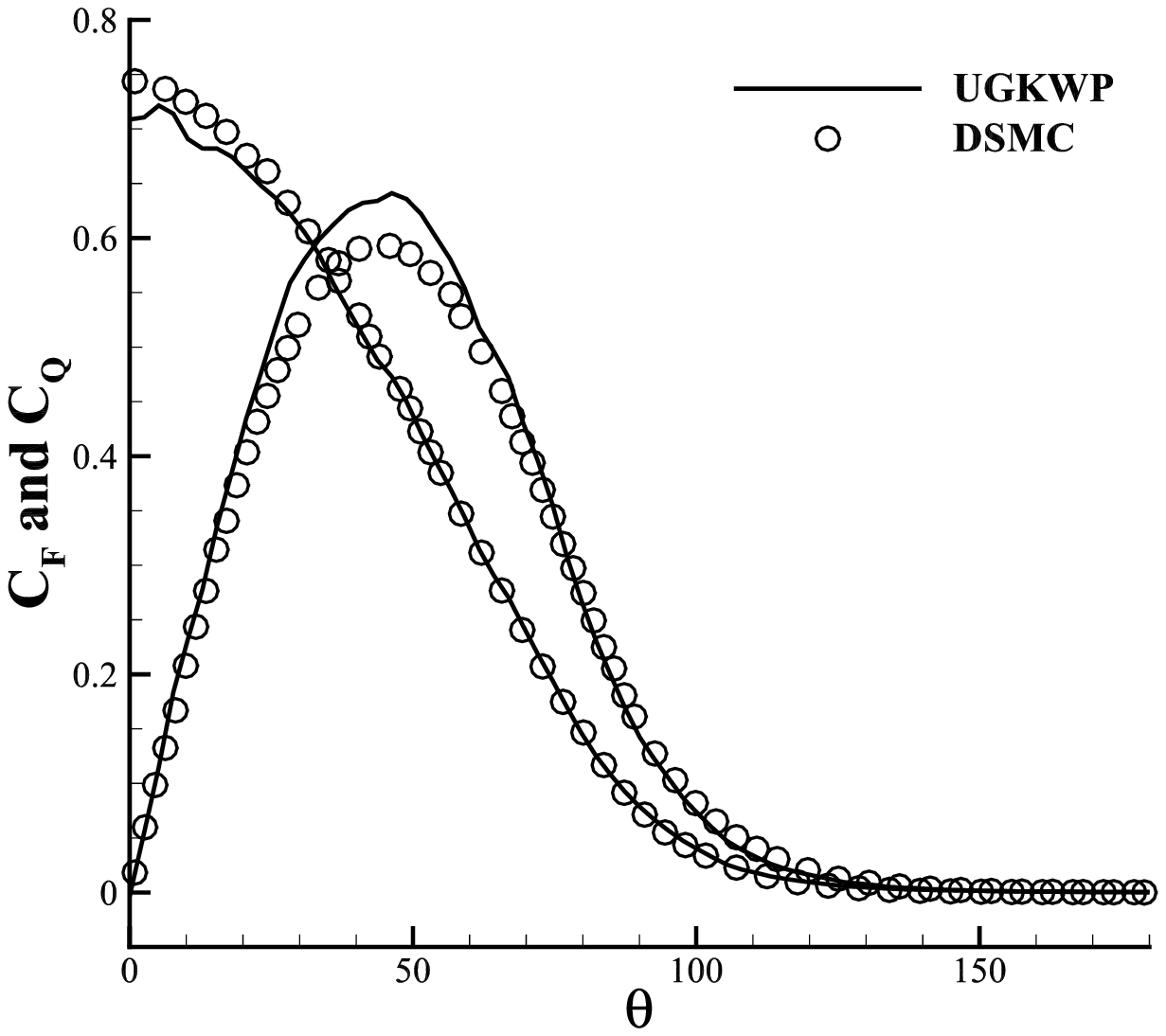}
    	}
	\caption{\label{cylinder-kn1} Hypersonic flow around a cylinder at ${\rm{Ma}}_{\infty}=9$, ${\rm{Kn}}_{\infty}=1$, and forward exothermic chemical reaction $O_2+N\rightleftharpoons NO+O$ is considered: (a) Gas mixture temperature contour, (b) mole fraction along the stagnation line, (c) gas mixture density along the stagnation line, (d) gas mixture velocity and temperature along the stagnation line, (e) pressure coefficient at the wall, (f) shear stress and heat flux coefficients at the wall.}
\end{figure}

\begin{figure}[H]
	\centering
	\subfigure[]{
			\includegraphics[width=0.3 \textwidth]{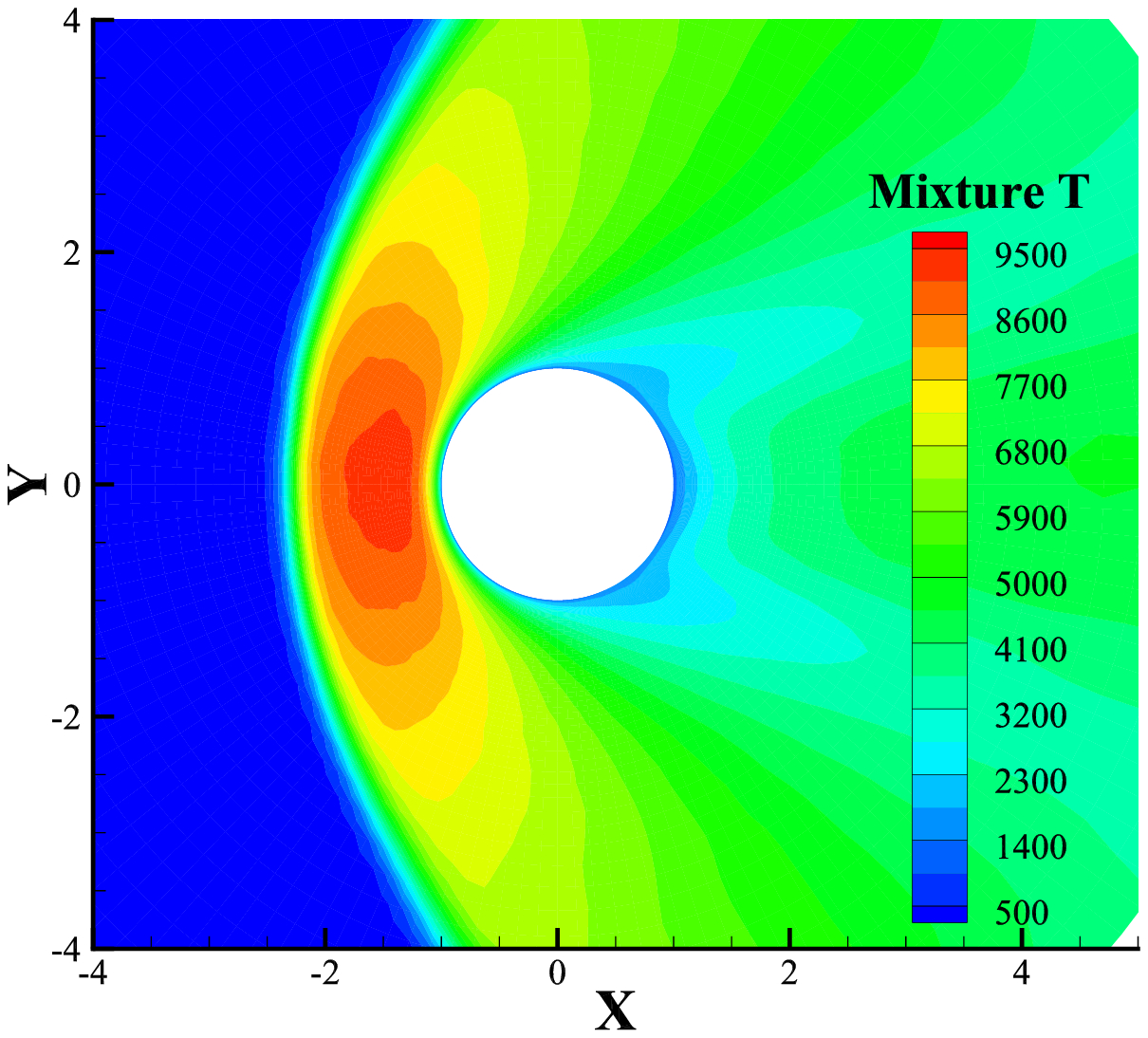}
		}
    \subfigure[]{
    		\includegraphics[width=0.3 \textwidth]{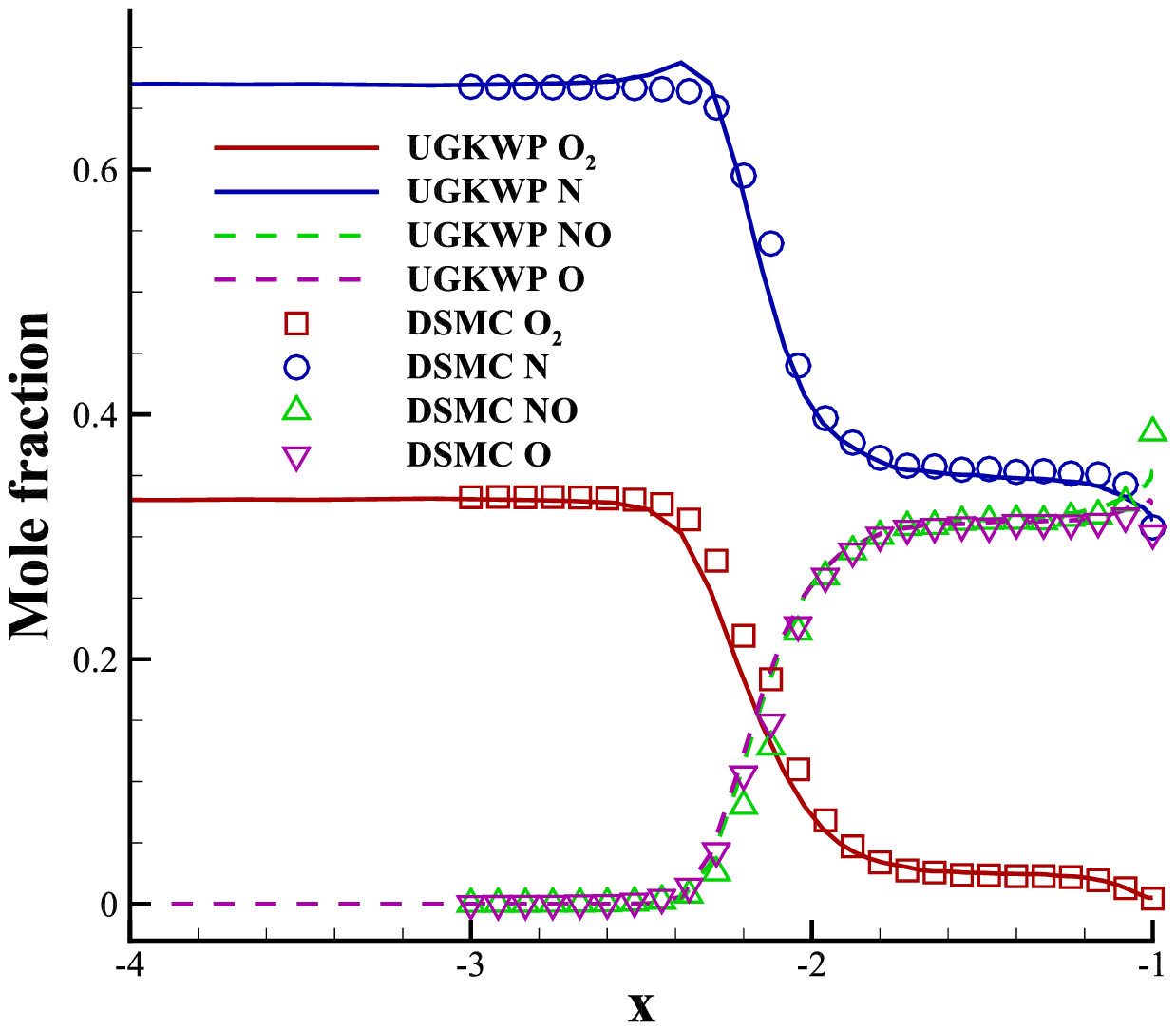}
    	}
    \subfigure[]{
    		\includegraphics[width=0.3 \textwidth]{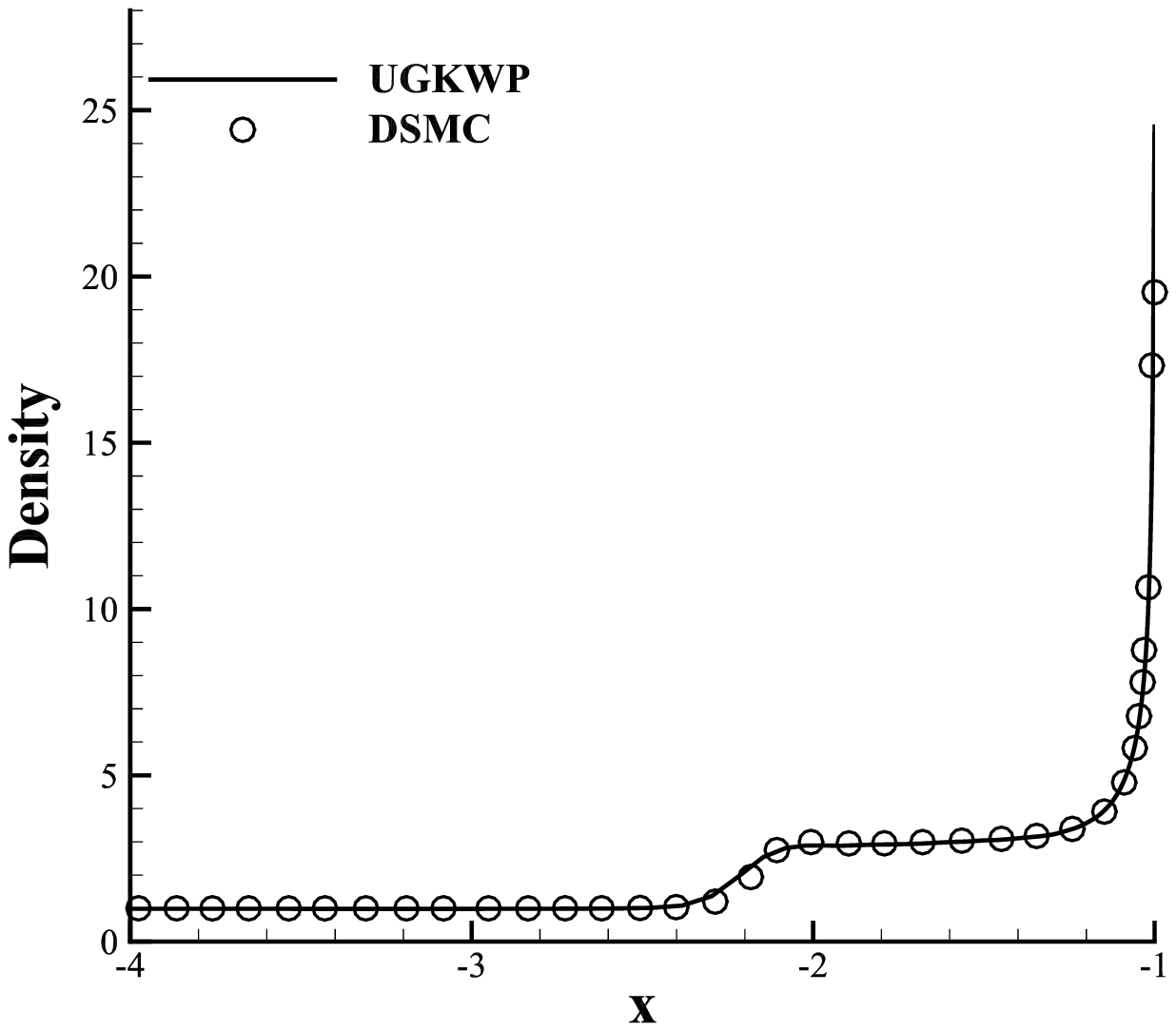}
    	}
    \\
    \subfigure[]{
			\includegraphics[width=0.3 \textwidth]{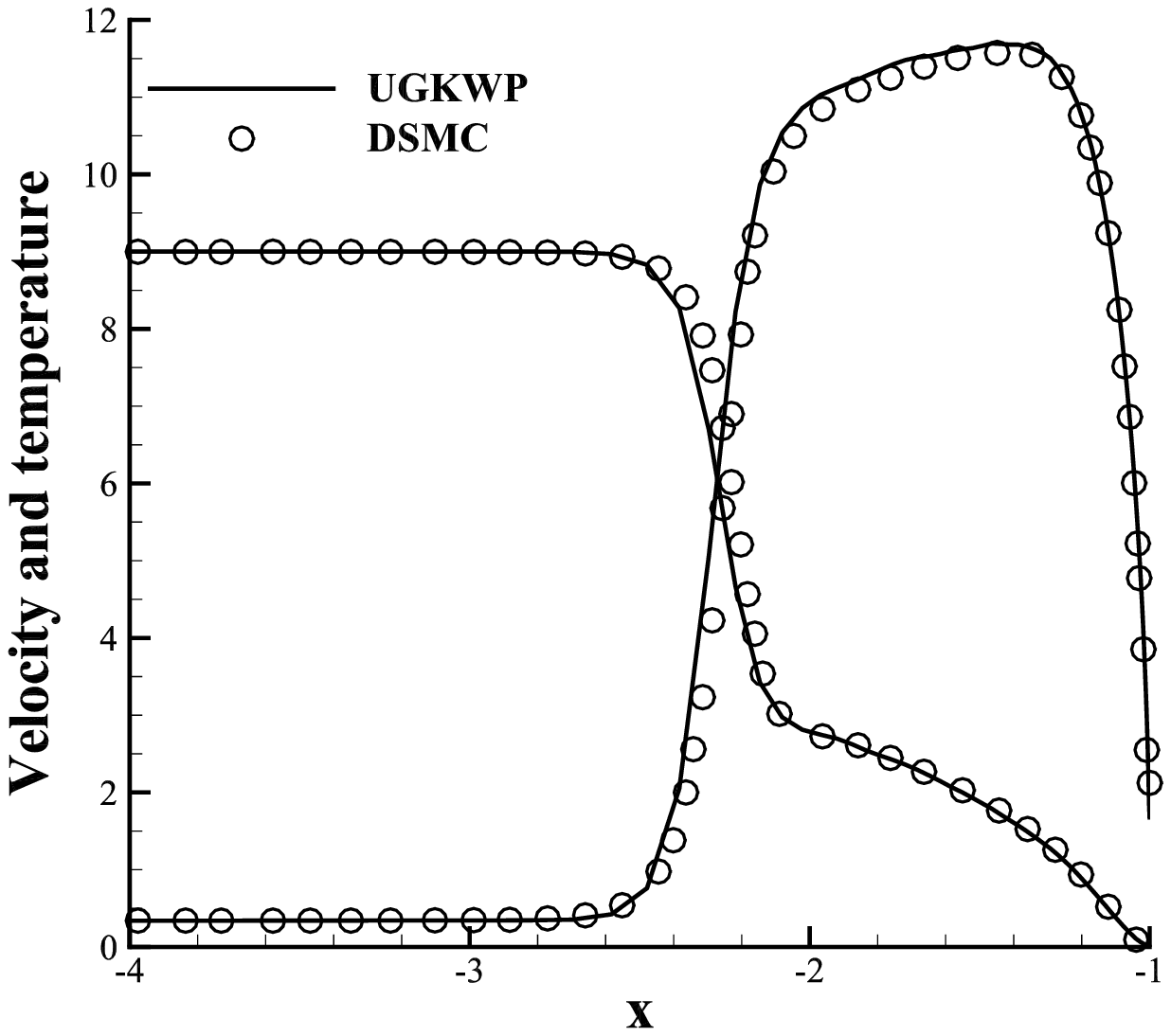}
		}
    \subfigure[]{
    		\includegraphics[width=0.3 \textwidth]{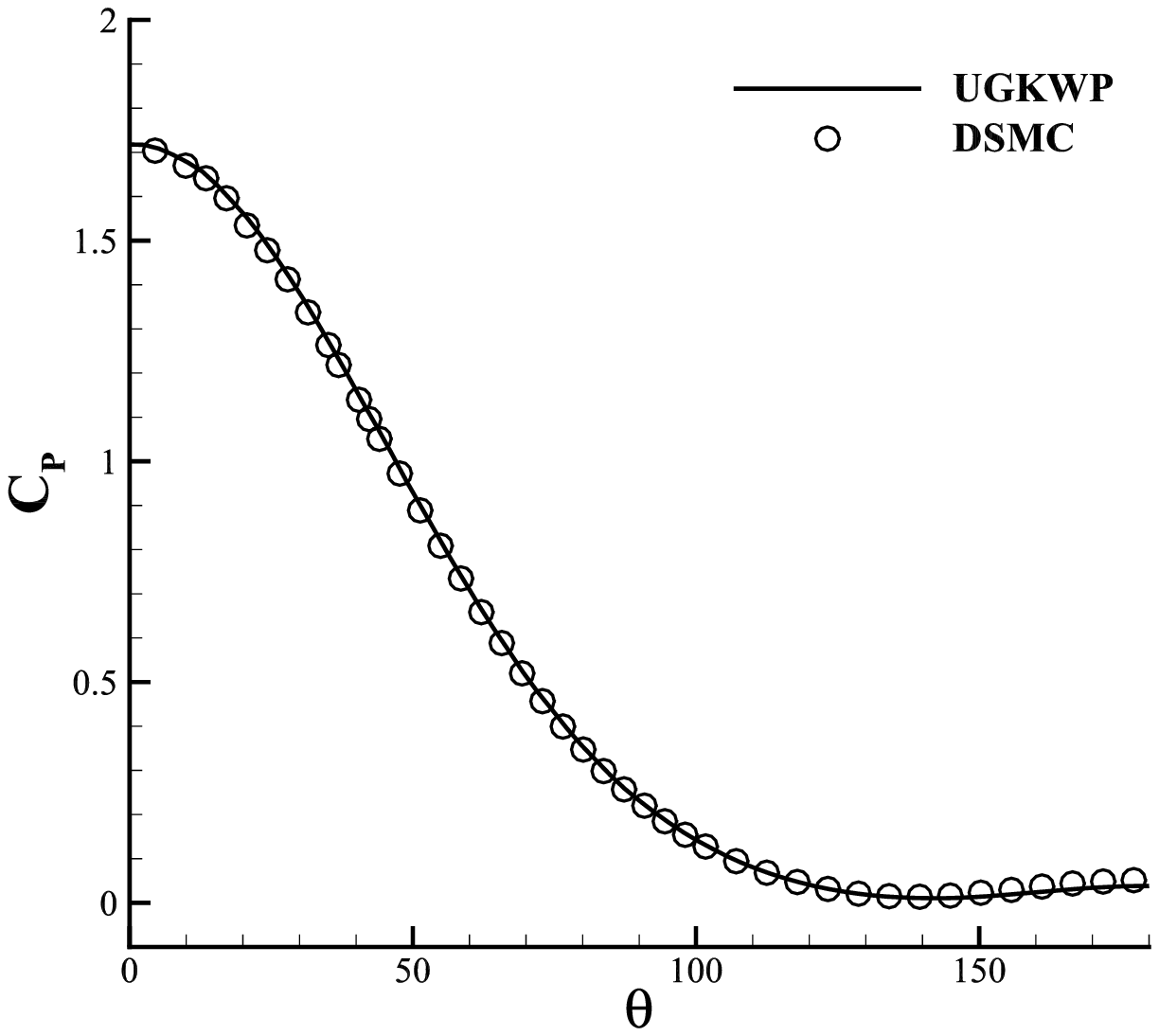}
    	}
    \subfigure[]{
    		\includegraphics[width=0.3 \textwidth]{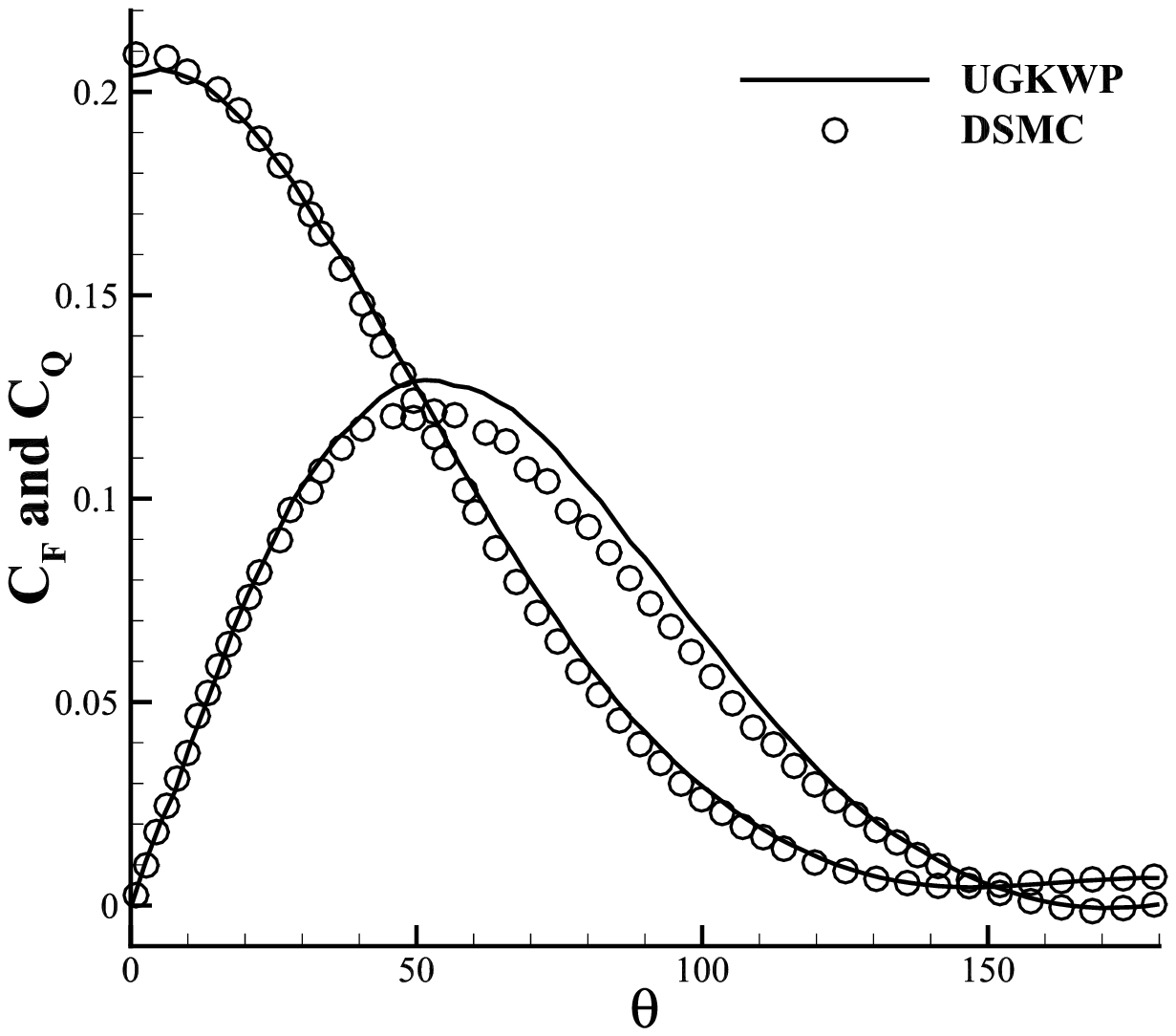}
    	}
	\caption{\label{cylinder-kn0.02} Hypersonic flow around a cylinder at ${\rm{Ma}}_{\infty}=9$, ${\rm{Kn}}_{\infty}=0.02$, and forward exothermic chemical reaction $O_2+N\rightleftharpoons NO+O$ is considered: (a) Gas mixture temperature contour, (b) mole fraction along the stagnation line, (c) gas mixture density along the stagnation line, (d) gas mixture velocity and temperature along the stagnation line, (e) pressure coefficient at the wall, (f) shear stress and heat flux coefficients at the wall.}
\end{figure}

Hypersonic flow past a cylinder is a representative multiscale problem spanning continuum to rarefied regimes. On the windward side, the strong high-$\rm{Ma}$ shock compresses the gas and drives the local Knudsen number toward the continuum, whereas expansion on the leeward side yields a more rarefied state. These features resemble the aerodynamics of near-space vehicles. The present section examines a broad range of ${\rm{Kn}}_{\infty}$ for validation. First we consider the ${\rm{Ma}}_{\infty}=9$ and ${\rm{Kn}}_{\infty}=0.1$ case, where,
\begin{equation}\label{eq:makn}
\begin{aligned}
&{\rm{Ma}}_{\infty}=U_{0,\infty}\left.\Big/\right.\sqrt{\frac{5k_BT_{0,\infty}}{3m_{0,\infty}}},\\
&{\rm{Kn}}_{\infty}=\frac{16}{5L_{\rm{ref}}}\sqrt{\frac{m_{0,\infty}}{2\pi k_BT_{0,\infty}}}\frac{\mu_{0,\infty}}{\rho_{0,\infty}},
\end{aligned}
\end{equation}
where reference length $L_{\rm{ref}}=1{\rm{m}}$ is the cylinder radius. The inflow mole fraction is set to be $\chi_{O_2}:\chi_{N}=1:2$. For dimensional quantities, inflow variables are set to be $\rho_{O_2,\infty}=3.1087\times10^{-7}{\rm{kg/m^3}}$, $\rho_{N,\infty}=2.7213\times10^{-7}{\rm{kg/m^3}}$, $U_{O_2,\infty}=U_{N,\infty}=3914.5{\rm{m/s}}$, and $T_{O_2,\infty}=T_{N,\infty}=273{\rm{K}}$, and wall temperature is set to be $800{\rm{K}}$. When exhibiting stagnation-line results, density, velocity and temperature are nondimensionalized by $\rho_{0,\infty}$, $\sqrt{\frac{5k_BT_{0,\infty}}{3m_{0,\infty}}}$, and wall temperature, respectively. The computational domain is an annular region with $140$ cells in the circumferential direction, and the height of first-layer mesh is set to be $0.01{\rm{m}}$. To begin with, we examine the chemically inert case as shown in Fig.~\ref{cylinder-kn0.1-binary}. When only multispecies effects are taken into account, the present method agrees well with DSMC. Then all cases of $\Delta E=0{\rm{J}}$, forward exothermic (original one), forward endothermic chemical reaction $O_2+N\rightleftharpoons NO+O$ are simulated. As shown in Fig.~\ref{cylinder-kn0.1-tx} and Fig.~\ref{cylinder-kn0.1-rhoutwall}, compared with $\Delta E=0{\rm{J}}$ case, the forward exothermic reaction has larger temperature peak value, further shock location, and larger heat flux peak value. For the forward endothermic reaction, it has lower temperature peak value, nearer shock location, and lower heat flux peak value. For all cases, pressure coefficients at the wall are similar, and shear stress coefficients have small deviations. Results of the present UGKWP method match well with the DSMC, while there are small deviations in the $C_Q$ result for the forward endothermic case, and $C_F$ results. Additionally, a more rarefied case ${\rm{Kn}}_{\infty}=1$ is tested as Fig.~\ref{cylinder-kn1}, and a more continuum one ${\rm{Kn}}_{\infty}=0.02$ is tested as Fig.~\ref{cylinder-kn0.02}. The height of first-layer mesh is decreased to $0.002{\rm{m}}$ for the ${\rm{Kn}}_{\infty}=0.02$ case. Because the flow is closer to the continuum regime, the chemical reactions proceed more completely, and $\chi_{O_2}$ approaches zero near the stagnation point. In contrast, for the ${\rm{Kn}}_{\infty}=1$ case the chemical reactions remain relatively weak, and the mole fraction of products accumulates only to about $10\%$ near the stagnation point. Compared with the DSMC, both simulations have accurate results.

\subsection{Shock structure}\label{sec:shock}
Shock structure is a typical 1D microscopic test of nonequilibrium models and numerical methods. Rarefied transport, multispecies effect, and chemistry interact strongly in this configuration. Unlike a hypersonic cylinder flow, where reactions stay weak in the cold freestream and become active only behind the shock, both upstream and downstream states here are hot enough for chemistry, with asymptotic equilibrium conditions prescribed by the law of mass action $k_b/k_f=\prod\limits_\beta^{CR}n_\beta\left.\Big/\right.\prod\limits_\beta^{CP}n_\beta$. Considering the chemical reaction effect, the Rankine--Hugoniot conditions are derived in Ref.~\cite{groppi3}, and the downstream parameters can be calculated by,
\begin{equation}\label{eq:pretopost}
\begin{aligned}
\frac{n_{0}^{+}}{n_{0}^{-}} &= 2\left(1-\frac{T_{0}^{-}}{T_{0}^{+}}\right)-\frac{-\Delta E}{k_BT_{0}^{+}}\Delta \chi + \sqrt{\left[2\left(1-\frac{T_{0}^{-}}{T_{0}^{+}}\right)-\frac{-\Delta E}{k_BT_{0}^{+}}\Delta \chi\right]^2+\frac{T_{0}^{-}}{T_{0}^{+}}},\\
U_{0}^{-} &= \sqrt{\frac{n_{0}^{-}k_BT_{0}^{-}}{\rho_{0}^{-}}}\sqrt{\frac{n_{0}^{+}}{n_{0}^{-}}\frac{1-(n_{0}^{+}/n_{0}^{-})(T_{0}^{+}/T_{0}^{-})}{1-(n_{0}^{+}/n_{0}^{-})}},\\
U_{0}^{+} &= U_{0}^{-}\frac{n_{0}^{-}}{n_{0}^{+}},
\nonumber
\end{aligned}
\end{equation}
where superscripts ``$-$'' and ``$+$'' denote the upstream and downstream states respectively, and $\Delta \chi$ is the increment of each product after the shock. The sound speed in this case is calculated as Ref.~\cite{groppi3},
\begin{equation}\label{eq:shockma}
\sqrt{\frac{\frac{5}{3}\frac{n_0k_BT_0}{\rho_0}\left[\sum\limits_{\alpha=1}^C{\frac{1}{\chi_{\alpha}}}+\frac{2}{5}\left(\frac{\Delta E}{k_BT_0}\right)^2\right]}{\sum\limits_{\alpha=1}^C{\frac{1}{\chi_{\alpha}}}+\frac{2}{3}\left(\frac{\Delta E}{k_BT_0}\right)^2}},
\nonumber
\end{equation}
and the upstream value is used to define the Mach number. For consistency with Ref.~\cite{groppi3}, the chemical reaction is set to be forward endothermic, $\Delta E=-2.2\times 10^{-19}{\rm{J}}$. Note that our convention for $\Delta E$ is reversed compared with Ref.~\cite{groppi3}. Through an iterative solution, the upstream and downstream parameters are obtained for ${\rm{Ma}}=3$ and ${\rm{Ma}}=5$ cases, as Tab.~\ref{tab-ma3} and Tab.~\ref{tab-ma5}.
\begin{table}[h]
\caption{\text{Upstream and downstream parameters of shock structure at ${\rm{Ma}}=3$}}\label{tab-ma3}
\centering
\begin{tabular}{*{9}{c}}
\toprule
    Gas &$\rho^-\left({\rm{kg/m^3}}\right)$ &$\rho^+\left({\rm{kg/m^3}}\right)$ &$\chi^-$ &$\chi^+$ &$U^-\left({\rm{m/s}}\right)$ &$U^+\left({\rm{m/s}}\right)$ &$T^-\left({\rm{K}}\right)$ &$T^+\left({\rm{K}}\right)$  \\
\midrule
    $O_2$ &$3.0278\times 10^{-7}$ &$2.2513\times 10^{-6}$ &$0.0570$ &$0.1487$ &$5490.5$  &$1927.0$  &$6000$  &$21568$ \\
	$N$   &$2.3250\times 10^{-7}$ &$1.2702\times 10^{-6}$ &$0.1000$ &$0.1917$ &$5490.5$  &$1927.0$  &$6000$  &$21568$ \\
    $NO$  &$2.4402\times 10^{-6}$ &$5.6511\times 10^{-6}$ &$0.4900$ &$0.3983$ &$5490.5$  &$1927.0$  &$6000$  &$21568$ \\
    $O$   &$9.3757\times 10^{-7}$ &$1.9771\times 10^{-6}$ &$0.3530$ &$0.2613$ &$5490.5$  &$1927.0$  &$6000$  &$21568$ \\
\bottomrule
\end{tabular}
\end{table}

\begin{figure}[H]
	\centering
	\subfigure[]{
			\includegraphics[width=0.22 \textwidth]{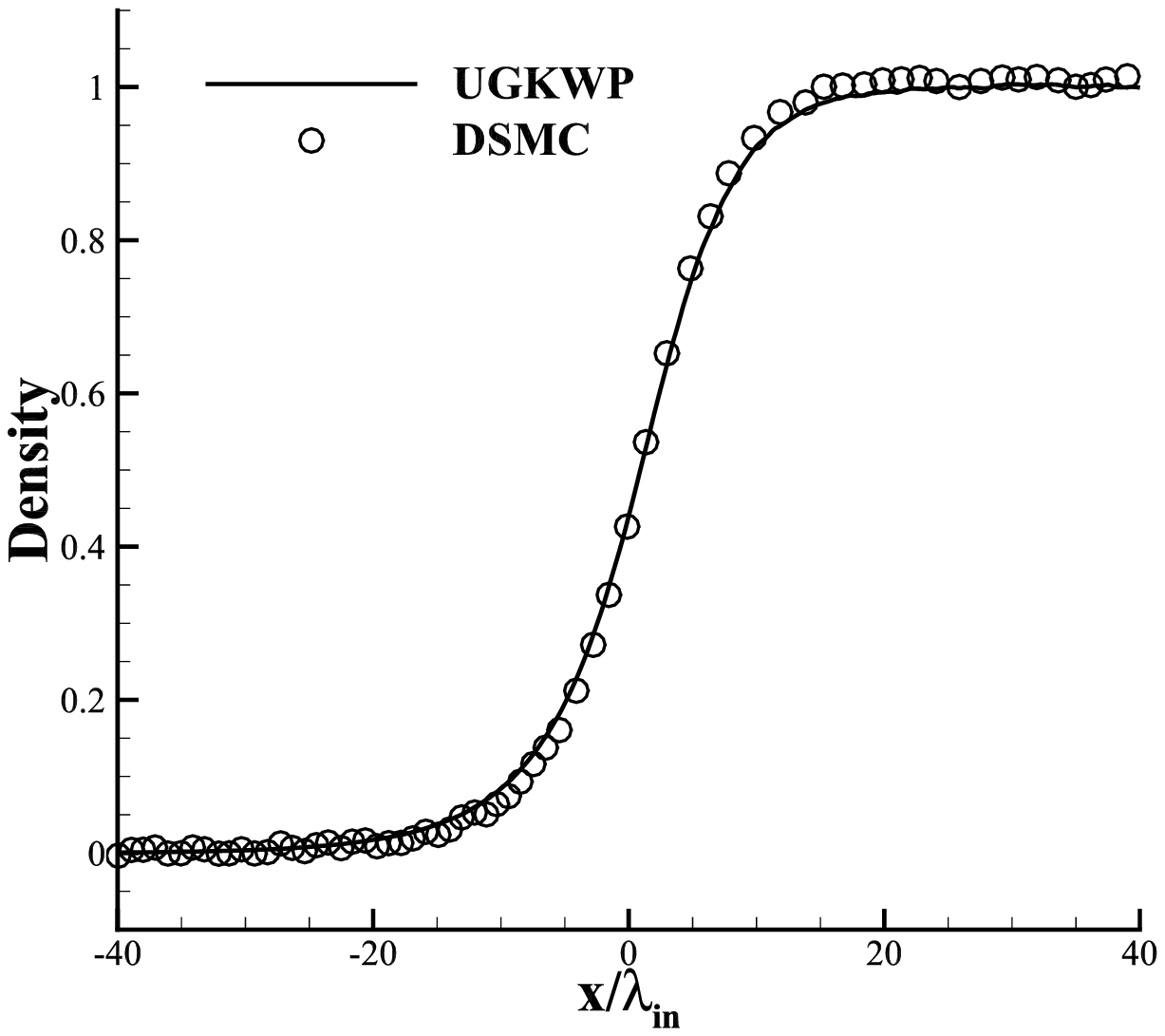}
		}
    \subfigure[]{
    		\includegraphics[width=0.22 \textwidth]{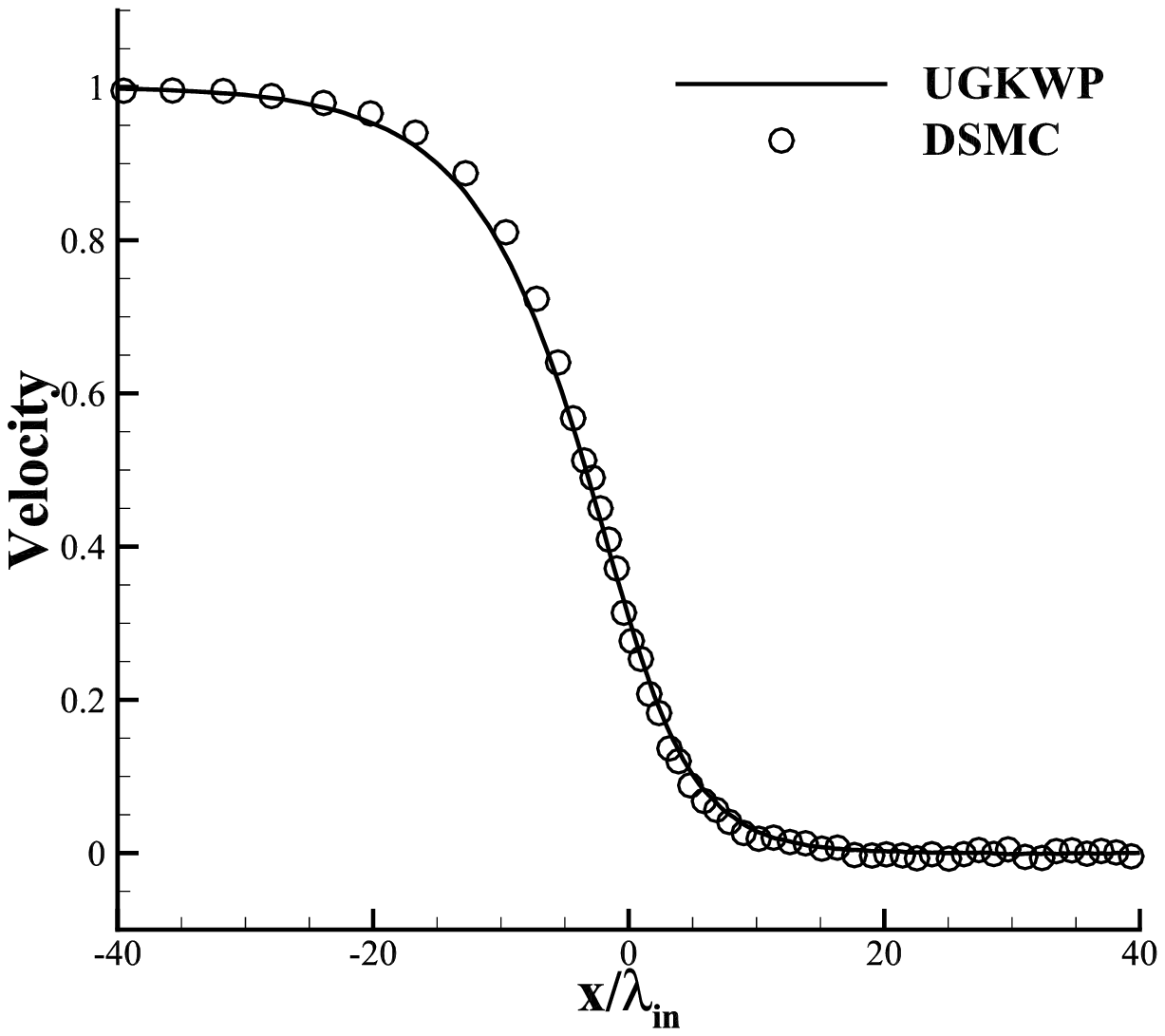}
    	}
    \subfigure[]{
    		\includegraphics[width=0.22 \textwidth]{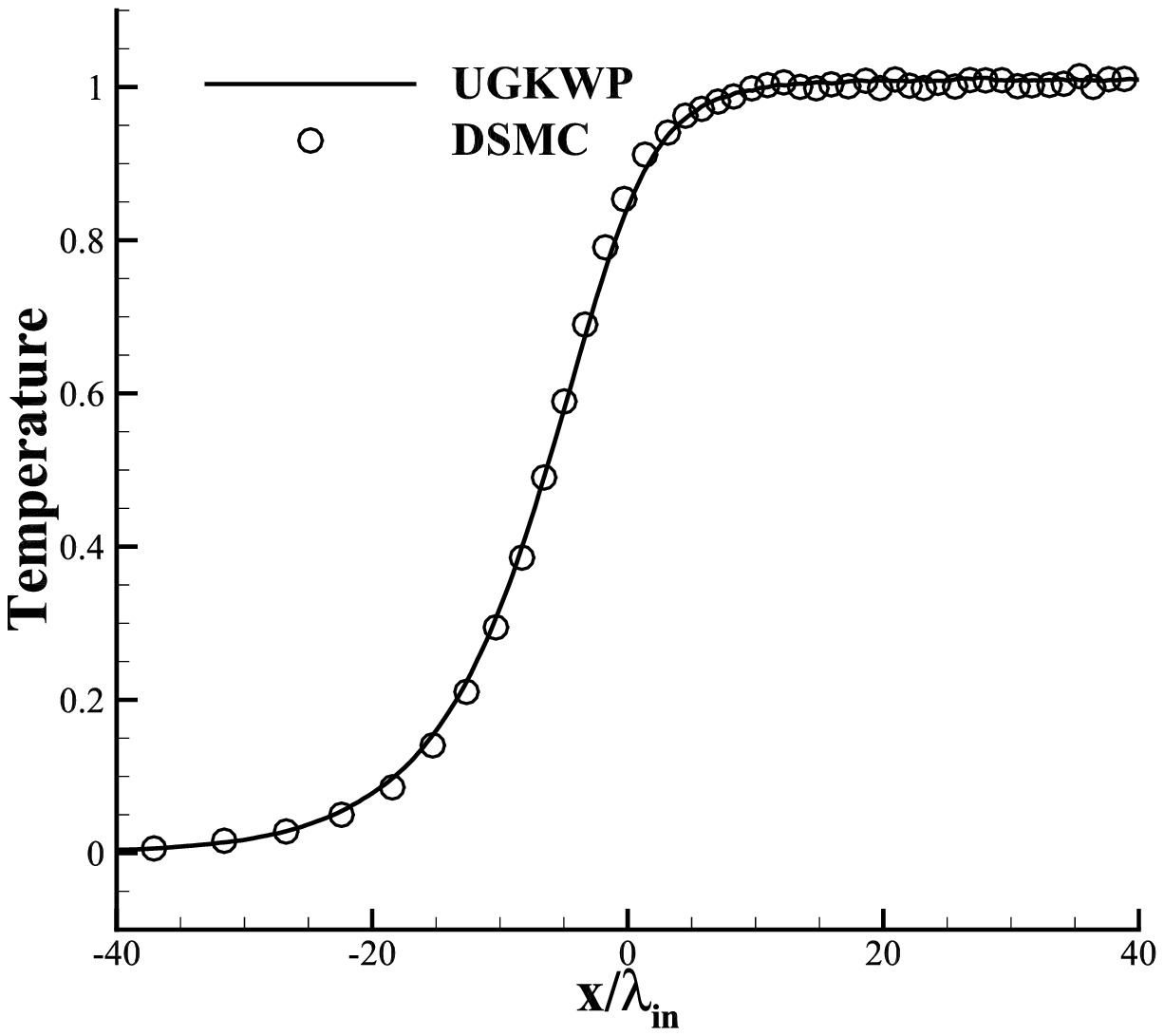}
    	}
    \subfigure[]{
			\includegraphics[width=0.22 \textwidth]{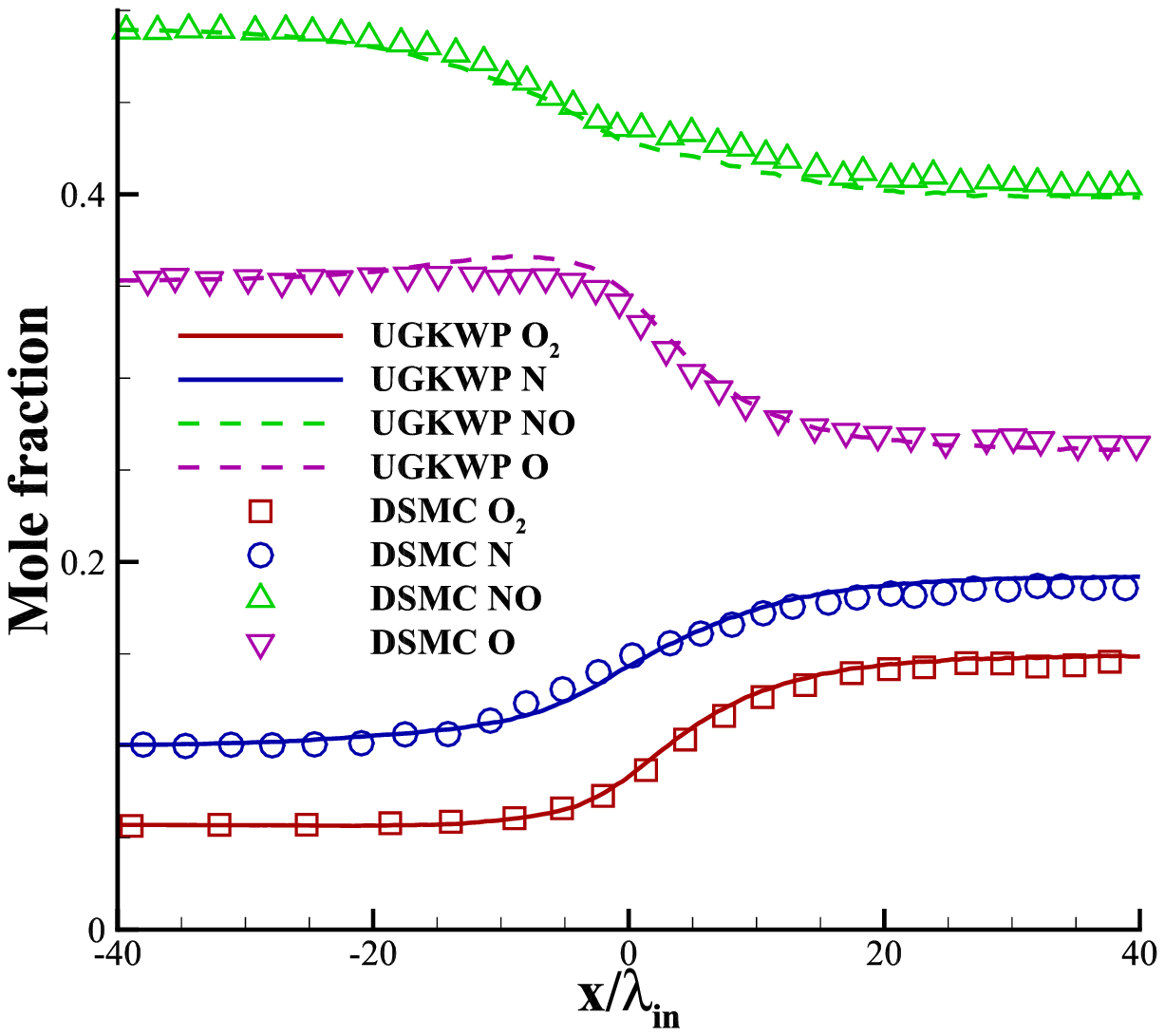}
		}
	\caption{\label{shock-ma3} Shock structure results at ${\rm{Ma}}=3$: (a) Gas mixture density profile, (b) gas mixture velocity profile, (c) gas mixture temperature profile, (d) species mole fraction profile.}
\end{figure}

\begin{figure}[H]
	\centering
	\subfigure[]{
			\includegraphics[width=0.22 \textwidth]{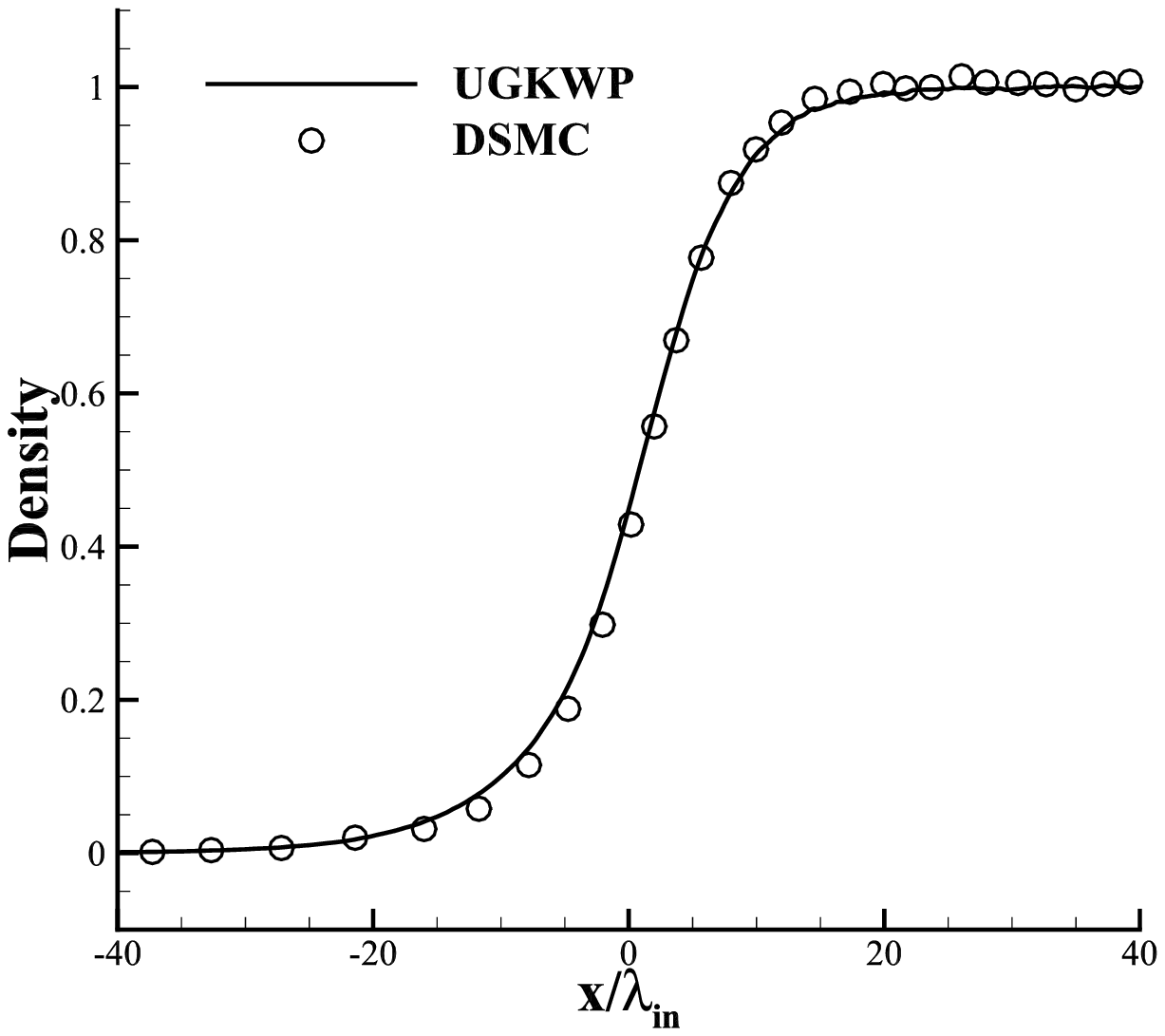}
		}
    \subfigure[]{
    		\includegraphics[width=0.22 \textwidth]{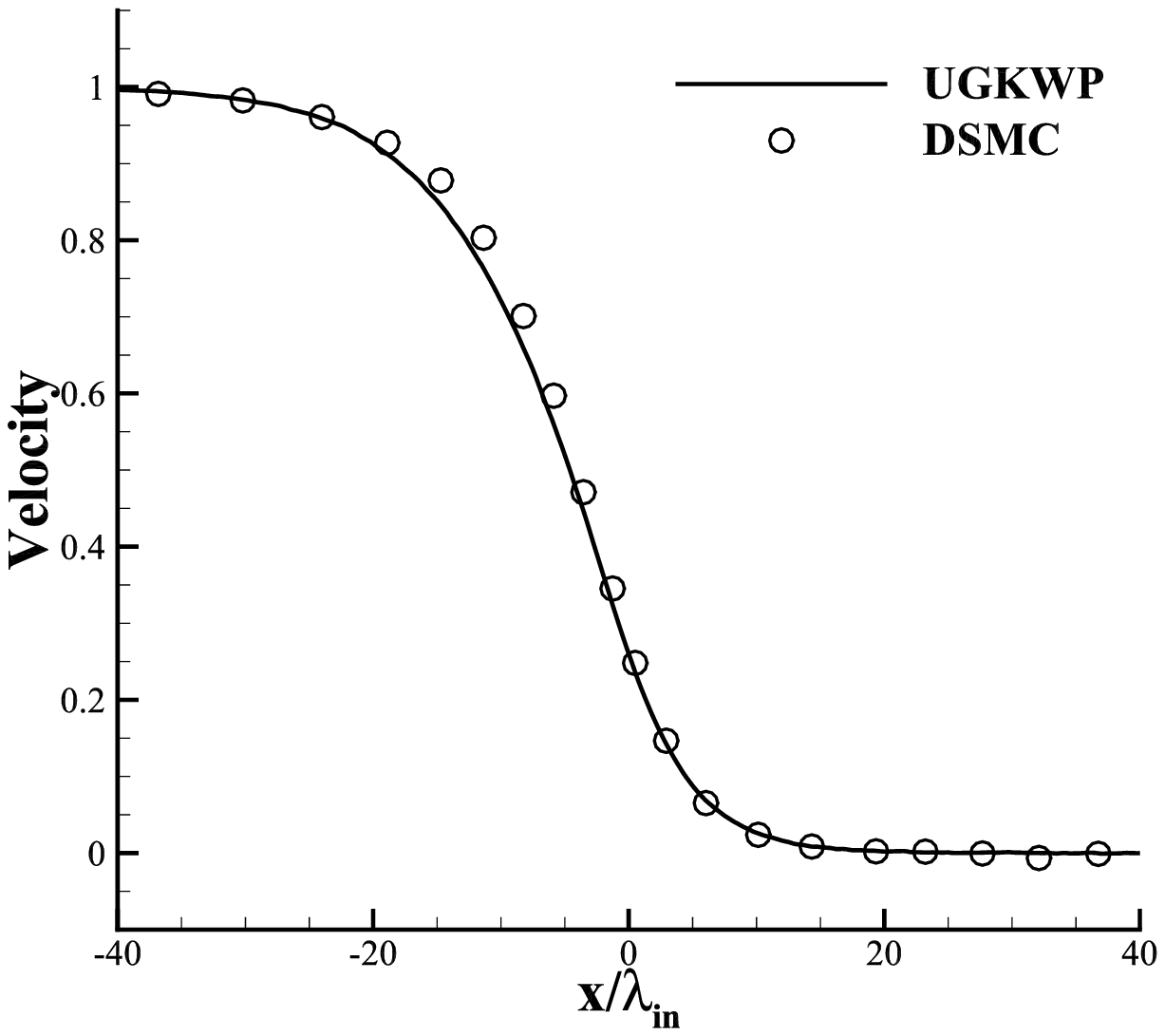}
    	}
    \subfigure[]{
    		\includegraphics[width=0.22 \textwidth]{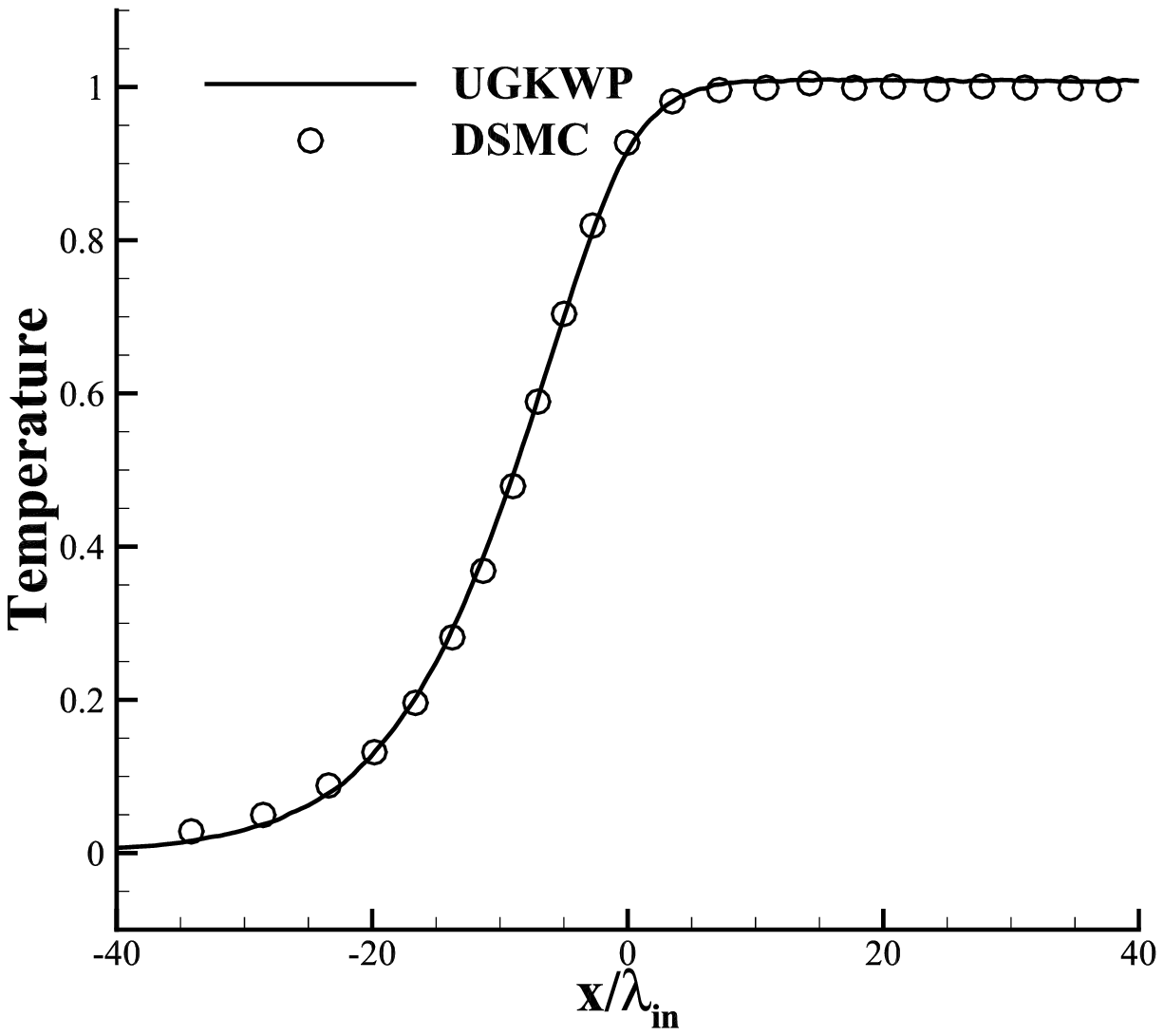}
    	}
    \subfigure[]{
			\includegraphics[width=0.22 \textwidth]{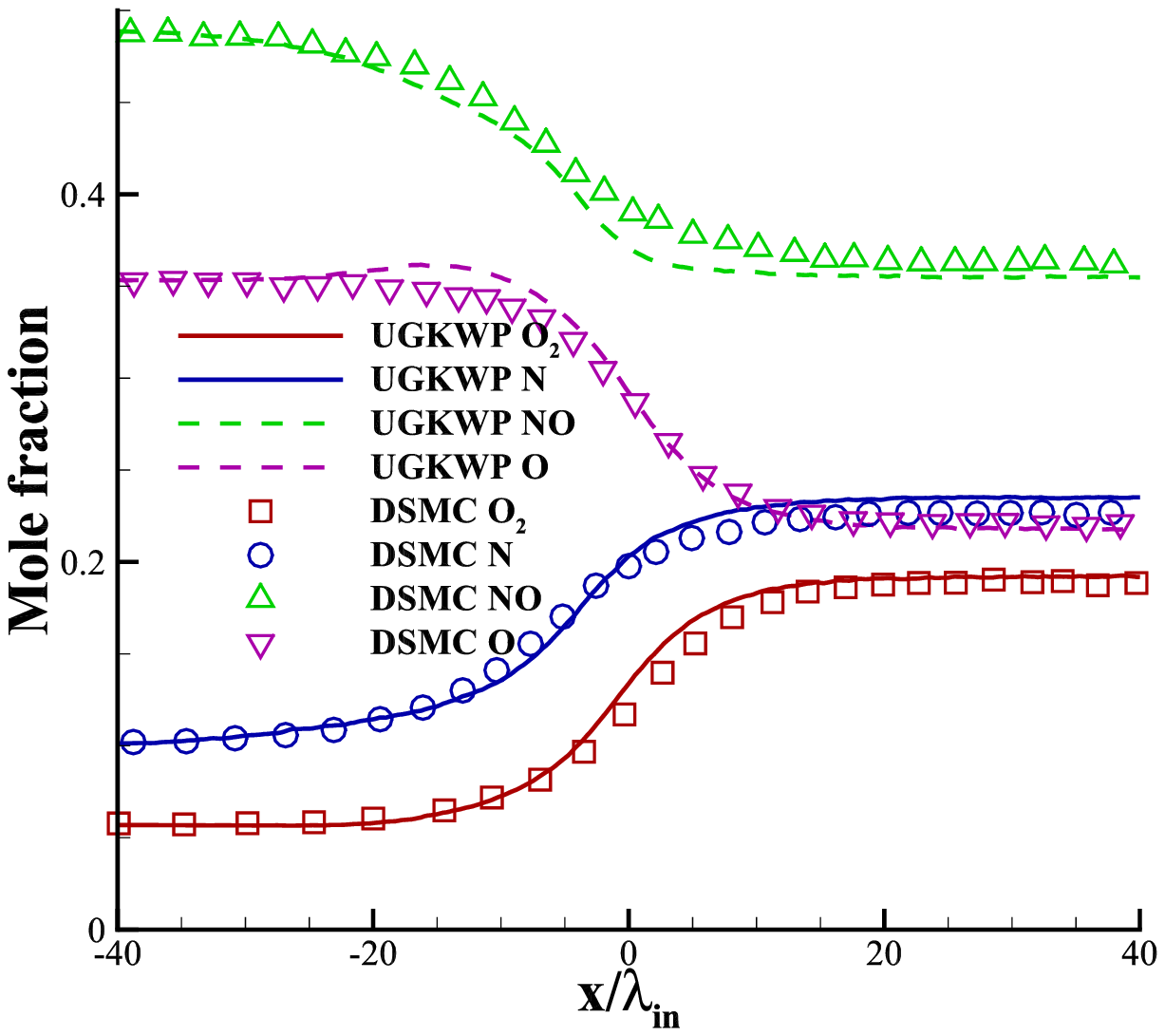}
		}
	\caption{\label{shock-ma5} Shock structure results at ${\rm{Ma}}=5$: (a) Gas mixture density profile, (b) gas mixture velocity profile, (c) gas mixture temperature profile, (d) species mole fraction profile.}
\end{figure}

\begin{table}[h]
\caption{\text{Upstream and downstream parameters of shock structure at ${\rm{Ma}}=5$}}\label{tab-ma5}
\centering
\begin{tabular}{*{9}{c}}
\toprule
    Gas &$\rho^-\left({\rm{kg/m^3}}\right)$ &$\rho^+\left({\rm{kg/m^3}}\right)$ &$\chi^-$ &$\chi^+$ &$U^-\left({\rm{m/s}}\right)$ &$U^+\left({\rm{m/s}}\right)$ &$T^-\left({\rm{K}}\right)$ &$T^+\left({\rm{K}}\right)$  \\
\midrule
    $O_2$ &$3.0278\times 10^{-7}$ &$3.5385\times 10^{-6}$ &$0.0570$ &$0.1918$ &$9154.0$  &$2635.7$  &$6000$  &$50420$ \\
	$N$   &$2.3250\times 10^{-7}$ &$1.8960\times 10^{-6}$ &$0.1000$ &$0.2348$ &$9154.0$  &$2635.7$  &$6000$  &$50420$ \\
    $NO$  &$2.4402\times 10^{-6}$ &$6.1435\times 10^{-6}$ &$0.4900$ &$0.3552$ &$9154.0$  &$2635.7$  &$6000$  &$50420$ \\
    $O$   &$9.3757\times 10^{-7}$ &$2.0128\times 10^{-6}$ &$0.3530$ &$0.2182$ &$9154.0$  &$2635.7$  &$6000$  &$50420$ \\
\bottomrule
\end{tabular}
\end{table}
The reference length is set to be the inflow mean free path calculated by parameters of $O_2$, $\lambda_{in}=0.0135877{\rm{m}}$. The cell size is set to be $\lambda_{in}/4$, and $\lambda_{in}$ is also used as the nondimensionalized spatial coordinate along the horizontal axis. The vertical coordinate is normalized by the upstream and downstream values so that it spans $\left[0,1\right]$. As shown in Fig.~\ref{shock-ma3} and Fig.~\ref{shock-ma5}, results of the present UGKWP match well with those of DSMC.

\subsection{Side jet on a three-dimensional blunt cone}\label{sec:sidejet}
\begin{figure}[H]
	\centering
	\subfigure[]{
			\includegraphics[width=0.5 \textwidth]{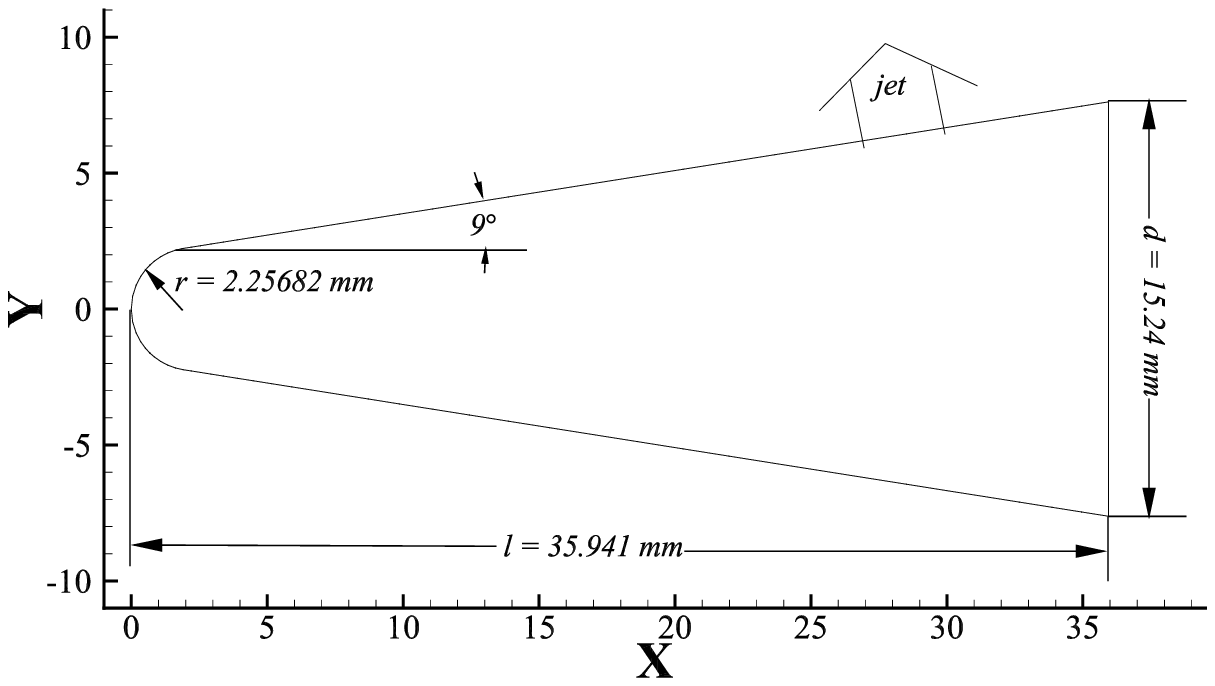}
		}
	\caption{\label{cone-shape} Cone geometric shape (dimensions in millimeters).}
\end{figure}

\begin{figure}[H]
	\centering
	\subfigure[]{
			\includegraphics[width=0.45 \textwidth]{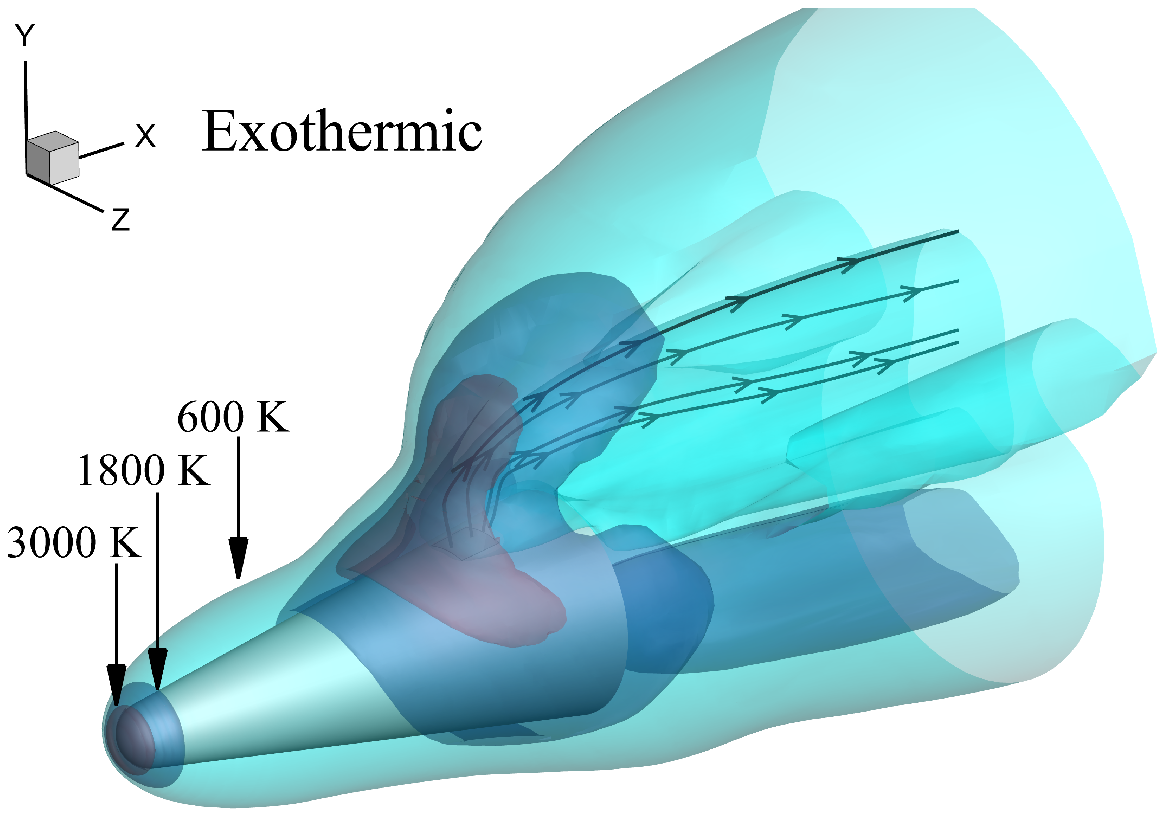}
		}
    \subfigure[]{
    		\includegraphics[width=0.45 \textwidth]{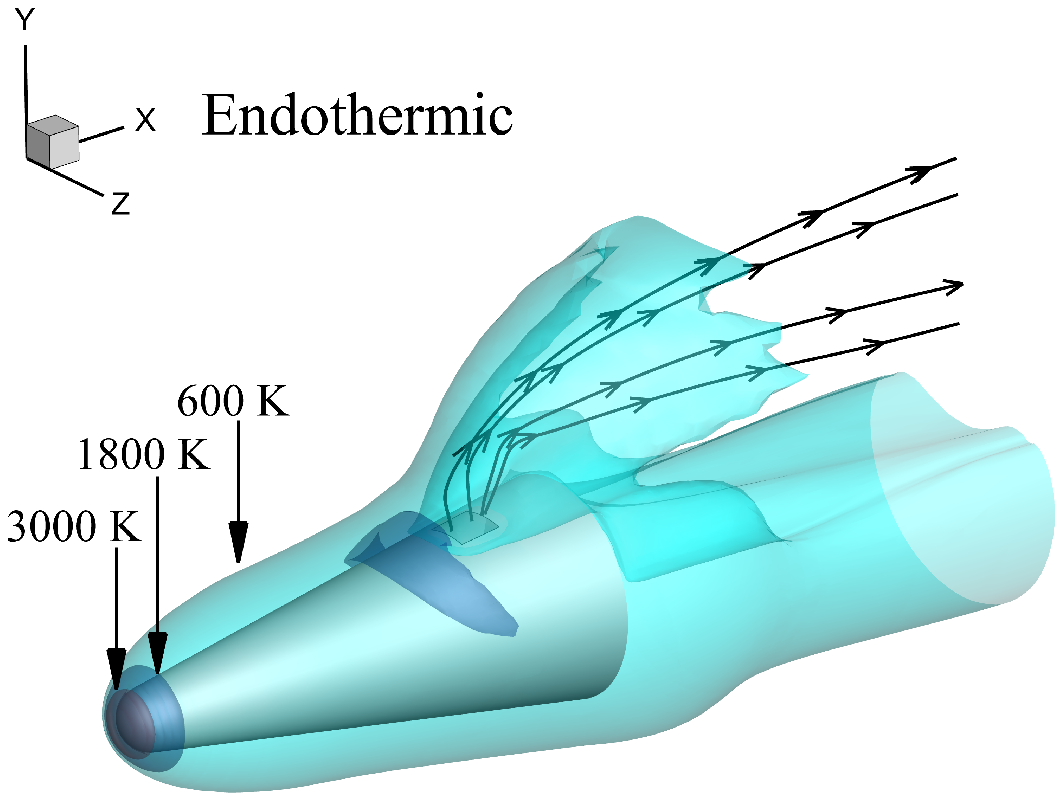}
    	}
	\caption{\label{cone9_tem} Temperature contours of side jet on a three-dimensional blunt cone: (a) Forward exothermic case, (b) forward endothermic case.}
\end{figure}

\begin{figure}[H]
	\centering
	\subfigure[]{
			\includegraphics[width=0.22 \textwidth]{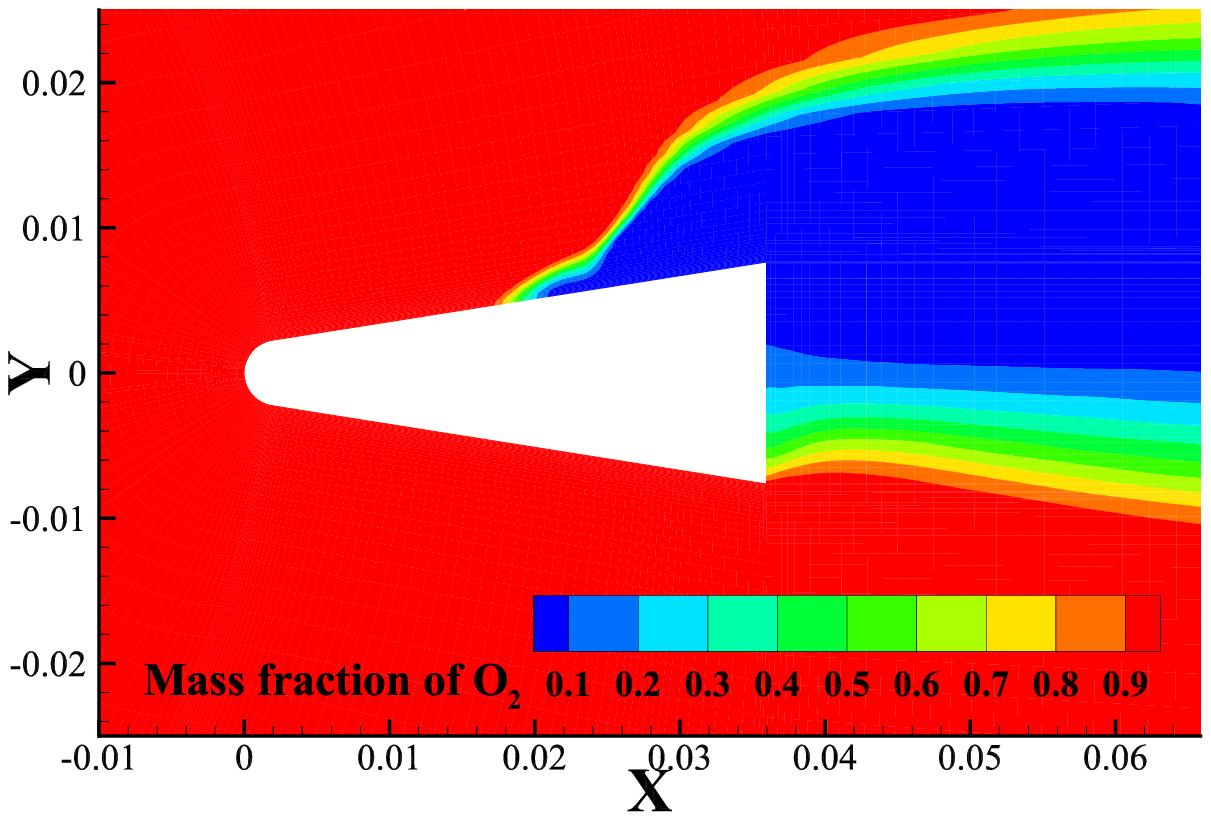}
		}
    \subfigure[]{
    		\includegraphics[width=0.22 \textwidth]{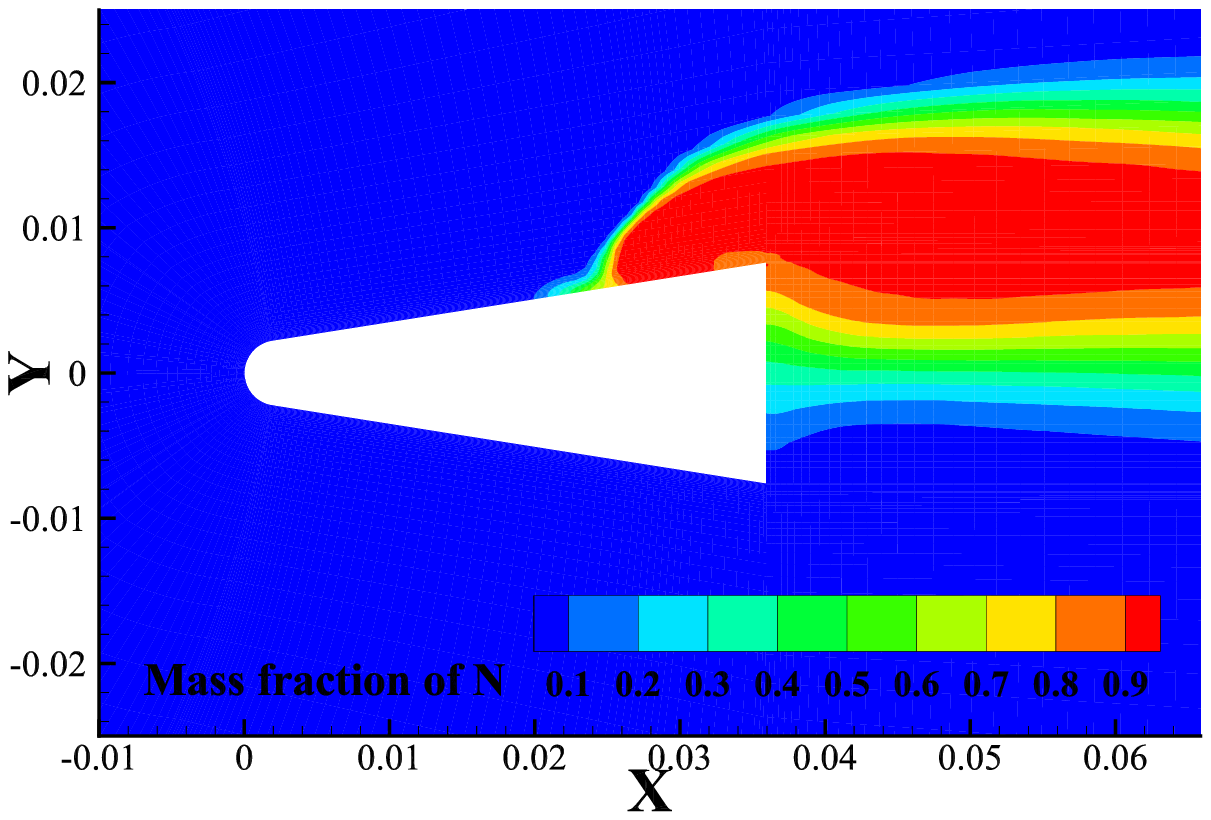}
    	}
    \subfigure[]{
    		\includegraphics[width=0.22 \textwidth]{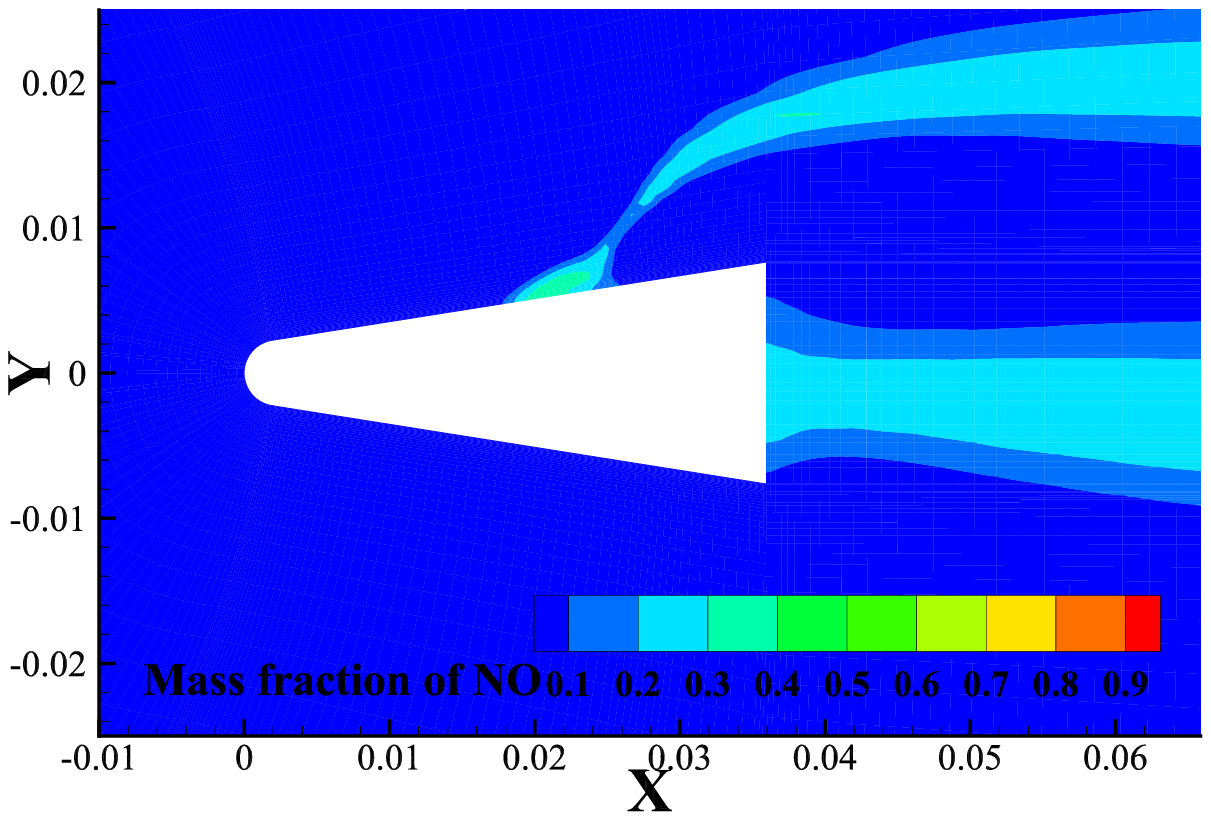}
    	}
    \subfigure[]{
			\includegraphics[width=0.22 \textwidth]{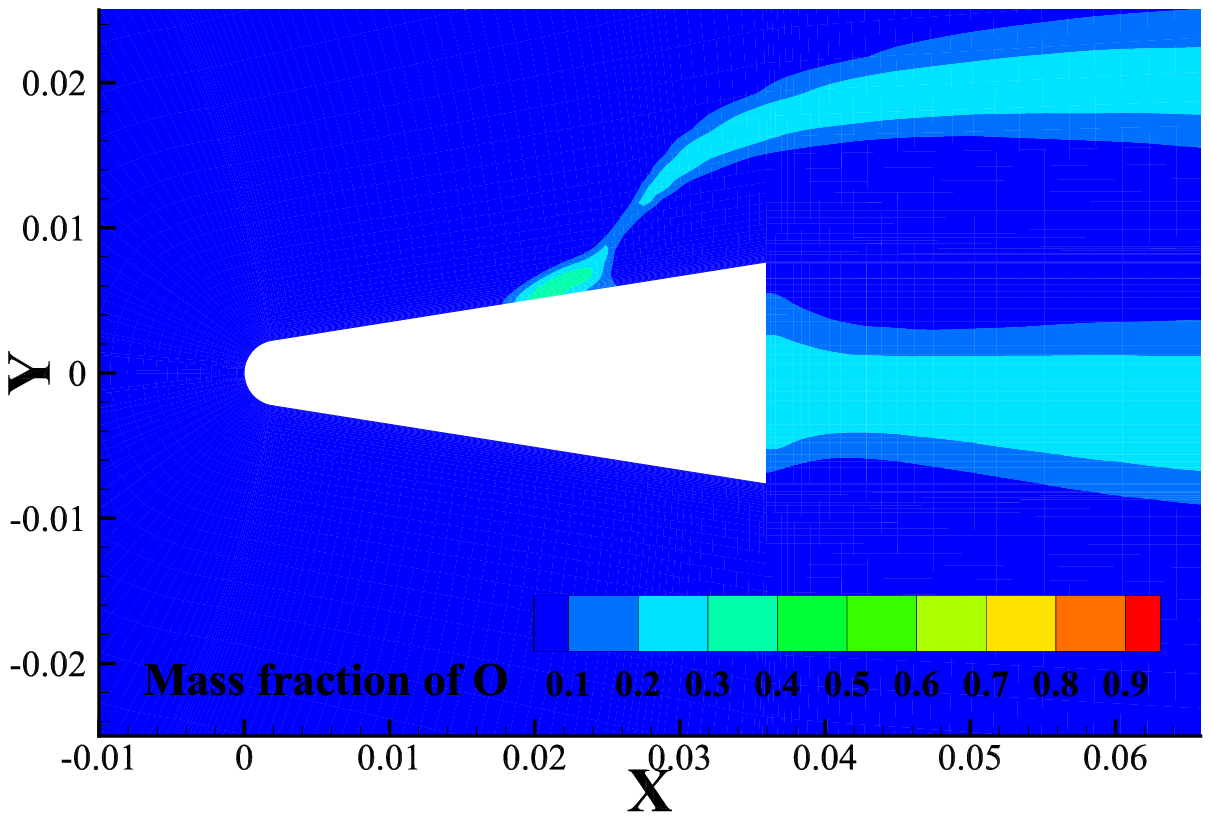}
		}
	\caption{\label{cone9_x_exothermic} Mole fraction contours of side jet on a three-dimensional blunt cone (forward exothermic case): (a) $O_2$, (b) $N$, (c) $NO$, (d) $O$.}
\end{figure}

\begin{figure}[H]
	\centering
	\subfigure[]{
			\includegraphics[width=0.22 \textwidth]{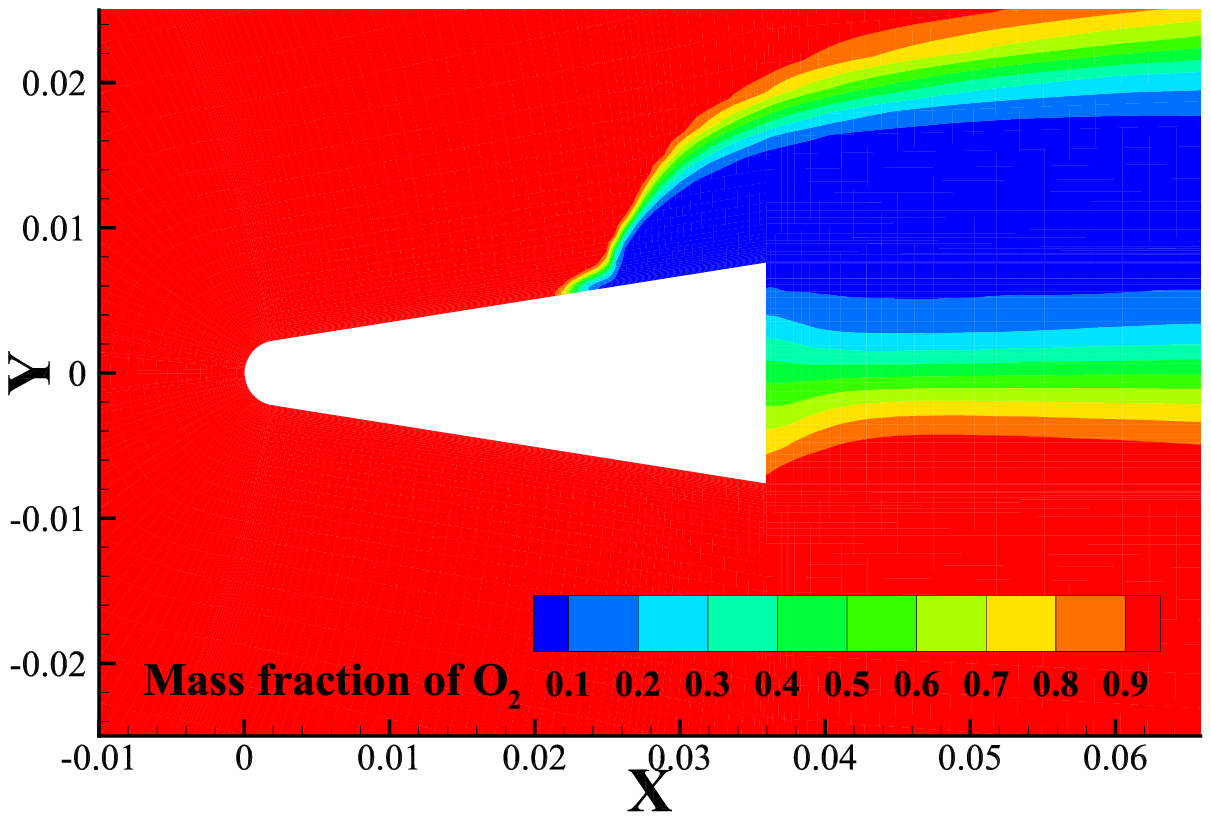}
		}
    \subfigure[]{
    		\includegraphics[width=0.22 \textwidth]{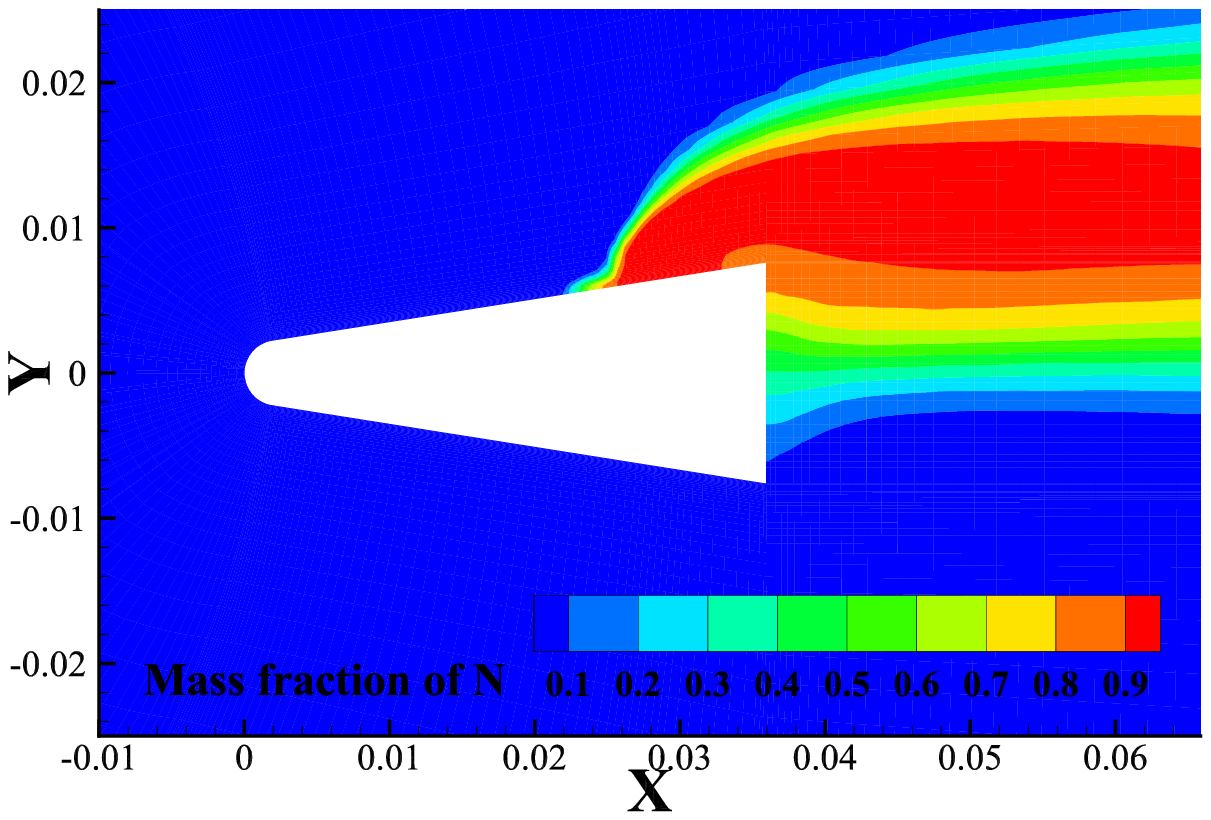}
    	}
    \subfigure[]{
    		\includegraphics[width=0.22 \textwidth]{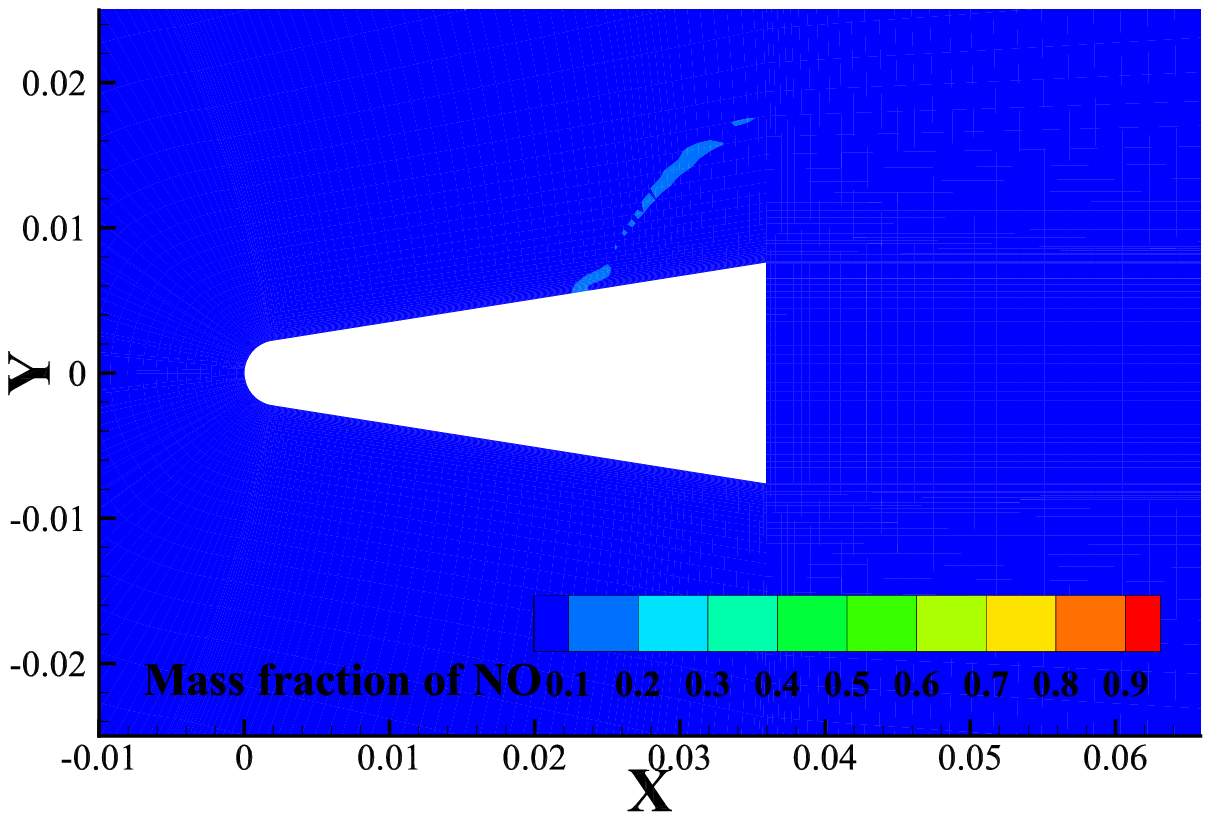}
    	}
    \subfigure[]{
			\includegraphics[width=0.22 \textwidth]{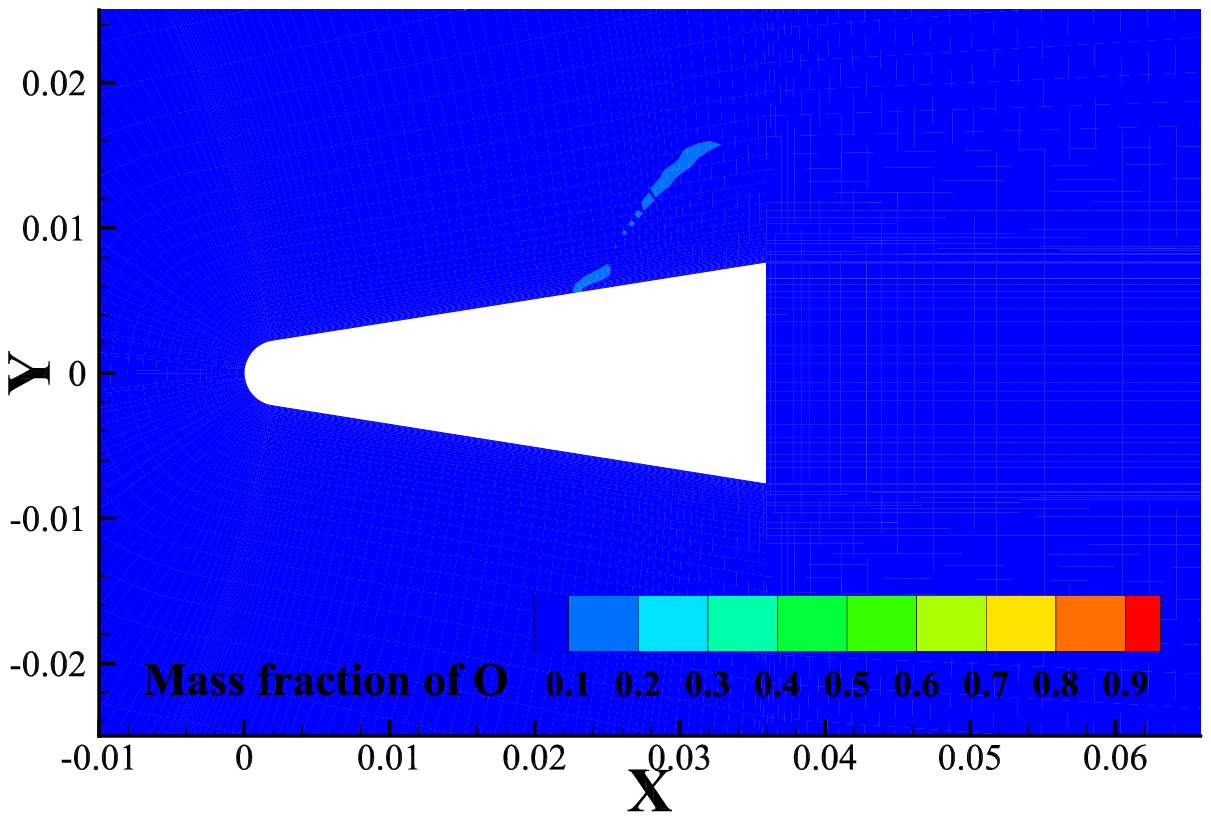}
		}
	\caption{\label{cone9_x_endothermic} Mole fraction contours of side jet on a three-dimensional blunt cone (forward endothermic case): (a) $O_2$, (b) $N$, (c) $NO$, (d) $O$.}
\end{figure}

\begin{figure}[H]
	\centering
	\subfigure[]{
			\includegraphics[width=0.3 \textwidth]{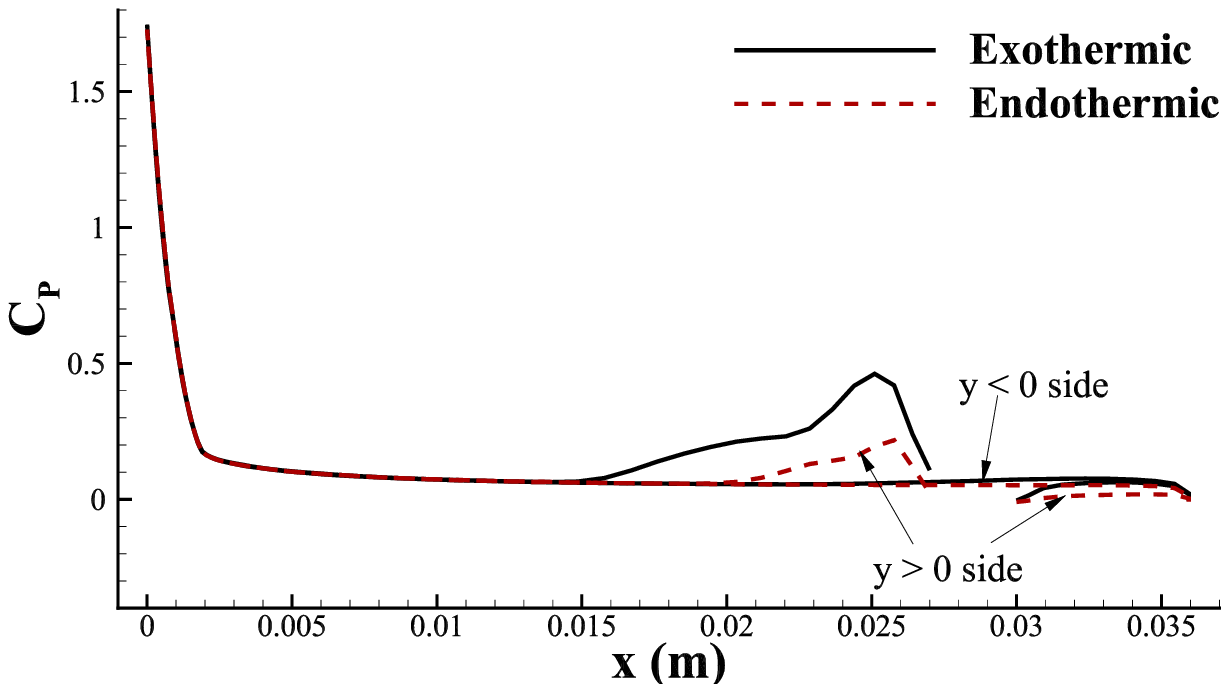}
		}
    \subfigure[]{
			\includegraphics[width=0.3 \textwidth]{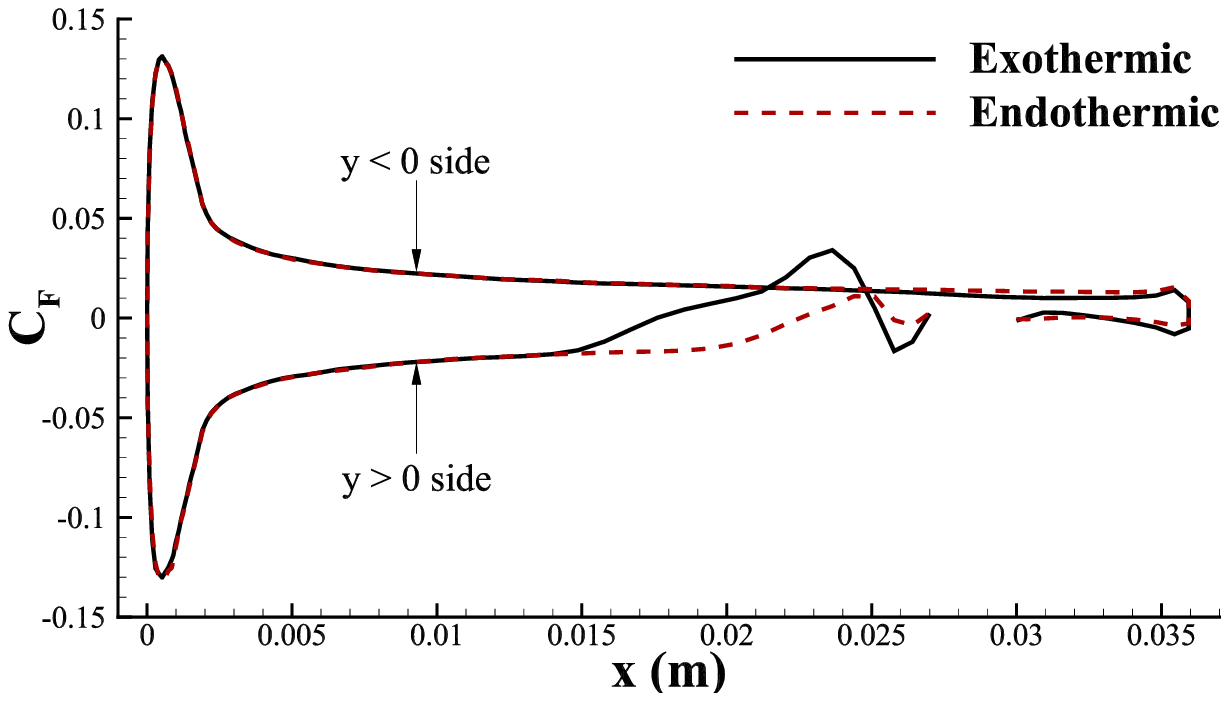}
		}
    \subfigure[]{
			\includegraphics[width=0.3 \textwidth]{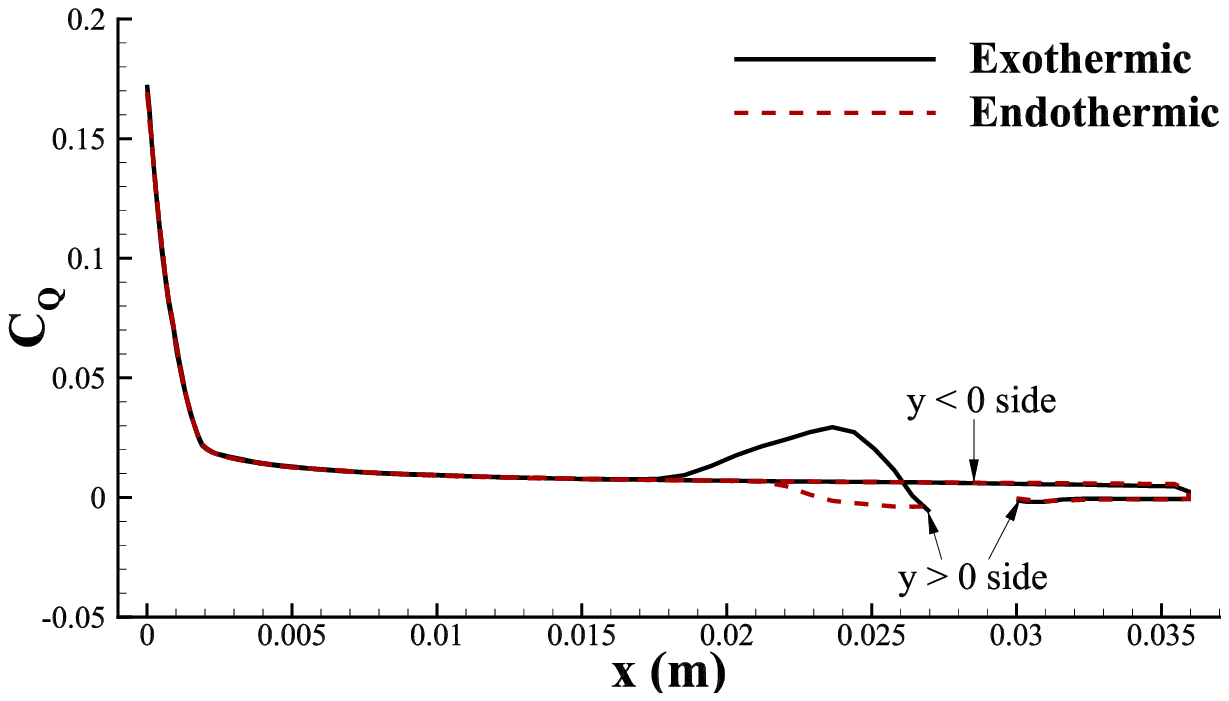}
		}
	\caption{\label{cone9_wall} Wall (a) pressure coefficient, (b) shear stress coefficient, and (c) heat flux coefficient along the symmetry-plane cut of side jet on a three-dimensional blunt cone.}
\end{figure}

In this section, to demonstrate the three-dimensional computational capability of the present code, a side-jet flow is simulated. The geometric shape of the blunt cone in Ref.~\cite{cone-exp} is taken as shown in Fig.~\ref{cone-shape}, which has been widely studied~\cite{cone-dsmc,cone-zr,wp-2nd}. As in Ref.~\cite{wp-2nd}, the farfield parameters are set to be ${\rm{Ma}}_{\infty}=10.15$, ${\rm{Kn}}_{\infty}=0.002$ (calculated by Eq.~\eqref{eq:makn}), whose reference length is the base diameter, and this condition approximately corresponds to an altitude of $70{\rm{km}}$. The farfield gas is set to be pure $O_2$. The farfield temperature is set to be $143.5{\rm{K}}$ and the wall temperature is set to be $800{\rm{K}}$. The square jet orifice extends from $x=27{\rm{mm}}$ to $x=30{\rm{mm}}$ over a circumferential span of $30^{\circ}$. Its direction is perpendicular to the z-axis and forms a $9^{\circ}$ angle with the y-axis. The jet flow is full $N$. Its temperature is the same as farfield, ${\rm{Ma}}=2$, and its density is $200$ times the farfield. For the mesh production, the height of first-layer mesh is set to be $0.05{\rm{mm}}$, and the total cell number is $309240$. In the simulation, both forward exothermic and forward endothermic cases are considered. The $3D$ temperature contours are shown in Fig.~\ref{cone9_tem}, and mole fraction contours on the symmetry plane are shown in Fig.~\ref{cone9_x_exothermic} and Fig.~\ref{cone9_x_endothermic}. In the forward exothermic case, a large high-temperature region forms because the heat released by the reaction substantially raises the local temperature, which in turn further accelerates the reaction rate. Product concentrations near the jet exit and in the wake are accordingly increased. By contrast, in the forward endothermic case the reactions absorb heat, so the reaction rate remains limited and the product mole fractions stay small. The surface pressure coefficients, shear stress coefficients and heat flux coefficients for both cases are also shown in Fig.~\ref{cone9_wall}. Near the jet exit and in the downstream region, $C_P$, $C_F$ and $C_Q$ are strongly affected by whether the forward reaction is endothermic or exothermic, whereas the cone nose remains almost unaffected. For $C_P$ and $C_F$, because the exothermic reaction is more intense, the magnitude of the change is larger. For $C_Q$, since the local temperature is strongly affected by the reaction, there is a region where the sign reverses.

\section{Conclusions}\label{sec:conclusion}
This paper establishes a framework for extending the multiscale UGKWP method to gas mixture flows with an elementary chemical reaction. Chemical reaction effects are first imposed on the macroscopic variables and then reflected in the wave-particle decomposition by excluding free transport particles from the chemical reaction, without requiring a fully resolved reactive kinetic model for the entire distribution function. For the multispecies effect, the kinetic model in Refs.~\cite{groppi-2,todorova-2} is further extended to a larger number of species ($C>2$). The accuracy of the present UGKWP method is demonstrated by numerical tests against DSMC for hypersonic cylinder flows and shock structures over a wide range of freestream ${\rm{Kn}}$ and considering all of chemically inert, forward exothermic, forward endothermic, and $\Delta E=0{\rm{J}}$ reactions. A three-dimensional hypersonic side jet flow over a blunt cone is further simulated to illustrate the capability of the code for three-dimensional reacting simulations. Overall, the comparisons confirm that the present method captures the coupled continuum-rarefied multiscale effect, multispecies effect, and chemical reaction effect with satisfactory fidelity. An important feature of the present approach is that stochastic particles already carry rich nonequilibrium information. On one hand, this property provides a natural pathway toward more elaborate chemical reaction models, in which reaction rates may depend on microscopic particle states beyond classical Arrhenius closures based solely on macroscopic temperature. On the other hand, effects of chemical reaction on the nonequilibrium distribution function can be treated in a wave-particle decoupled approach. Free transport particles need not participate in reactions, analogous to DSMC, and under more complex conditions, reactions may alter the free transport time of particles. Meanwhile, the dominant chemical influence is applied to the wave part that accounts for multiple collisions. For future work, diatomic thermal nonequilibrium in the rotational and vibrational modes should be incorporated, so that internal energy relaxation can be coupled with chemical reaction in a more physically consistent manner. More complicated chemical reaction source terms will also be developed within this wave-particle framework.

\section*{Appendix A: Extension of an advanced kinetic multispecies model for $C>2$ case}
This model is an approximated extension of the binary-species model in Refs.~\cite{groppi-2,todorova-2} to larger species number. The only differences from binary-species model in Ref.~\cite{todorova-2} lie on the formulas to calculate $\hat{\boldsymbol{U}}_{0}$ and $\vartheta_{\alpha}$ in Eq.~\eqref{eq:groppi}. In the binary-species model, they are set to be $\hat{\boldsymbol{U}}_{0}=\boldsymbol{U}_{0}$ and $\vartheta_{\alpha}=\frac{5m_0}{3A^{\ast}\sum\limits_{\beta=1}^2{m_{\beta}}}$, while in the current model Eq.~\eqref{eq:u2} and Eq.~\eqref{eq:theta} are used. Because these are the sole two distinctions, the Chapman-Enskog expansion results in Ref.~\cite{todorova-2} can be directly used. In the continuum region, the momentum equation and energy equation of the current model are the same as the binary-species model, so that the viscosity coefficient and heat conduction coefficient of gas mixture can also recover the NS equations, and our work concentrates on the mass equation and diffusion effect. As in Ref.~\cite{todorova-2}, through Chapman-Enskog expansion, the mass equation is,
\begin{equation}
\frac{\partial\rho_{\alpha}}{\partial t}+\frac{\partial}{\partial x}\left(\rho_{\alpha}U_0+F_{\alpha}^{{\rm{d-CE}}}\right)=0,
\nonumber
\end{equation}
where we consider only the one-dimensional case without loss of generality, and,
\begin{equation}\label{eq:appendix1}
F_{\alpha}^{{\rm{d-CE}}}=-\frac{\mu_0}{\vartheta_{\alpha}n_0}\frac{\partial n_{\alpha}}{\partial x}+\frac{\mu_0}{\vartheta_{\alpha}n_0}\frac{\rho_{\alpha}}{\rho_0}\frac{\partial n_0}{\partial x}-\frac{\mu_0}{\vartheta_{\alpha}n_0}\frac{\rho_{\alpha}}{T_0}\left(\frac{1}{m_{\alpha}}-\frac{n_0}{\rho_0}\right)\frac{\partial T_0}{\partial x},
\end{equation}
is the mass diffusion flux. Meanwhile, as in Ref.~\cite{birdbook}, the NS-scale mass diffusion flux is,
\begin{equation}\label{eq:appendix2}
F_{\alpha}^{{\rm{d-NS}}}=\frac{n_0^2}{\rho_0}\sum\limits_{\beta\neq\alpha}^{C}m_{\alpha}m_{\beta}\mathcal{D}_{\alpha\beta}\zeta_{\beta},
\end{equation}
where the Soret effect is not considered here, $\mathcal{D}_{\alpha\beta}$ is the diffusion coefficient between species $\alpha$ and $\beta$, and,
\begin{equation}\label{eq:appendix3}
\zeta_{\beta}=\frac{\partial}{\partial x}\left(\frac{n_{\beta}}{n_0}\right)+\left(\frac{n_{\beta}}{n_0}-\frac{n_{\beta}m_{\beta}}{\rho_{0}}\right)\frac{\partial\ln p_0}{\partial x},
\end{equation}
where the external force is not considered. In the binary-species case, by setting $\hat{\boldsymbol{U}}_{0}=\boldsymbol{U}_{0}$ and $\vartheta_{\alpha}=\frac{5m_0}{3A^{\ast}\sum\limits_{\beta=1}^2{m_{\beta}}}$, $F_{\alpha}^{{\rm{d-CE}}}=F_{\alpha}^{{\rm{d-NS}}}$ can be derived, so that Chapman-Enskog expansion result is consistent with the NS-scale equation. However, for a larger number of species, the situation becomes far more complex, and some approximations have to be introduced. First, to factor the gas mixture viscosity out of the summation, we assume that the viscosity between any two species equals the gas mixture viscosity, that the binary $\mathcal{D}_{\alpha\beta}$ is used to approximate the multispecies case which is a first-order approximation according to Ref.~\cite{birdbook}, and that $A^{\ast}_{\alpha\beta}$ takes the same value for any pair of species, with $A^{\ast}=1.11$ given in Ref.~\cite{todorova-2}. Under these assumptions, we can employ the relation between viscosity coefficient and diffusion coefficient for binary mixture~\cite{birdbook} as follows,
\begin{equation}\label{eq:appendix4}
\mu_{0}=\frac{5}{3}\frac{m_{\alpha}m_{\beta}}{m_{\alpha}+m_{\beta}}\frac{n_0\mathcal{D}_{\alpha\beta}}{A^{\ast}}.
\end{equation}
By substituting Eq.~\eqref{eq:appendix4} into Eq.~\eqref{eq:appendix2}, we have,
\begin{equation}\label{eq:appendix5}
F_{\alpha}^{{\rm{d-NS}}}\approx\frac{3\mu_0A^{\ast}}{5}\frac{n_0}{\rho_0}\sum\limits_{\beta\neq\alpha}^{C}\left(m_{\beta}+m_{\alpha}\right)\zeta_{\beta}.
\end{equation}
Combining Eq.~\eqref{eq:appendix5} with Eq.~\eqref{eq:appendix3}, we have,
\begin{equation}\label{eq:appendix6}
\begin{aligned}
F_{\alpha}^{{\rm{d-NS}}}\approx&\frac{3\mu_0A^{\ast}}{5}\frac{n_0}{\rho_0}\Bigg[\frac{\partial\rho_{\alpha}}{\partial x}\left(-\frac{2}{n_0}\right)+\frac{\partial n_0}{\partial x}\left(\frac{m_{\alpha}\rho_{\alpha}}{\rho_0 n_0}-\sum\limits_{\beta\neq\alpha}^{C}\frac{m_{\beta}\rho_{\beta}}{\rho_0 n_0}\right)\\
+&\frac{\partial\rho_0}{\partial x}\left(\frac{1}{n_0}\right)+\frac{\partial T_0}{\partial x}\left(-\frac{2\rho_{\alpha}}{T_0n_0}+\frac{m_{\alpha}\rho_{\alpha}}{T_0\rho_0}+\frac{m_0}{T_0}-\sum\limits_{\beta\neq\alpha}^{C}\frac{m_{\beta}\rho_{\beta}}{\rho_0T_0}\right)
\Bigg].
\end{aligned}
\end{equation}
When the mass ratio is not extreme, we employ an expansion in small deviations and neglect second-order terms as,
\begin{equation}
\begin{aligned}
&\frac{\sum\limits_{\beta=1}^Cm_{\beta}\rho_{\beta}}{\sum\limits_{\beta=1}^{C}\rho_{\beta}}=\frac{\sum\limits_{\beta=1}^C\left(m_{\beta}^2n_0\chi_{\beta}\right)}{\sum\limits_{\beta=1}^C\left(n_{\beta}m_{\beta}\right)}
=n_0\frac{\sum\limits_{\beta=1}^C\left(\chi_{\beta}m_{\beta}^2\right)}{\sum\limits_{\beta=1}^C\left(n_{\beta}m_{\beta}\right)}\\
&=n_{0}\frac{\sum\limits_{\beta=1}^C\left\{\left[\frac{m_0}{C}+\left(\chi_{\beta}m_{\beta}-\frac{m_0}{C}\right)\right]\left[\sum\limits_{\gamma=1}^C\frac{m_{\gamma}}{C}+\left(m_{\beta}-\sum\limits_{\gamma=1}^C\frac{m_{\gamma}}{C}\right)\right]\right\}}
{\sum\limits_{\beta=1}^C\left\{\left[\frac{n_0}{C}+\left(n_{\beta}-\frac{n_0}{C}\right)\right]\left[\sum\limits_{\gamma=1}^C\frac{m_{\gamma}}{C}+\left(m_{\beta}-\sum\limits_{\gamma=1}^C\frac{m_{\gamma}}{C}\right)\right]\right\}}\\
&=n_0\frac{\sum\limits_{\beta=1}^C\left[\frac{m_0}{C}\sum\limits_{\gamma=1}^C\frac{m_{\gamma}}{C}
+\underbrace{\frac{m_0}{C}\left(m_{\beta}-\sum\limits_{\gamma=1}^C\frac{m_{\gamma}}{C}\right)}_{\text{vanishes upon summation}}
+\underbrace{\left(\chi_{\beta}m_{\beta}-\frac{m_0}{C}\right)\sum\limits_{\gamma=1}^C\frac{m_{\gamma}}{C}}_{\text{vanishes upon summation}}
+\underbrace{\left(\chi_{\beta}m_{\beta}-\frac{m_0}{C}\right)\left(m_{\beta}-\sum\limits_{\gamma=1}^C\frac{m_{\gamma}}{C}\right)}_{\text{2nd-order term}}\right]}
{\sum\limits_{\beta=1}^C\left[\frac{n_0}{C}\sum\limits_{\gamma=1}^C\frac{m_{\gamma}}{C}
+\underbrace{\frac{n_0}{C}\left(m_{\beta}-\sum\limits_{\gamma=1}^C\frac{m_{\gamma}}{C}\right)}_{\text{vanishes upon summation}}
+\underbrace{\left(n_{\beta}-\frac{n_0}{C}\right)\sum\limits_{\gamma=1}^C\frac{m_{\gamma}}{C}}_{\text{vanishes upon summation}}
+\underbrace{\left(n_{\beta}-\frac{n_0}{C}\right)\left(m_{\beta}-\sum\limits_{\gamma=1}^C\frac{m_{\gamma}}{C}\right)}_{\text{2nd-order term}}\right]}\\
&\approx n_0\frac{\sum\limits_{\beta=1}^C\left[\frac{m_0}{C}\left(\sum\limits_{\gamma=1}^C\frac{m_{\gamma}}{C}\right)\right]}
{\sum\limits_{\beta=1}^C\left[\frac{n_0}{C}\left(\sum\limits_{\gamma=1}^C\frac{m_{\gamma}}{C}\right)\right]}=n_0\frac{m_0}{n_0}=m_0,
\nonumber
\end{aligned}
\end{equation}
where $\gamma$ is a third label of species, to differentiate from $\alpha$ and $\beta$. Consequently,
\begin{equation}
\sum\limits_{\beta\neq\alpha}^{C}\frac{m_{\beta}\rho_{\beta}}{\rho_0n_0}\approx\frac{m_0}{n_0}-\frac{m_{\alpha}\rho_{\alpha}}{\rho_0n_0},
\qquad
\sum\limits_{\beta\neq\alpha}^{C}\frac{m_{\beta}\rho_{\beta}}{\rho_0T_0}\approx\frac{m_0}{T_0}-\frac{m_{\alpha}\rho_{\alpha}}{\rho_0T_0},
\nonumber
\end{equation}
which can be substituted into Eq.~\eqref{eq:appendix6}, then it can be derived that,
\begin{equation}
\begin{aligned}
F_{\alpha}^{{\rm{d-NS}}}\approx&\frac{3\mu_{0}A^{\ast}}{5}\frac{n_0}{\rho_0}\left[\frac{\partial\rho_{\alpha}}{\partial x}\left(-\frac{2}{n_0}\right)+\frac{\partial n_0}{\partial x}\left(\frac{2m_{\alpha}\rho_{\alpha}}{\rho_0n_0}\right)+\frac{\partial m_0}{\partial x}+\frac{\partial T_{0}}{\partial x}\left(-\frac{2\rho_{\alpha}}{T_0n_0}+\frac{2m_{\alpha}\rho_{\alpha}}{T_0\rho_0}\right)\right].
\nonumber
\end{aligned}
\end{equation}
Upon further neglecting the spatial derivatives of $m_0$, the final approximate form of $F_{\alpha}^{{\rm{d-NS}}}$ is given by,
\begin{equation}\label{eq:appendix7}
F_{\alpha}^{{\rm{d-NS}}}\approx\frac{3\mu_{0}A^{\ast}}{5}\frac{n_0}{\rho_0}\left[\frac{\partial\rho_{\alpha}}{\partial x}\left(-\frac{2}{n_0}\right)+\frac{\partial n_0}{\partial x}\left(\frac{2m_{\alpha}\rho_{\alpha}}{\rho_0n_0}\right)+\frac{\partial T_{0}}{\partial x}\left(-\frac{2\rho_{\alpha}}{T_0n_0}+\frac{2m_{\alpha}\rho_{\alpha}}{T_0\rho_0}\right)\right].
\end{equation}
Comparing Eq.~\eqref{eq:appendix7} with Eq.~\eqref{eq:appendix1}, it is found that Eq.~\eqref{eq:theta} is the condition for $F_{\alpha}^{{\rm{d-NS}}}$ and $F_{\alpha}^{{\rm{d-CE}}}$ to be consistent. Meanwhile, to satisfy the momentum conservation of righthand-side relaxation term,
\begin{equation}
\begin{aligned}
&\sum\limits_{\alpha=1}^{C}\int_{\mathbb{R}^3} \boldsymbol{u}\frac{g_{\alpha}-f_{\alpha}}{\tau_0} {\rm d}\boldsymbol{u}
=\frac{1}{\tau_0}\sum\limits_{\alpha=1}^{C}\rho_{\alpha}\left(\tilde{\boldsymbol{U}}_{\alpha}-\boldsymbol{U}_{\alpha}\right)\\
=&\frac{1}{\tau_0}\sum\limits_{\alpha=1}^{C}\rho_{\alpha}\vartheta_{\alpha}\left(\hat{\boldsymbol{U}}_0-\boldsymbol{U}_{\alpha}\right)
=\frac{1}{\tau_0}\sum\limits_{\alpha=1}^{C}\frac{5\rho_{\alpha}}{6A^{\ast}}\frac{m_0}{m_{\alpha}}\left(\hat{\boldsymbol{U}}_0-\boldsymbol{U}_{\alpha}\right)\\
=&\frac{5m_0}{6A^{\ast}\tau_0}\sum\limits_{\alpha=1}^{C}n_{\alpha}\left(\hat{\boldsymbol{U}}_0-\boldsymbol{U}_{\alpha}\right)
=\frac{5m_0}{6A^{\ast}\tau_0}\left(n_0\hat{\boldsymbol{U}}_0-\sum\limits_{\alpha=1}^{C}n_{\alpha}\boldsymbol{U}_{\alpha}\right)=\boldsymbol{0},
\nonumber
\end{aligned}
\end{equation}
Eq.~\eqref{eq:u2} is the solution of $\hat{\boldsymbol{U}}_0$. It is also an approximation of $\boldsymbol{U}_0$ when the mass ratio is not extreme, as follows,
\begin{equation}
\begin{aligned}
&\boldsymbol{U}_0=\frac{\sum\limits_{\beta=1}^Cm_{\beta}n_{\beta}\boldsymbol{U}_{\beta}}{\sum\limits_{\beta=1}^Cm_{\beta}n_{\beta}}\\
=&\frac{\sum\limits_{\beta=1}^{C}\left\{\left[\sum\limits_{\gamma=1}^C\frac{m_{\gamma}}{C}+\left(m_{\beta}-\sum\limits_{\gamma=1}^C\frac{m_{\gamma}}{C}\right)\right]
\left[\sum\limits_{\gamma=1}^C\frac{n_{\gamma}\boldsymbol{U}_{\gamma}}{C}+\left(n_{\beta}\boldsymbol{U}_{\beta}-\sum\limits_{\gamma=1}^C\frac{n_{\gamma}\boldsymbol{U}_{\gamma}}{C}\right)\right]\right\}}
{\sum\limits_{\beta=1}^{C}\left\{\left[\sum\limits_{\gamma=1}^C\frac{m_{\gamma}}{C}+\left(m_{\beta}-\sum\limits_{\gamma=1}^C\frac{m_{\gamma}}{C}\right)\right]
\left[\frac{n_0}{C}+\left(n_{\beta}-\frac{n_0}{C}\right)\right]\right\}}\\
\approx&\frac{\sum\limits_{\beta=1}^{C}\left[\left(\sum\limits_{\gamma=1}^C\frac{m_{\gamma}}{C}\right)\left(\sum\limits_{\gamma=1}^C\frac{n_{\gamma}\boldsymbol{U}_{\gamma}}{C}\right)\right]}
{\sum\limits_{\beta=1}^{C}\left[\left(\sum\limits_{\gamma=1}^C\frac{m_{\gamma}}{C}\right)\frac{n_0}{C}\right]}=\frac{\sum\limits_{\gamma=1}^{C}n_{\gamma}\boldsymbol{U}_{\gamma}}{n_0}=\hat{\boldsymbol{U}}_0.
\nonumber
\end{aligned}
\end{equation}

The derivation of multispecies kinetic models remains a frontier topic. For single relaxation models, we recommend the recent work in Ref.~\cite{pfeiffermodel}. The core of the present work is to construct a multiscale UGKWP framework for chemical reaction flows, while the multispecies kinetic modeling itself awaits further improvement. The extension in Appendix A involves rather many approximations, while it still performs well in current numerical simulations.

\section*{Acknowledgements}
The current research is supported by the National Key R$\&$D Program of China (Grant No. 2022YFA1004500), the National Natural Science Foundation of China (92371107), and the Hong Kong Research Grants Council (16208324). Helpful discussion with Dr. Z. Pu is gratefully acknowledged.

\section*{Declaration of competing interest}
The authors declare that they have no known competing financial interests or personal relationships that could have appeared to influence the work reported in this paper.

\section*{Data availability}
The data that support the findings of this study are available from the corresponding author upon reasonable request.

\section*{Reference}
\bibliography{wp-elementary-ref}

\end{document}